\documentclass[aps,prd,reprint,superscriptaddress,noeprint,nofootinbib]{revtex4-2}
\usepackage{amsmath,amsfonts}
\usepackage{graphicx}
\usepackage[caption=false]{subfig}
\usepackage[dvipsnames,usenames]{xcolor}
\usepackage{fontawesome5}
\usepackage{aas_macros}

\usepackage[T1]{fontenc}
\newcommand*{\dittostraight}{---\textquotedbl---}

\usepackage{orcidlink}
\hypersetup{colorlinks,citecolor=blue}
\usepackage{nameref}

\newcommand{\0}[1]{{\rm #1}}

\newcommand{\Q}{{\rm QSE}}

\newcommand{\srcnuye}{Q_{\nu,e}}
\newcommand{\srcnumom}{Q_{\nu}^i}
\newcommand{\srcnutau}{Q_{\nu,\tau}}
\newcommand{\srcgravmom}{G^i}
\newcommand{\srcgravtau}{G_{\tau}}
\newcommand{\srcrhineye}{R_{\beta,e}}
\newcommand{\srcrhinetau}{R_{\beta,\tau}}
\newcommand{\srcrhinexn}{R_{n}}
\newcommand{\srcrhinexp}{R_{p}}
\newcommand{\srcrhinexa}{R_{\alpha}}
\newcommand{\srcrhinexh}{R_{h}}
\newcommand{\srcrhinexi}{R_{i}}
\newcommand{\srcrhinerest}{R_{\mathrm{rest}}}
\newcommand{\srcrhineah}{R_A}
\newcommand{\barm}{\bar{m}}
\newcommand{\tilm}{\widetilde{m}}
\newcommand{\erest}{e_{\mathrm{rest}}}
\newcommand{\etherm}{e_{\mathrm{therm}}}
\newcommand{\eint}{e_{\mathrm{int}}}
\newcommand{\DEheat}{\Delta E_{\mathrm{heat}}}

\newcommand{\dd}{\mathrm{d}}
\newcommand{\MeVc}{$\mathrm{MeV}/c^2$}
\newcommand{\MeVk}{$\mathrm{MeV}/k_B$}

\graphicspath{{./}{figures/}}

\begin{document}

\title{$r$-process heating implementation in hydrodynamic simulations with neural networks}


\date{\today}

\author{Oliver Just\,\orcidlink{0000-0002-3126-9913}}
\affiliation{GSI Helmholtzzentrum {f\"ur} Schwerionenforschung, Planckstra{\ss}e 1, D-64291 Darmstadt, Germany}
\affiliation{Astrophysical Big Bang Laboratory, RIKEN Cluster for Pioneering Research, 2-1 Hirosawa, Wako, Saitama 351-0198, Japan}

\author{Zewei Xiong\,\orcidlink{0000-0002-2385-6771}}
\affiliation{GSI Helmholtzzentrum {f\"ur} Schwerionenforschung, Planckstra{\ss}e 1, D-64291 Darmstadt, Germany}

\author{Gabriel Mart\'{i}nez-Pinedo\,\orcidlink{0000-0002-3825-0131}}
\affiliation{GSI Helmholtzzentrum {f\"ur} Schwerionenforschung,
  Planckstra{\ss}e 1, D-64291 Darmstadt, Germany}
\affiliation{Institut {f\"ur} Kernphysik (Theoriezentrum), Fachbereich
  Physik, Technische Universit{\"a}t Darmstadt,
  Schlossgartenstra{\ss}e 2, D-64289 Darmstadt, Germany}

\begin{abstract}
  Neutron-rich outflows in neutron-star mergers (NSMs) or other explosive events can be subject to substantial heating through the release of rest-mass energy in the course of the rapid neutron-capture (r-) process. This r-process heating can potentially have a significant impact on the dynamics determining the velocity distribution of the ejecta, but due to the complexity of detailed nuclear networks required to describe the r-process self-consistently, hydrodynamic models of NSMs often neglect r-process heating or include it using crude parametrizations. In this work, we present a conceptually new method, RHINE, for emulating the r-process and concomitant energy release in hydrodynamic simulations via machine-learning algorithms. The method requires the evolution of only a few additional quantities characterizing the composition, of which the nuclear rates of change are obtained at each location and time step from neural networks trained by a large set of trajectories from full nuclear-network calculations. The scheme is tested by comparing spherically symmetric wind simulations and long-term simulations of NSMs using RHINE with post-processing results from nucleosynthesis calculations, showing agreement in the released heating energy to within $\lesssim 10\,\%$. In our NSM models on average about $2.3\,$MeV, $0.7\,$MeV, and $2.1\,$MeV are released per baryon in dynamical ejecta, NS-torus ejecta, and black-hole (BH) torus ejecta, respectively. The strongest velocity boost is observed for BH-torus ejecta, which also become $40\,\%$ more massive with r-process heating. The nucleosynthesis yields are only mildly affected by r-process heating, but the kilonova gets significantly brighter once the BH-torus ejecta become visible. RHINE can be readily implemented in existing hydrodynamics codes using pre-trained machine-learning data and routines for source-term prediction that we provide online.
\end{abstract}

\maketitle

\section{Introduction}\label{sec:introduction}

The interpretation of multi-messenger observations of binary neutron-star mergers (NSMs), such as AT2017gfo/GW170817 \citep[e.g.][]{Abbott2017b, Smartt2017s, Tanaka2017t}, as well as investigations of their role for galactic archaeology \citep[e.g.][]{Frebel2018a,Lombardo2025a} and galactic chemical enrichment \citep[e.g.][]{Shen2015, vandeVoort2015, Kobayashi2020a} rely heavily on predictions made by theoretical models of the (neutrino-/magneto-)hydrodynamic processes occurring during and after the merger \citep[see, e.g.,][for reviews]{Baiotti2017a, Duez2018y, Shibata2019c, Bernuzzi2020b, Janka2022a, Foucart2023b, Siegel2022k, Rosswog2024b}, of the rapid neutron-capture (or r-) process nucleosynthesis operating in the ejected material \citep[e.g.][]{Horowitz2018a, Arnould2020f, Cowan2021origin, Holmbeck2023c}, and of the kilonova signal released from the outflows \citep[e.g.][]{metzger2010electromagnetic, Roberts2011, Metzger2019a, Burns2020a}. The last two decades have seen tremendous progress in each of these fields regarding the included physics ingredients and the degree of self-consistency, and models have finally become available that predict the nucleosynthesis yields and kilonova signal using directly the results from multi-dimensional hydrodynamic simulations \citep[e.g.][]{Kasen2015,Miller2019a,Kawaguchi2021a,just2023end,Klion2022a,shingles2023self,Combi2023b,Curtis2024a,Magistrelli2024b}.

Apart from powering the electromagnetic transient at late times, energy release from nuclear reactions can have a sizable impact on the dynamics of the ejected material, in particular when the kinetic energy of this material is comparable to the nuclear binding energy. Already before onset of r-process nucleosynthesis, i.e. while material is still in nuclear statistical equilibrium (NSE), recombination of free nucleons into $\alpha$-particles (releasing about 7\,MeV per baryon per $\alpha$-particle created) can significantly boost the matter ejection process in post-merger accretion disks \cite{Lee2005, Fernandez2013b, Siegel2018c, Haddadi2023a, Fernandez2024a}. Subsequently, however, several more MeV per baryon can be liberated during the formation of heavier elements in the course of the r-process. This r-process heating can affect the amount of the ejected material as well as its velocity and geometric distribution, composition, and kilonova signal \cite[e.g.][]{Rosswog2014, Fernandez2015c, Just2015a, Wu2016a, Desai2019a, Klion2022a, Foucart2021a, Darbha2021a, Foucart2021a, Kawaguchi2021a, Sneppen2023b}, while it may also have an impact on the gravitationally bound material falling back onto the central remnant on longer timescales \cite[e.g.][]{Rosswog2007c, Metzger2010, Ishizaki2021a, Musolino2024e}.

R-process heating remains a physics ingredient that is difficult to tackle in hydrodynamical simulations. One reason is the sheer computational cost required to follow r-process nucleosynthesis self-consistently, which requires solving an evolution equation for each of the several thousands of isotopes along the path of all expanding fluid elements. Another reason is connected to the relatively long characteristic timescale of r-process heating of $\sim$\,1\,s, which is challenging to cover in multi-dimensional simulations, in particular because the long-term evolution of NSM remnants depends sensitively on effects related to neutrino transport and turbulent angular-momentum transport that both are computationally cumbersome. For these reasons, r-process nucleosynthesis calculations are usually completely decoupled from hydrodynamical simulations, in the sense that they are conducted after the fact along outflow trajectories extracted from simulations that do not ``know'' about r-process nucleosynthesis. This post-processing approach efficiently avoids the enormous computational expenditures of a solver coupling hydrodynamics with detailed nuclear networks, but it ignores the dynamical feedback from r-process heating and thus is intrinsically inconsistent. Depending on the amount of energy released and the original kinetic energy of the ejecta, this feedback can potentially have a significant impact on the outflow trajectories and thereby affect the nucleosynthesis yields and the kilonova signal.

Several attempts have been made to incorporate the impact of r-process heating in hydrodynamical models. Starting with the pioneering work by \citet{Rosswog2014}, the most often adopted method \citep[][]{Fernandez2015c, Desai2019a, Klion2022a, Ishizaki2021a, Darbha2021a, Foucart2021a, Sneppen2023b} consists of adding a source term to the energy equation (or, in the case of \citet{Foucart2021a}, to other equations as well) that parametrizes r-process heating as a function of the global evolution time, tacitly assuming that all of the ejecta material is launched at, and expands on, similar timescales. While this assumption is more likely to be justified in simulations describing just the early dynamical ejecta component launched at times $t\ll 1\,$s, it is conceivably more problematic for post-merger ejecta that are released on secular timescales of up to seconds. Probably similarly crude are treatments that parametrize the heating rate as function of the temperature \cite{Just2015a,Wu2016a}. Moreover, having in mind only a rough estimate of the possible dynamical impact, many of the aforementioned studies (an exception being Ref.~\cite{Foucart2021a}) assume the same global values for the total amount of energy released per baryon and thus tend to be inconsistent with the local nucleosynthesis conditions in the ejecta.

A more sophisticated treatment of r-process heating was employed in recent long-term simulations by the Potsdam-Kyoto group~\citep[e.g.][]{Kawaguchi2021a, Kawaguchi2024c}, in which outflow material extracted from NS-NS (or NS-BH) merger simulations is injected into the computational domain (starting at a radius of several thousand kilometers) with heating rates provided by previously conducted post-processing nucleosynthesis calculations. Although potentially quite accurate, this method may be time-intensive (as it requires altogether three calculation steps, two hydro-simulations plus one nucleosynthesis calculation), it neglects heating below the injection radius (which may be significant for slow ejecta), and it assumes extrapolated fluid trajectories for the nucleosynthesis calculation outside of the injection radius.

The recent simulations reported in~\citet{Magistrelli2024b} map ejecta from multi-dimensional simulations at about 30\,ms after merger onto 1D radial rays and follow the hydrodynamic evolution in a Lagrangian fashion independently along each ray with the inclusion of a detailed nuclear network in each grid cell. This method represents an important step towards a fully self-consistent treatment of nucleosynthesis and hydrodynamics, but it may suffer from the fact that fluid motion is constrained to radial directions only, which could both over- or underestimate the effects resulting from r-process heating.

Very recently, ~\citet{Ma2025c} presented simulations of post-merger disks in which r-process heating was described by a newly-developed method based on co-evolving a large number of tracer particles with the fluid and evaluating the released heat through a sophisticated cloud-in-cell method. Their scheme appears to produce heating rates broadly consistent with post-processing nucleosynthesis calculations, however, the additional evolution of particles and their coupling to the quantities evolved on a grid may introduce a considerable degree of complexity and computational demand.

In this paper we present a new scheme that was developed to overcome the problematic aspects mentioned above, i.e. which is more consistent with full nuclear-network calculations than simple parametrizations, does not require multiple calculation steps, and can be used in multi-dimensional simulations without a major increase of the computational expense. The scheme, called RHINE (\textbf{R}-process \textbf{H}eating \textbf{I}mplementation in hydrodynamic simulations with \textbf{NE}ural networks), makes use of machine-learning (ML) models, more specifically deep-learning neural networks \citep[e.g.][]{Rosenblatt1958a, LeCun1989a, Hornik1989a, Cybenko1989a, Schmidhuber2014a}, in order to approximate the rates of change of evolved quantities characterizing the baryonic composition. While over the past years ML models have been used more and more frequently in computational astrophysics -- e.g. for gravitational-wave analysis \citep[][]{Soultanis2025a}, evolution and explosions of stars \citep[][]{Mirouh2018a,Hendriks2019a,Tsang2022a, Maltsev2025a}, or approximating moment closure relations and flavor oscillations of neutrinos \citep{Harada2022a, abbar2024physics, richers2024asymptoticstate} -- only a few works so far adopt ML models in the context of nucleosynthesis \citep[e.g.][]{Fan2022a, Li2024a, Grichener2025a, Saito2025a}. To our knowledge, the current study is the first in which ML models are used as approximators of nuclear-reaction rates in multi-dimensional hydrodynamical simulations.

The paper is structured as follows: Sect.~\ref{sec:description-scheme} will provide a detailed description of the  scheme including the evolved equations and the ML models utilized. In Sect.~\ref{sec:applications} the scheme is applied in wind-outflow test problems and in long-term simulations of NSMs and it is compared against full nuclear network calculations and simulations not including r-process heating. In Sect.~\ref{sec:summary-conclusions} the results are summarized. The Appendices provide supplementary information about the ML models and the initial conditions used for the wind-outflow test problems.

We will adopt CGS units throughout the paper with standard symbols denoting the speed of light ($c$), Boltzmann constant ($k_B$), gravitational constant ($G$), and reduced Planck constant ($\hbar$). 

\section{Description of the scheme}\label{sec:description-scheme}

This section details the working principle of RHINE starting with some basic estimates (Sect.~\ref{sec:expect-dynam-impact}) and the motivation of the scheme (Sect.~\ref{sec:design-goals-rhine}), followed by a summary of the evolved physics equations (Sect.~\ref{sec:evolution-equations}) and EOS treatment (Sect.~\ref{sec:extens-equat-state}). In Sect.~\ref{sec:pred-source-terms} the neural networks are explained that emulate the nuclear-reaction rates required during the evolution.

\subsection{Expected dynamical impact of r-process heating}\label{sec:expect-dynam-impact}

\begin{figure}
    \centering
    \includegraphics[width=\columnwidth]{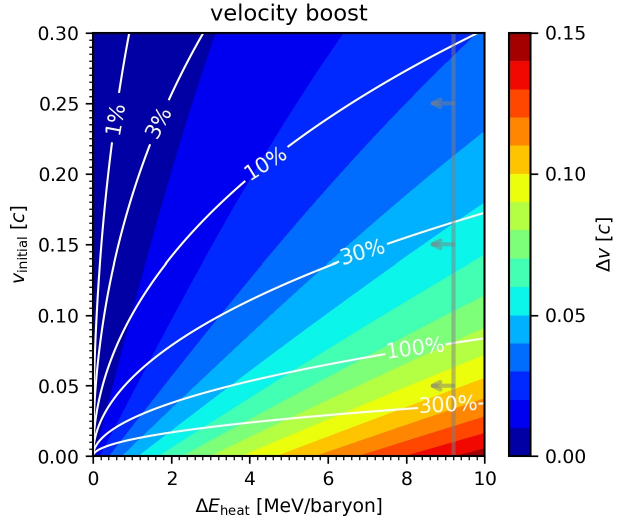}
    \caption{\label{fig:vboost} Velocity boost $\Delta v=v_{\rm final}-v_{\rm initial}$ expected to result from heating material moving initially with velocity $v_{\rm initial}$ by the amount of energy per baryon $\Delta E_{\rm heat}$ assuming perfect conversion into kinetic energy. The white curves denote lines of constant relative velocity change $\Delta v/v_{\rm initial}$ by the indicated percentage. The grey line denotes the physical upper limit $\Delta E^{\rm max}_{\rm heat}$ of Eq.~(\ref{eq:emax}).}
\end{figure}

Before considering the specifics of RHINE, it is instructive to get a broad idea of the dynamical impact expected to result from the release of nuclear energy in expanding ejecta. Depositing an amount of energy per baryon $\DEheat$ into material initially expanding with velocity $v_{\rm initial}$ will accelerate the material until it reaches a new velocity $v_{\rm final}$. 
Assuming complete conversion into kinetic energy and that the energy originated due to the difference in rest mass between initial and final composition, energy conservation implies that
\begin{align}\label{eq:econs}
W_{\text{final}} m c^2= W_{\text{initial}} (m c^2 +\DEheat)
\end{align}
where $W_{\text{final/initial}}=1/\sqrt{1-v^2_{\text{final/initial}}/c^2}$ are the (final or initial) Lorentz factors and $m$ a representative baryon mass (approximately equal to the atomic mass unit $m_u \approx 931.5$\,\MeVc). The velocity boost $\Delta v=v_{\rm final}- v_{\rm initial}$ computed through Eq.~(\ref{eq:econs}) quantifies the expected dynamical impact of r-process heating for a given amount of released heat per baryon and baseline velocity (i.e. velocity that would result without heating) and is illustrated in Fig.~\ref{fig:vboost}, which will later serve as a useful reference for interpreting the simulation results in Sect.~\ref{sec:applications}.

For a fixed value of $\DEheat$, the velocity boost is not constant but decreases for higher values of $v_{\rm initial}$, because the kinetic energy grows nearly quadratically with velocity (at least if $v/c$ is significantly smaller than unity). Thus, slower material will be affected more significantly by r-process heating than faster material. For instance, $5\,$MeV/baryon injected into material moving with $0.05\,c$ (such as BH-torus ejecta \cite[e.g.][]{Fernandez2013b, Just2015a, Siegel2017b, Miller2019a, Fujibayashi2020a, Haddadi2023a}) would induce a relative velocity boost by $\Delta v/v_{\rm initial}\sim 100\,\%$. Hence, r-process heating in the range of several MeV can have an important dynamical impact which, at least for slow ejecta, cannot be ignored for a reliable prediction of the final velocity distribution and the kilonova signal. On the other hand, for faster ejecta components, r-process heating becomes less important, as for instance material expanding with $\sim$\,0.2$\,c$ (as is typical for dynamical ejecta \cite[e.g.][]{Bauswein2013, Hotokezaka2013b, Radice2016a, Foucart2016}) would only become about $10\,\%$ faster when heated with the same amount of energy.

Obviously, $\Delta v$ also grows with the amount of released r-process heat, however, $\DEheat$ cannot attain arbitrarily large values. The upper limit of $\DEheat$ that can possibly be released by nuclear reactions if free neutrons would recombine into the nucleus with the smallest mass per nucleon, $^{56}$Fe, is
\begin{align}\label{eq:emax}
  \DEheat^{\rm max} \approx 939.6\,\mathrm{MeV}-930.4 \, \mathrm{MeV} = 9.2\,\mathrm{MeV} \, .
\end{align}
The fraction of energy actually used for heating and accelerating the ejecta during just the r-process, i.e. after leaving nuclear statistical equilibrium (NSE) at a temperature of $T\approx 5$--10~GK, is in many cases smaller than $\DEheat^{\rm max}$, because, first, partial recombination already takes place during NSE (cf. discussion of $\tilm_{\rm NSE}$ in Sect.~\ref{sec:evolution-equations}), and second, some 10's of percent of the liberated rest-mass energy is lost to neutrinos emitted in $\beta$-decays. A more realistic maximum value of $\DEheat$ is therefore closer to 7{--}8\,MeV,  which however is only reached by material that is very neutron-rich at the onset of r-process nucleosynthesis.

\subsection{Design goals of RHINE}\label{sec:design-goals-rhine}

The overall aim of RHINE is to capture nuclear energy release during the r-process in hydrodynamic simulations as consistently with full nuclear networks as possible by adding only as much numerical and physical complexity to the evolution scheme as necessary. Specifically, the scheme was developed with the following goals in mind: 1) The total energy per baryon $\DEheat$ released in a simulation using RHINE should agree reasonably well -- i.e. with relative errors of $\mathcal{O}(10)\,\%$ or less -- with results obtained when post-processing the same simulation with a full nuclear network. In particular, the resulting $\DEheat$ should reproduce the (significant) dependence on the initial electron fraction (cf. discussion of $\tilm_{\rm NSE}$ in Sect.~\ref{sec:evolution-equations}) and must always obey $\DEheat<\DEheat^{\rm max}$. 2) Not only the integrated energy but also the time dependence of energy deposition should be consistent. This is a particularly non-trivial aspect, because material can be ejected on vastly different timescales during a merger, with a significant fraction ejected at late times comparable to the r-process timescale of $\sim$\,1\,s. Many previous implementations of r-process heating assumed for all ejecta the same dependence of the heating rate on the evolution time, $t$, which is a poor approximation for outflows launched at late times. 3) The scheme should be self-contained and readily applicable to any NSM simulation, meaning that no previous post-processing step is necessary for obtaining the heating rates along the outflow trajectories. 4) The scheme should rely only on standard advection-reaction equations, and the source terms on the right-hand side of these equations should be entirely local functions of the evolved variables that are straightforward to evaluate at runtime. No implicit time stepping involving matrix inversions is required, no explicit dependence on global parameters (such as evolution time or total ejecta mass) exists, and no additional scheme is needed that advects Lagrangian tracer particles in order to memorize ejecta properties at earlier times. 5) The scheme should be computationally lightweight in the sense that the efficiency of a production run is impacted by no more than a factor of about two\footnote{Admittedly, we have not invested significant efforts into optimizing the performance of the scheme so far, because the current increase of the wall-clock time per iteration step of 33\,\% (108\,\%) with (without) neutrino transport measured for the NSM production runs discussed in this study was sufficient for our purposes. However, the efficiency can likely be improved further by tuning the structure of the adopted neural networks (cf. Sect.~\ref{sec:pred-source-terms}) used for inference of the source terms.}.

\subsection{Evolved equations and basic concept of RHINE}\label{sec:evolution-equations}

We now outline the working method of RHINE, starting off by summarizing the evolved equations. The following presentation assumes special relativistic hydrodynamics but the extension to Newtonian or general relativistic hydrodynamics or to magneto-hydrodynamics is straightforward. The evolution equations describing conservation of baryon number, electron-lepton number, momentum, and energy are given, respectively, by:
\begin{subequations}\label{eq:hdevo}
\begin{align}
   \partial_t D + \nabla_j(D v^j)  &= 0  \, , \\
   \partial_t (D Y_e) + \nabla_j(D Y_e v^j) &= \srcrhineye + \srcnuye \, ,  \label{eq:yeevo} \\
   \partial_t S^i + \nabla_j(S^i v^j + P\delta^{ij}) &= \srcnumom + \srcgravmom \, ,  \label{eq:sievo} \\
    \partial_t \tau + \nabla_j( v^j \tau + v^j P ) &= \srcrhinetau + \srcnutau + \srcgravtau \,  , \label{eq:tauevo}
\end{align}
\end{subequations}
where the evolved quantities $D= \rho W$, $S^i = \rho h W^2 v^i$, and $\tau = \rho h W^2c^2 - P - \rho W c^2$ are functions of the baryonic mass density $\rho$, three-velocity $v^i$, Lorentz factor $W = 1/\sqrt{1- v^2/c^2}$, gas pressure $P$, and specific enthalpy $h = (\erest+\etherm+P)/(\rho c^2)$ with $\erest$ and $\etherm$ being the rest-mass energy density and the thermal (plus degeneracy) energy density of the fluid as measured in the frame comoving with the fluid, respectively. The baryonic mass density is defined as $\rho= m_u n_B $ with $n_B$ being the comoving-frame number density of baryons and $m_u$ a (generally arbitrary) conversion constant that we choose to be the atomic mass unit. The electron fraction is given by $Y_e=(n_{e^-}-n_{e^+})/n_B$ as function of the comoving-frame number densities of electrons ($n_{e^-}$) and positrons ($n_{e^+}$). The source terms on the right-hand side of Eqs.~(\ref{eq:hdevo}) are the protonization- and neutrino-cooling rates associated with $\beta$-decay in outflows undergoing r-process nucleosynthesis ($\srcrhineye$ and $\srcrhinetau$, respectively), the exchange of lepton number, momentum, and energy with neutrinos through reactions not associated with r-process nucleosynthesis ($\srcnuye$, $\srcnumom$, and $\srcnutau$, respectively\footnote{These source terms need to be provided by some additional neutrino scheme. In the simulations discussed in this study, the M1 two-moment neutrino scheme described in \citet{Just2015b} is used for that purpose.}), as well as gravitational momentum- and energy-exchange ($\srcgravmom$ and $\srcgravtau$, respectively).

\begin{figure*}
    \centering
    \includegraphics[width=\textwidth]{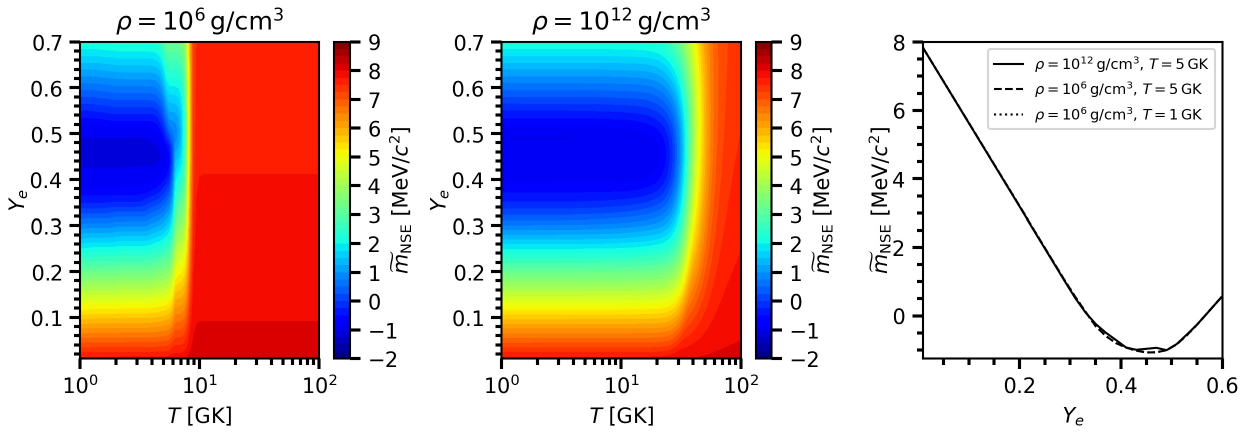}
    \caption{\label{fig:massex} Average mass excess per baryon (cf. Eq.~(\ref{eq:mtil})) resulting for a composition in NSE \cite{hix2006thermonuclear}. The two left contour plots show $\tilm_{\rm NSE}$ in the plane of temperature and electron fraction at fixed densities $\rho=10^{6}\,$g\,cm$^{-3}$ and $10^{12}\,$g\,cm$^{-3}$, and the right panel depicts $\tilm_{\rm NSE}$ as a function of $Y_e$ for different temperatures and densities. At low temperatures, $T\lesssim 5-8\,$GK, the mass excess becomes nearly independent of temperature and density but a steep function of $Y_e$, decreasing from $\tilm_{\rm NSE}\sim 8 \,$\MeVc at $Y_e\approx 0$ to $\tilm_{\rm NSE}\sim -1$\,\MeVc at $Y_e\approx 0.45$.}
\end{figure*}

Apart from the new terms $\srcrhineye$ and $\srcrhinetau$ discussed below, Eqs.~(\ref{eq:hdevo}) are formally identical to the conventional hydrodynamics equations used in studies not attempting to capture r-process heating. Notably, r-process heating (or generally any nuclear heating) does not enter the energy equation, Eq.~(\ref{eq:tauevo}), in the form of a source term, because the evolved energy variable, $\tau$, includes both the rest-mass energy and thermal energy. It enters instead through a reduction of the rest-mass energy density, $\erest$, which causes, because of energy conservation (modulo changes related to the source terms), a corresponding increase of the thermal energy density, $\etherm$. The rest-mass energy density can be written as\footnote{Following common convention, we subtract the rest-mass energy density of electron-positron pairs, $2m_e c^2n_{e^+}$, from $\erest$ and attribute it to $\etherm$.}
\begin{align}\label{eq:erest}
  \erest/c^2 & = m_e (n_{e^-}+n_{e^+}-2n_{e^+}) + \sum_i m_i n_i   \nonumber \\
  & = (m_e Y_e + \barm) n_B \, ,            
\end{align}
where $i$ runs over all baryonic species with nuclear masses $m_i$ and number densities $n_i$, $m_e$ is the electron mass, and $\barm$ the average mass per baryon. The difference between the actual and the baryonic rest-mass energy per baryon can then be expressed as
\begin{align}\label{eq:massex}
  \frac{\erest- \rho c^2}{n_B c^2} = m_e Y_e+\barm-m_u = \tilm \, ,
\end{align}
which defines the average mass excess per baryon, $\tilm$. This quantity typically attains values between about 8\,\MeVc (for free nucleons) and $-1$\,\MeVc (for strongly bound nuclei). Hence, in order to capture heating from nuclear energy release in hydrodynamic simulations, knowledge of $\tilm$ is needed. Under NSE conditions, $\tilm$ is just a function of $\rho,T,Y_e$ and is depicted in Fig.~\ref{fig:massex} for exemplary conditions of $\rho$ and $T$. As can be seen, the values of $\tilm_{\rm NSE}$ freeze out (i.e. become independent of $T$ for smaller $T$) near NSE-threshold temperatures around 5{--}8\,GK. This means that an NSE treatment below these temperatures will not allow material to release any further energy. Figure~\ref{fig:massex} also shows that the freeze-out values of $\tilm_{\rm NSE}$ depend mainly on $Y_e$ and only weakly on $\rho$. Initially more neutron-rich material is thus projected to release a greater amount of heating energy per baryon during the r-process than less neutron-rich material. Given the wide range of $Y_e$ conditions in NSM ejecta, this immediately suggests that the amount of r-process heat dumped into the ejecta can vary significantly depending on the ejecta component and even within a single ejecta component. The dependence of $\tilm_{\rm NSE}$ on $Y_e$ was also pointed out by \cite{Foucart2021a} and used by \cite{Haddadi2023a} to construct a simplified treatment of r-process heating by increasing $Y_e$ with time in a parametrized manner while keeping the composition in NSE.

In addition to $\tilm$, knowledge is also required of the $\beta$-decay related source terms $\srcrhineye$ and $\srcrhinetau$. The former describes the increase of $Y_e$ towards $\sim 0.4$ during the r-process, while the latter captures the energy loss to neutrinos, which reduces the effective heating rate by a few tens of percent typically.

Ideally, knowledge of the entire composition is needed to determine $\tilm, \srcrhineye$,~and $\srcrhinetau$ fully self-consistently. However, this would require a large nuclear network to be coupled to the hydrodynamics equations, resulting in thousands of additional advection-reaction equations for the abundances of each isotope. The approach of RHINE is instead to only advect a small number of quantities characterizing the most relevant properties of the local composition and to evaluate the corresponding rates of change using ML models trained by data from full nuclear networks. The set of additional quantities should be small enough for the overall computational cost of RHINE to remain comparable to the original hydrodynamic evolution, while it needs to be large enough for RHINE to enable a -- compared to a detailed nuclear network evolution -- reasonably consistent description of r-process heating and EOS properties in outflowing material.

The additional evolution equations used by RHINE track the local value of $\tilm$ itself,
\begin{align}\label{eq:mtilevo}
   \partial_t (D \tilm) + \nabla_j(D \tilm v^j) =  \srcrhinerest \, ,
\end{align}
the mass fractions of free neutrons, $X_n$, free protons, $X_p$, helium, $X_\alpha$, and all remaining nuclei (subsumed under ``heavy nuclei'' hereafter), $X_h$,
\begin{align}\label{eq:xievo}
   \partial_t (D X_i) + \nabla_j(D X_i v^j) = \srcrhinexi \, ,
\end{align}
where $i=n,p,\alpha,h$, as well as the average mass number of heavy nuclei, $A_h$,
\begin{align}\label{eq:ahevo}
   \partial_t (D A_h) + \nabla_j(D A_h v^j) = \srcrhineah \, .
\end{align}
As usual, mass fractions and relative abundances are defined as
\begin{subequations}\label{eq:xiyi}
\begin{align}
   X_i &= \frac{A_i m_u n_i}{\sum_j A_j m_u n_j} = \frac{A_i n_i}{n_B}  \, , \label{eq:xi} \\
   Y_i &= \frac{n_i}{\sum_j A_j n_j} = \frac{n_i}{n_B} = \frac{X_i}{A_i}       \label{eq:yi}
\end{align}
\end{subequations}
for any species $i$ with number density $n_i$, where the sum runs over all nucleons and nuclei, and the mass number, $A_i=Z_i+N_i$, equals the number of protons plus neutrons. For the $i=h$ species representing heavy nuclei the mass fraction, abundance, average mass number, average proton number, and average mass are defined, respectively, as:
\begin{subequations}\label{eq:heavyav}
\begin{align}
   X_h &= \sum\limits_{j\neq n,p,\alpha} X_j \, , \label{eq:xh} \\
   Y_h &= \sum\limits_{j\neq n,p,\alpha} Y_j \, , \label{eq:yh} \\
   A_h &= \frac{1}{Y_h}\sum\limits_{j\neq n,p,\alpha} A_jY_j \, , \label{eq:ah} \\
   Z_h &= \frac{1}{Y_h}\sum\limits_{j\neq n,p,\alpha} Z_jY_j \, , \label{eq:zh} \\
   m_h &= \frac{1}{Y_h}\sum\limits_{j\neq n,p,\alpha} m_jY_j \, , \label{eq:mh}
\end{align}
\end{subequations}
while in this case the sums only run over all heavy nuclei. Note that Eqs.~(\ref{eq:xiyi}) are also valid for the compound species $i=h$. With these definitions the average mass excess per baryon can be written as:
\begin{align}\label{eq:mtil}
  \tilm & = m_e Y_e - m_u + \sum\limits_{i} m_i Y_i \nonumber \\
        & = \sum\limits_{i} \tilm_i X_i  \, ,
\end{align}
where
\begin{align}\label{eq:mtili}
  \tilm_i = (m_i+Z_i m_e-A_i m_u)/A_i
\end{align}
is the individual mass excess of species $i$, which is commonly found in compilations of experimental~\cite{Wang.Huang.ea:2021} and theoretical masses~\cite{moller1995nuclear}, per baryon. Equation~(\ref{eq:mtil}) allows to retrieve the average mass of heavy nuclei, $m_h$, from the evolved quantities. Additional useful relations are given by the constraints of mass- and charge-balance,
\begin{subequations}\label{eq:mc}
\begin{align}
   1    &= \sum_j X_j     = X_n + X_p + X_\alpha + X_h     \, , \label{eq:mb} \\
   Y_e  &= \sum_j Z_j Y_j = Y_p + 2Y_\alpha + Z_h Y_h  \, , \label{eq:cb}
\end{align}
\end{subequations}
respectively, which must be fulfilled at any time (see Appendix~\ref{sec:ensur-phys-cons} for more information). Equation~(\ref{eq:cb}) provides the average charge number of heavy nuclei, $Z_h$, for given evolved quantities.

\begin{figure}
    \centering
    \includegraphics[width=\columnwidth]{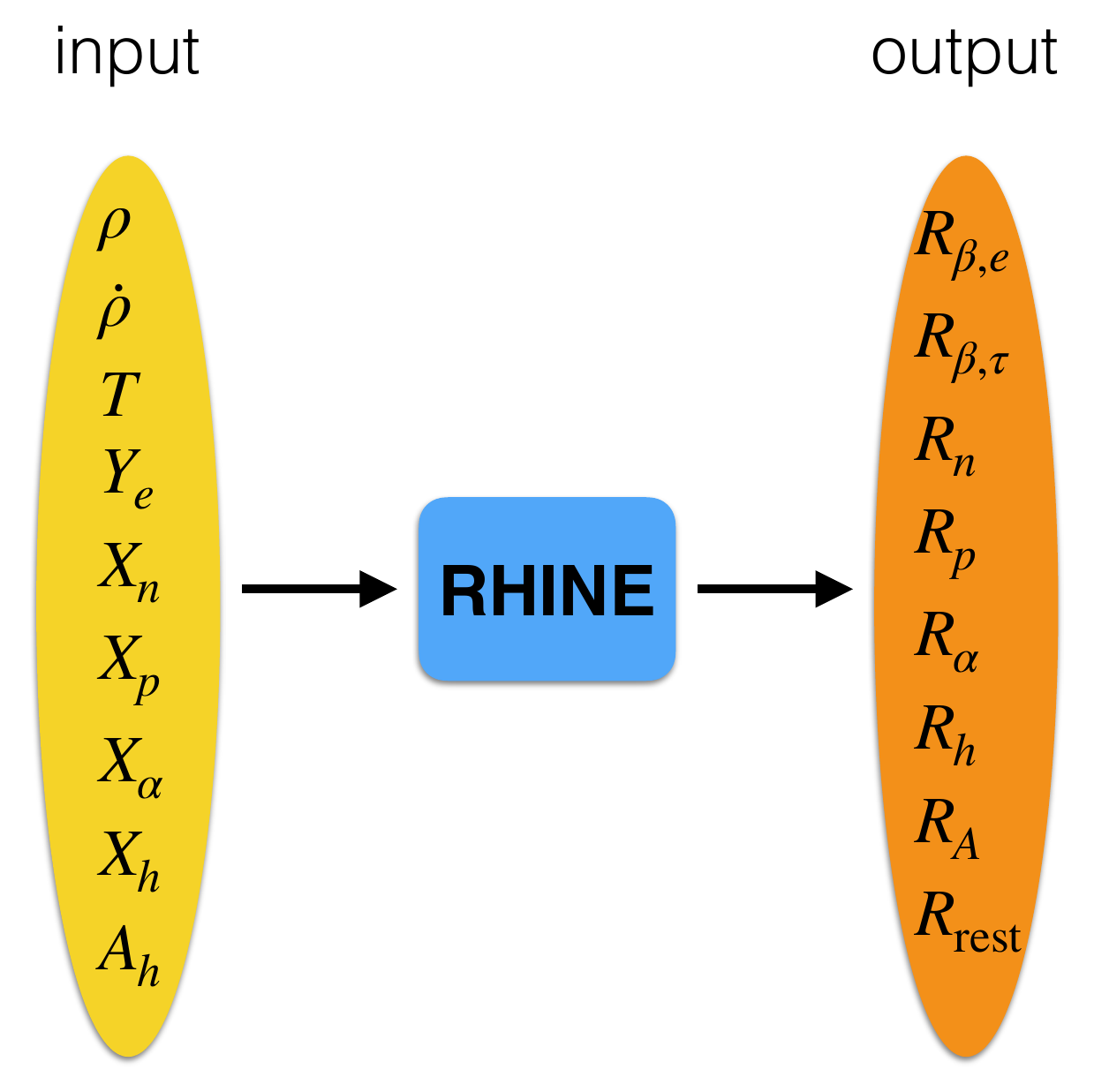}
    \caption{\label{fig:sketch} Sketch illustrating the concept of RHINE. In addition to the conventional set of hydrodynamics equations, Eqs.~(\ref{eq:hdevo}), RHINE requires the evolution of four mass fractions, $X_n,X_p,X_\alpha,X_h$, the average mass number of heavy nuclei, $A_h$, and the average mass excess per baryon, $\tilm$ (cf. Eqs.~(\ref{eq:xievo}),~(\ref{eq:ahevo}),~and~(\ref{eq:mtilevo}), respectively). Using the evolved quantities shown on the left as input variables, RHINE provides at each time step and at each location the source terms connected to r-process heating shown on the right utilizing neural networks trained by full nuclear-network calculations.}
\end{figure}

In order to evolve the six additional equations, Eqs.~(\ref{eq:mtilevo})--(\ref{eq:ahevo}), the rates $\srcrhinerest$,~$\srcrhinexn$,~$\srcrhinexp$,~$\srcrhinexa$,~$\srcrhinexh$,~and~$\srcrhineah$ are required. Together with the $\beta$-decay related rates, $\srcrhineye$ and $\srcrhinetau$, eight source terms are needed at each location and (partial) time step in order to evolve the overall system. RHINE infers these source terms from ML models as functions of the local evolved variables, as illustrated in the sketch of Fig.~\ref{fig:sketch}. It turns out that evolving just the above four baryonic species together with $\tilm$ and $A_h$ provides enough information for the ML models to predict reasonably accurate source terms. This is likely related to the fact that the mass excess per nucleon, or equivalently the nuclear binding energy per nucleon, changes smoothly with $A$ and $Z$ for heavy nuclei. Hence, $\tilm$ depends mainly on $A_h$ and $Z_h$ for a wide range of conditions in material undergoing the r-process (see, e.g., accuracy of ML model 1 in Table~\ref{tab:parameters}). In principle, more heavy species, or more degrees of freedom characterizing their distribution, could be added to the scheme to improve the accuracy of the ML models. However, this would increase the complexity of the overall scheme and its computational expense, and we found that the current set of evolved quantities is large enough for the simulations to produce results that are already in very good agreement with nuclear network calculations (cf. Sect.~\ref{sec:applications}).

This summarizes the basic concept of the scheme. The ML-based prediction of the source terms will be described in Sect.~\ref{sec:pred-source-terms}.

\subsection{Extension of the equation of state}\label{sec:extens-equat-state}

Before turning to the source terms, we address one more requirement of RHINE, namely the ability of the EOS to deal with non-NSE conditions and with low temperatures and densities. Conventional simulation codes often assume NSE everywhere and use 3D EOS tables that are limited at some density and temperature (e.g. at $\sim 10^{3}\,$g\,cm$^{-3}$ and 0.1\,\MeVk, respectively, in the case of the SFHO EOS of Ref.~\cite{Steiner2013}). In NSE, the composition, and therefore the quantities $P$ and $\etherm$, can be interpolated from 3D tables for given $\rho, T, Y_e$. Under non-NSE conditions, however, the EOS depends explicitly on the composition, i.e. the mass fractions of all nuclei, $X_i$, are independent degrees of freedom. In the current implementation of RHINE in the neutrino-hydrodynamics code Aenus-ALCAR \citep{Obergaulinger2008a, Just2015b}, before applying the EOS we first flag each zone of the computational grid as being either in NSE or non-NSE, depending on the temperature of the last time step (see Sect.~\ref{sec:infer-source-terms} for the explicit criterion). If a zone is flagged as NSE, all EOS-related variables are read from a table as functions of $\rho, T, Y_e$, overwriting the local values of $\tilm, X_i$,~and~$A_h$ that are advected via Eqs.~(\ref{eq:mtilevo})--(\ref{eq:ahevo}). These equations become active only once a zone is flagged as non-NSE, in which case a different (``low-density EOS'') routine is called that depends explicitly on the compositional variables $\tilm, X_i$,~and~$A_h$. Our low-density EOS (H.-Th.~Janka, private communication) adopts the same assumptions as the well-known Helmholtz-EOS by \citet{timmes2000accuracy}, namely non-relativistic, non-degenerate nucleons and nuclei, arbitrarily relativistic and arbitrarily degenerate electrons and positrons, and a thermal photon bath, optionally with the inclusion of Coulomb-lattice corrections. The explicit formulae for the various components of $P$ and $\etherm$ can be found, e.g., in \citet{Timmes1999}. Fortunately, most quantities computed by this EOS depend only on the average properties of heavy nuclei, i.e. on $X_h, A_h$,~and~$Z_h$, and therefore do not require a more detailed knowledge of the composition. The only two exceptions are the chemical potential, $\mu_h$, and the entropy density, $S_h$, of heavy nuclei, which depend on the statistical weights, $g$, of individual nuclei. We ignore this dependence and compute the two quantities for the four evolved baryonic species $i$ as:
\begin{subequations}\label{eq:muhsh}
\begin{align}
   \frac{\mu_i-m_i c^2}{k_B T}    &= \ln\left[\frac{n_i}{g_i} \left(\frac{2\pi\hbar^2}{m_ik_B T}\right)^{3/2}\right] \, , \\
   S_i &= Y_i n_B k_B \left(\frac{5}{2}- \frac{\mu_i-m_i c^2}{k_B T}\right) \, .
\end{align}
\end{subequations}
with $g_i=2,2,1,1$ for $i=n,p,\alpha,h$, respectively.

\subsection{Prediction of source terms}\label{sec:pred-source-terms}

\begin{figure}
  \centering
  \includegraphics[width=0.99\columnwidth]{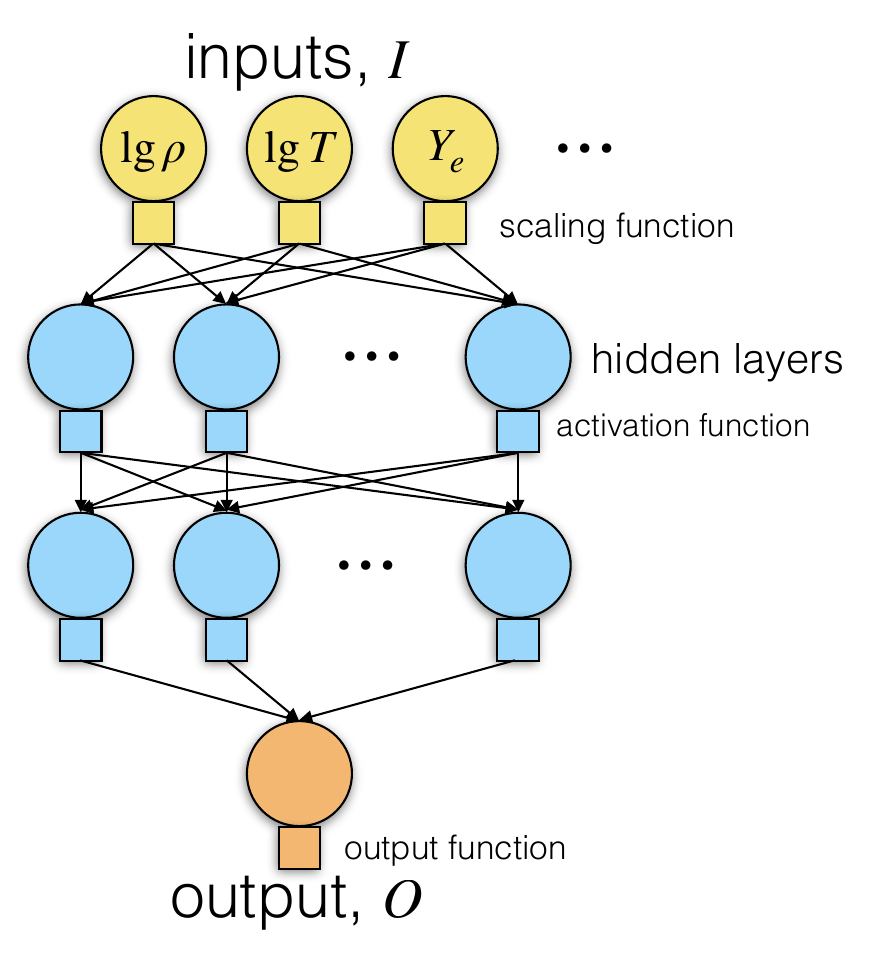}
    \caption{\label{fig:nnarch} Schematic picture illustrating the architecture of a multilayer perceptron (MLP) neural network used to fit and predict the source terms related to the r-process in RHINE. Starting with a set of input quantities, $I$, the MLP computes an output quantity, $O$, by passing through a sequence of hidden layers of perceptrons, where the output of a perceptron in one layer is used as input for all perceptrons in the next layer. See main text for details.}
\end{figure}

The key component of RHINE is the estimation of the eight source terms $R_k$ on the right-hand sides of the evolution equations presented in Sect.~\ref{sec:evolution-equations} by making use of machine-learning (ML) models. The ML models serve the purpose of regression functions in that each ML model provides the estimate of some desired quantity (e.g. the rate of change of $Y_e$) as a function of the evolved quantities (say, $\rho,T, Y_e$,~and~$A_h$).

In contrast to classical fitting approaches, neural networks are not based on, or motivated by, some underlying model or theory but built entirely from empirical observations, with the only objective being the ability to reproduce a set of reference (``training'') data as accurately as possible. This obviates the need for manually searching for suitable types of fit functions, e.g. polynomials, because ML schemes can learn arbitrarily shaped patterns quasi-automatically during the training process. Moreover, instead of a few fit parameters typically used in conventional fits, our ML models consist of $\mathcal{O}(1000)$ parameters and therefore can fit complex functions more accurately.

In Sect.~\ref{sec:mult-perc} the basic structure of the ML models and their construction is outlined, while Sect.~\ref{sec:infer-source-terms} describes the way how they are used to infer the source terms $R_k$ during the evolution. More details regarding the construction of the ML models (data preparation, training and testing, as well as measures to ensure physical consistency) will be presented in Appendix~\ref{sec:addit-inform-about}. The performance of the overall scheme will be examined in Sect.~\ref{sec:applications} by comparing the results of simulations using RHINE with full nuclear-network results.

\subsubsection{Fitting data with multilayer perceptrons}\label{sec:mult-perc}
 
\begingroup
\begin{table*}[t]
\begin{ruledtabular}
\centering
\caption{\label{tab:mlmodels} List of ML models used by RHINE to infer the r-process related source terms during the hydrodynamical evolution. The columns provide for each ML model its number, the output quantity, the input quantities, the function used to convert the output of the last perceptron into its original physical range, the type of ML model, the thermodynamic regime(s) in which it is applied, and additional conditions that must be fulfilled at the current point in order for the ML model to be applied. If these conditions are not met, the corresponding rate is assume to vanish. The upper-limiting function is defined as up$(x,\epsilon)=-\frac{1}{2}\ln(\exp[-2(x-\epsilon)]+1)+\epsilon$, the scaling and rescaling functions are given by Eqs.~(\ref{eq:sc_func})~and~(\ref{eq:rsc_func}), respectively, and $\lg x$ denotes the logarithm of $x$ with base 10.}
\begin{tabular}{clllccl}
ML model & output $O$         & input quantities $I$                               & output function $O(x)$                        & ML type        & regime  & only applied if                   \\\hline
 1       & $\tilm$            & $A_h, z_h$                                         & $xY_h+\sum\limits_{i=n,p,\alpha} \tilm_i X_i$ & regression     & both    &                                   \\\hline
 2       & $\lg Y_n^\Q$       & $\lg\rho,\lg T,Y_e$                                & up$(\0{rsc}(x),0)$                            & regression     & QSE     &                                   \\
 3       & $\lg Y_p^\Q$       & $\lg\rho,\lg T,Y_e$                                & up$(\0{rsc}(x),0)$                            & regression     & QSE     &                                   \\
 4       & $\lg Y_\alpha^\Q$  & $\lg\rho,\lg T,Y_e$                                & up$(\0{rsc}(x),-2\lg2)$                      & regression     & QSE     &                                   \\
 5       & $\lg Y_h^\Q$       & $\lg\rho,\lg T,Y_e$                                & rsc$(x)$                                      & regression     & QSE     &                                   \\
 6       & $A_h^\Q$           & $\lg\rho,\lg T,Y_e$                                & rsc$(x)$                                      & regression     & QSE     &                                   \\
 7       & $z_h^\Q$           & $\lg\rho,\lg T,Y_e$                                & rsc$(x)$                                      & regression     & QSE     &                                   \\\hline
 8       & $f_\nu$            & $\lg\rho,\lg T, Y_e, A_h$                          & $0.5\,\0{sigmoid}(x)$                         & regression     & dynamic & $\dd \tilm/\dd t<0$               \\
 9       & $\lg \dot{Y}_e$    & $\lg\rho,\lg T, Y_e, A_h$                          & rsc$(x)$                                      & regression     & dynamic &                                   \\
 10      & $\lg (-\dot{Y}_n)$ & $\lg\rho,\lg T, Y_e, A_h,\lg\lambda$               & rsc$(x)$                                      & regression     & dynamic & $\dot\rho<0$                      \\
 11      & $\dot{Y}_h>0$?     & $\lg\rho,\lg T,Y_e,A_h,\lg\lambda,\lg Y_n,\lg Y_h$ & $\0{sigmoid}(x)$                              & classification & dynamic & $\dot\rho<0$                      \\
 12      & $\lg\dot{Y}_h$     & $\lg\rho,\lg T,Y_e,A_h,\lg\lambda,\lg Y_h$         & rsc$(x)$                                      & regression     & dynamic & $\dot\rho<0$ \& $\dot{Y}_h>0$     \\
 13      & $\dot{A}_h\neq 0$? & $\lg\rho,\lg T,Y_e,A_h,\lg\lambda,\lg Y_n$         & $\0{sigmoid}(x)$                              & classification & dynamic & $\dot\rho<0$                      \\
 14      & $\dot{A}_h>0$?     & $\lg\rho,\lg T,Y_e,A_h,\lg\lambda,\lg Y_n$         & $\0{sigmoid}(x)$                              & classification & dynamic & $\dot\rho<0$ \& $\dot{A}_h\neq 0$ \\
 15      & $\lg\dot{A}_h$     & $\lg\rho,\lg T,Y_e,A_h,\lg\lambda,\lg Y_n$         & rsc$(x)$                                      & regression     & dynamic & $\dot\rho<0$ \& $\dot{A}_h>0$     \\
 16      & $\lg(-\dot{A}_h)$  & $\lg\rho,\lg T,Y_e,A_h,\lg\lambda,\lg Y_n$         & rsc$(x)$                                      & regression     & dynamic & $\dot\rho<0$ \& $\dot{A}_h<0$     \\
\end{tabular}
\end{ruledtabular}
\end{table*}
\endgroup

For predicting the rates of change contained in the source terms, $R_k$, we utilize ``multilayer perceptrons'' (MLPs), which represent a class of deep neural networks suitable for regression and classification problems \citep[e.g.][]{LeCun1989a, Hornik1989a, Cybenko1989a, Schmidhuber2014a}. A detailed introduction into MLPs is out of the scope of this work, but can be found in, e.g., \citet{Goodfellow2016a}. For each quantity to predict, an independent neural network is used. In Fig.~\ref{fig:nnarch} the structure of an MLP is schematically illustrated. For a set of input quantities $I$ the MLP computes an output quantity $O$ by passing through a number of $N_{\rm hid}$ hidden layers of perceptrons (or neurons). Each perceptron (except in the input layer) carries a vector of weights, $w_n$ (with $n$ running over all perceptrons from the previous layer), and a bias $b$. With these, the output value $x^{\rm out}$ of this perceptron is computed from its input values $x^{\rm in}_n$ as
\begin{align}\label{eq:perout}
  x^{\rm out} = f_{\rm act}\left(\sum\limits_{n} w_n x_n^{\rm in}+b\right) \, ,
\end{align}
where $f_{\rm act}$ is an activation function that is chosen to be the ELU (exponential linear unit) function,
\begin{align}\label{eq:fact}
f_{\rm act}(x) =
\begin{cases}
x, & \text{if } x \geq 0 \\
 e^x - 1, & \text{if } x < 0
\end{cases}
\end{align}
for all perceptrons in the hidden layers. The output value of these perceptrons is then used as input value for all connected perceptrons in the next layer, and so on, until reaching the perceptron in the output layer, of which the output is given by $O\left(\sum_{n} w_n x_n^{\rm in}+b\right)$ with some output function $O(x)$ that is chosen differently for each ML model (see, e.g., Eq.~(\ref{eq:rsc_func}) below and Table~\ref{tab:mlmodels}).

The ML models are trained through an optimization procedure based on stochastic gradient descent, in which the weights and biases of all perceptrons are varied in a way that the loss function, 
\begin{align}\label{eq:lossfunc}
\mathcal{L} = \frac{1}{N}\sum\limits_{j=1}^N \left[O(I_j)-O_{\rm target}(I_j)\right]^2 \, ,
\end{align}
is minimized. This function measures the mean squared difference between the outputs $O(I_j)$ predicted by the ML model for a set of $N$ input points from the training dataset, $I_j$, and the ``true'' target values $O_{\rm target}(I_j)$ for the same input data, i.e. the results from full nuclear-network calculations.

We summarize the ML models adopted by RHINE in Table~\ref{tab:mlmodels}. All of our ML models adopt $N_{\rm hid}=2$ hidden layers and the same number of perceptrons per hidden layer, $N_{\rm perc}=50$ or $30$ (see Table~\ref{tab:parameters}). The set of input quantities consists of $\rho, T,Y_e$ in all cases and additionally $A_h,Y_n,Y_h,$~or~$\lambda$ in some cases. The quantity $\lambda$ is computed from the inverse expansion timescale,
\begin{align}\label{eq:lambda0}
  \lambda_0 = \frac{1}{\rho}\frac{\dd\rho}{\dd t} = -\nabla_j v^j 
\end{align}
(where we used the continuity equation ignoring relativistic effects), as
\begin{align}\label{eq:lambda}
  \lambda = \0{sgn}(\lambda_0)\max\left(\left|\lambda_0\right|,0.3\,\mathrm{s}^{-1}\right) \, .
\end{align}
The restriction to expansion timescales shorter than $\sim 3\,$s is motivated by the fact that typically only a very small fraction of the ejecta expands on expansion timescales longer than that and, therefore, that such conditions are poorly resolved by the ML models while being overall less relevant for the dynamics.

In order to ensure a balanced impact of all input quantities, these should be of the same order of magnitude. To this end, logarithmic values are used for some input and output quantities that range over several orders of magnitude (see Table~\ref{tab:mlmodels}), and in addition all input quantities are scaled into the interval $(-1,1)$ by using
\begin{equation}\label{eq:sc_func}
    \0{sc}(I) = 2 (I-I^{\rm min})/(I_{\rm max}-I_{\rm min})-1 \, ,
\end{equation}
where $I_{\rm min}$ and $I_{\rm max}$ are the $0.01$ and $0.99$ quantiles\footnote{The $p$-quantile of a set of $n$ numbers is given by the number that is greater than the fraction $p n$ of these numbers and lower than the remaining $(1-p)n$ numbers.} of the training dataset for this input quantity. For related reasons, the output function $O(x)$ is used to scale the result of the output perceptron back to a physical scale, for instance by using the rescale function,
\begin{align}\label{eq:rsc_func}
    \0{rsc}(x) =  O_{\rm min}+(O_{\rm max}-O_{\rm min})(x+1)/2 
\end{align}
(with $O_{\rm min/max}$ being the $0.01/0.99$ quantiles of the training dataset for the target output quantity), as one possible choice for $O(x)$; see Table~\ref{tab:mlmodels} for the output function adopted for each ML model.

Our set of ML models not only consists of regression models, which approximate numerical relationships between input and output quantities, but also classification models, which are used to predict whether a certain condition is fulfilled or not (see Table~\ref{tab:mlmodels} and Sect.~\ref{sec:infer-source-terms} for the particular conditions). We adopt the same MLP architecture also for classification models but use the sigmoid function, 
\begin{align}\label{eq:sig_func}
    \0{sigmoid}(x) =  \frac{1}{1+e^{-x}}\, ,
\end{align}
as output function $O(x)$ in these cases, assuming that the condition is fulfilled when $O(x)>0.5$ and not fulfilled otherwise.

The ML models are trained on trajectories of dynamical ejecta taken from \citet{collins2023radiative} (which also have been adopted in \citet{shingles2023self}) and of post-merger ejecta taken from models sym-n1-a6 and asy-n1-a6 of \citet{just2023end} using the detailed nuclear network developed in \citet{mendoza2015nuclear}. The network calculations on the dynamical ejecta were conducted using reaction rates for neutron capture, photo-dissociation and fission based on the FRDM mass model \cite{moller1995nuclear} as well as the $\beta$-decay rates from Ref.~\cite{moller2003new}, while the HFB21 mass model \cite{goriely2010further} and the $\beta$-decay rates from Ref.~\cite{marketin2016largescale} were used for the post-merger ejecta\footnote{The circumstance that the training data is not coherently assembled with the same nuclear physics input may seem worrisome, but the corresponding uncertainty is likely of similar magnitude, if not subdominant, compared to the intrinsic prediction uncertainties carried by the ML models.} More details about the procedure of training and testing the ML models and a discussion of their accuracy will be provided in Appendix~\ref{sec:addit-inform-about}.

\subsubsection{Inference of source terms}\label{sec:infer-source-terms}

\begin{figure}
    \centering
    \includegraphics[width=\columnwidth]{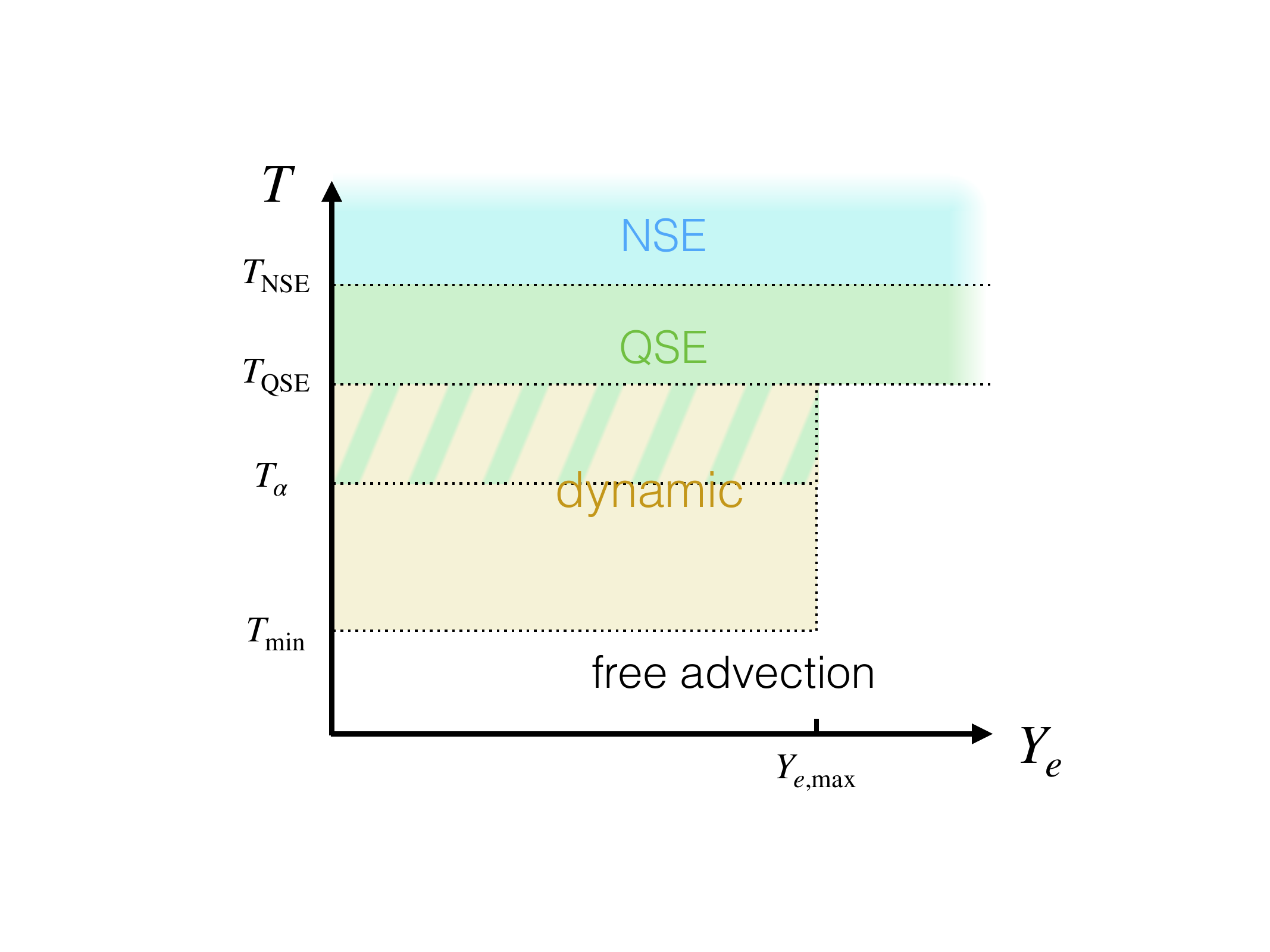}
    \caption{\label{fig:regime} Illustration of the four thermodynamic regimes in the domain of electron fraction, $Y_e$, and temperature, $T$. Nuclear statistical equilibrium (NSE) is assumed to hold above temperatures of $T_{\rm NSE}=7\,$GK. Below this temperature and up to $T_{\rm QSE}=5\,$GK, we assume quasi-statistical equilibrium (QSE) and use ML models 2{--}7 to obtain the current composition. In the dynamic regime, which prevails at temperatures $T_{\rm min}=0.1\,\mathrm{GK}<T<T_{\rm QSE}$ for $Y_e<Y_{e,\rm max}=0.45$, no equilibrium conditions apply and the equations for $\tau, Y_e, X_n$,~and~$X_h$ are integrated with the r-process related rates using ML models 8{--}16. For the abundance of $\alpha$-particles, $Y_\alpha$, we assume QSE to hold down to $T_\alpha=3\,$GK, and $\dot{Y}_\alpha=0$ at lower temperatures. For all remaining conditions we assume vanishing source terms $R_k$, i.e. free advection.}
\end{figure}

This section describes how the source terms $R_k$ are obtained from the ML models, which are needed to advance the full system of hydrodynamic equations, Eqs.~(\ref{eq:hdevo}),~(\ref{eq:mtilevo}){--}(\ref{eq:ahevo}), from an old instance of time, $t^{\rm old}$, to a new one, $t^{\rm new}=t^{\rm old}+\Delta t$. For simplicity, we assume a 1st-order time-integration scheme in this presentation, in which case the input quantities entering the ML models, listed in Table~\ref{tab:mlmodels}, are equal to the quantities defined at the old time step. An extension to higher-order (e.g. Runge-Kutta) schemes is realized by using quantities defined at some intermediate time, obtained after advancing the quantities to some partial time step, as input quantities for the ML models. We will first look at Lagrangian rates of change, $\dot{\mathcal{A}}=\dd \mathcal{A}/\dd t$, of various quantities $\mathcal{A}$ measured in the comoving frame and then convert these rates into the desired Eulerian-frame rates, $R_k$.

The treatment depends on the thermodynamic regime, which is selected based on the local temperature $T$ and electron fraction $Y_e$; see Fig.~\ref{fig:regime} for a schematic overview of the regimes. At high temperatures, $T>T_{\rm NSE}=7\,$GK, NSE is assumed to hold. Here, the r-process related source terms $R_k$ are not needed, because the composition (i.e. $X_i$ and $A_h$) is fully determined by $\rho,T,Y_e$, and $\tilm$ is given by $\tilm_{\rm NSE}$ provided by the NSE-EOS table. 

The region at intermediate temperatures, $T_{\rm QSE}=5\,\mathrm{GK}<T<T_{\rm NSE}$, and arbitrary $Y_e$ is denoted as the QSE (quasi-statistical equilibrium) regime. As was pointed out by \citet{woosley1973explosive,meyer199648ca,meyer1998theory}, the composition in this temperature range may significantly deviate from that in NSE, e.g. because triple-$\alpha$ reactions can become less efficient. Depending on the detailed conditions, $Y_h$ can either become higher or lower compared to its value in NSE, which correspondingly affects $A_h$. 

In this QSE regime we use the ML models 2{--}6 to estimate based on the local $\rho, T, Y_e$ (obtained from $D, DY_e, S^i,$~and~$\tau$ evolved via Eqs.(~\ref{eq:hdevo})) the abundances (i.e. not their rates of change) of the four evolved species, $Y_i$, and the average number of heavy nuclei, $A_h$. The rate of change of these quantities needed for Eqs.~(\ref{eq:xievo})~and~(\ref{eq:ahevo}) is then computed as:
\begin{subequations}\label{eq:NSErates}
\begin{align}
   \dot{Y}_i &= \frac{Y_i^{\rm QSE}-Y_i^{\rm old}}{\Delta t}   \, , \\
   \dot{A}_h &= \frac{A_h^{\rm QSE}-A_h^{\rm old}}{\Delta t}   \, , 
\end{align}
\end{subequations}
where the superscript ``old'' denotes the values of these quantities before the current time step. One more ML model, namely number 7 in Table~\ref{tab:mlmodels}, is used to ensure that charge balance, cf. Eq.~(\ref{eq:cb}), is fulfilled; see Appendix~\ref{sec:ensur-phys-cons} for details. Since in the QSE phase the abundance of heavy nuclei is still small, the $\beta$-decay related source terms are ignored in the QSE regime, i.e. $\srcrhineye =\srcrhinetau=0$.

The main part of the energy from r-process activity is assumed to be released in the temperature range $T_{\rm min}=0.1\,\mathrm{GK}<T<T_{\rm QSE}$ and only for neutron-rich matter with $Y_e<Y_{e,\mathrm{max}}=0.45$. In this region, called ``dynamic regime'', the composition evolves dynamically based on the actual reaction rates. In this regime, ML model 9 provides the rate by which $\beta$-decay neutrinos are emitted, $\dot{Y}_e>0$, ML model 10 the rate of neutron captures, $\dot{Y}_n<0$, and ML models 11{--}16 the rates of change for $Y_h$ and $A_h$. The abundance of heavy nuclei, $Y_h$, is assumed only to increase, not decrease. However, $\dot{Y}_h$ can be arbitrarily small before and after the phase of efficient neutron capture. Since we want to avoid wasting resources on modeling irrelevant conditions, we first call the classification model 11, which tells us whether $\dot{Y}_h$ is of significant size or very close to zero. In the former case, ML model 12 is called to predict $\dot{Y}_h$ while otherwise $\dot{Y}_h$ is set to zero. A similar strategy is used for $A_h$, but here, since $A_h$ can both increase or decrease (e.g. through fission), two classification models and two regression models are needed. First, classification model 13 is used to decide whether $\dot{A}$ is very close to zero. If true, $\dot{A}$ is assumed to vanish, and if not, ML model 14 provides the sign of $\dot{A}_h$, based on which ML models 15 and 16 are selected to predict the corresponding value of $\dot{A}_h$ or $-\dot{A}_h$. For the remaining two species, protons and $\alpha$-particles, no rates are directly predicted by ML models. For $\alpha$-particles, we were unable to construct an ML model that predicts the rate $\dot{Y}_\alpha$ accurately enough for the $\alpha$ abundance to reproduce the full nuclear network results consistently. This is likely due to the sharp dependence of the $\alpha$-related reaction rates on conditions that we cannot resolve with our reduced set of four effective species. In order to deal with this problem, we assume $\alpha$-particles to remain close to QSE conditions down to a temperature of $T_\alpha=3\,$GK (instead of 5\,GK for all other species) and freeze out afterwards. Specifically, we set 
\begin{align}\label{eq:ydotalpha}
  \dot{Y}_\alpha= \frac{Y_\alpha^{\rm QSE}-Y_i^{\rm old}}{10^{-4}\,\rm s}   \, .
\end{align}
when $T_\alpha<T\leq T_{\rm QSE}$ and $\dot{Y}_\alpha=0$ when $T<T_{\alpha}$. This seemingly crude treatment of $\alpha$-particles can be justified by the argument that their abundances tend to be subdominant \citep[e.g.][]{perego2022production, Sneppen2024e} in neutron-rich conditions with significant r-process heating (e.g. heating by more than $1\,$MeV per baryon is only expected for $Y_e\lesssim 0.3$; cf. Fig.~\ref{fig:massex}). We found that this recipe produces reasonably good agreement with nuclear network results (cf. Appendix~\ref{sec:training-testing} for a discussion of the accuracy of each ML model). Hence, the impact of this approximation on the kilonova should likewise be minor. Finally, for protons, of which the abundance is typically very small once leaving NSE/QSE and thus of little relevance, the abundance is assumed to drop to zero roughly within one expansion timescale $\lambda^{-1}$, i.e.
\begin{align}\label{eq:ydotp}
  \dot{Y}_p= -Y_p^{\rm old}\lambda   \, .
\end{align}

A noteworthy restriction of the above treatment in the dynamic regime is that ML models 10{--}16 describing $\dot{Y}_n, \dot{Y}_h,$~and~$\dot{A}_h$ are only used if the material is currently expanding, i.e. if $\dot{\rho}<0$, whereas these rates are set to zero otherwise\footnote{Without this restriction, we found artificially strong cooling in individual fallback regions, which is likely related to poor sampling of fallback conditions by the training data, resulting in a bad performance of the ML models for these conditions.}. In principle, weakly bound material (such as in the outer layers of the accretion disk) can fall back and undergo photo-disintegration reactions before being ejected again (or accreted onto the BH) later. However, since most of this material reaches NSE or QSE conditions when falling back, which resets its composition, we assume the overall impact of this limitation to be less problematic. Regardless of the sign of $\dot\rho$, $Y_e$ is always evolved in the dynamic regime (using ML model 9) to account for the effect of $\beta$-decays.

The conversion of rest-mass energy into thermal and neutrino energy is described by using ML model 1, which for a given average baryon number, $A_h$, and charge fraction, $z_h=Z_h/A_h$ (computed via Eq.~(\ref{eq:cb})), of heavy nuclei predicts the quantity $\tilm$. The corresponding rate of change is then given by
\begin{align}\label{eq:mdot}
  \dot{\tilm}= \frac{\tilm-\tilm^{\rm old}}{\Delta t}   \, .
\end{align}
We intentionally do not directly predict the rate of change, $\dot{\tilm}$, by an ML model, in order to keep the value of $\tilm$ as consistent with the current composition as possible. In particular, this ensures that $\tilm$ can never exceed its physical bounds, $-1\lesssim \tilm/(1\,\mathrm{MeV}/c^2)\lesssim 8$, and therefore avoids the risk of overestimating the effective heating rate.

From the total energy-release rate $\dot{\tilm}c^2$ a relative fraction $f_\nu$ is lost to neutrinos emitted in $\beta$-decays, which do not contribute to heating up the fluid. This quantity is predicted by ML model~8 in RHINE which, however, is only evaluated in the case of $\dot{\tilm}<0$ (corresponding to net energy release), while $f_\nu$ is set to zero otherwise. In the underlying nuclear-network data used for training ML model~8 we estimate this fraction as:
\begin{align}
 f_\nu = \begin{cases}
\displaystyle{\frac{\sum_k f_{\nu,k} Q_{k} \lambda_{\beta^-,k}}{|\dot{\tilde m}^{\rm net}| c^2}}  & \text{if }\dot{\tilm}^{\rm net}<0 \, ,  \\
 0 & \text{otherwise \, ,}
\end{cases} 
\end{align}
where $f_{\nu,k}$ is the ratio between the average neutrino energy and the $Q$-value, $Q_{k}$, for a specific reaction labeled by $k$, $\lambda_{\beta^-,k}$ the corresponding reaction rate, and $\dot{\tilm}^{\rm net}$ the rate of change of the average baryon mass induced by all nuclear reactions along the nuclear-network trajectory. We take the average neutrino energy from the table provided by \citet{marketin2016largescale}, while the quantities $Q_{k}$ and $\lambda_{\beta^-,k}$ depend on the adopted nuclear mass model (see Appendix~\ref{sec:addit-inform-about} for more details on the training data). The value of $f_\nu$ typically lies in the range $\sim 0.05\ldots0.2$. This quantity appears to be notoriously difficult to fit by ML models, at least given the set of input quantities that are available during our evolution scheme. A larger number of quantities providing more information about the composition of heavy nuclei seems to be needed in order to resolve the neutrino emission rate as function of the evolved quantities more accurately. Thus, in some cases we find relatively large errors for the final amount of energy emitted in neutrinos (cf. $E_\nu$ in Sect.~\ref{sec:applications}). However, considering that neutrino losses contribute only a sub-dominant fraction to the total energy released, the impact of this uncertainty on the ejecta dynamics, which lie in the main focus of our study, is relatively small.

If the current fluid element is in the QSE or in the dynamic regime, where the abundances were (partially) predicted from ML models, one more step is needed to ensure that the new composition implied by the source terms fulfills mass balance and charge balance, i.e. Eqs.~(\ref{eq:mc}), to which end some of the predicted rates have to be slightly modified. This step is necessary because the ML models for the various rates are unaware of each other and each rate is predicted with an intrinsic fitting error. The details of this procedure are presented in Appendix~\ref{sec:ensur-phys-cons}.

In the dynamic regime or in the QSE regime, the last step consists of computing the source terms $R_k$ for the conserved variables from the comoving-frame rates presented above. This is done as follows:
\begin{align}\label{eq:rkrates}
  \begin{pmatrix}
    \srcrhineye \\ \srcrhinetau \\ \srcrhinexn \\ \srcrhinexp \\ \srcrhinexa \\ \srcrhinexh \\
    \srcrhineah \\ \srcrhinerest
  \end{pmatrix}
  = D 
  \begin{pmatrix}
    \dot{Y}_e \\ f_\nu\dot{\tilm}/m_u \\ \dot{Y}_n \\ \dot{Y}_p \\ 4\dot{Y}_\alpha \\
    \dot{X}_h \\ \dot{A}_h \\ \dot{\tilm}
  \end{pmatrix} \, ,
\end{align}
where the rate of change of $X_h$ is computed as $\dot{X}_h=\dot{A}_h Y_h+A_h\dot{Y}_h + \Delta t \dot{A}_h\dot{Y}_h$.

In the fourth and last thermodynamic regime of RHINE, the free-advection regime present at temperatures of $T<0.1\,$GK for $Y_e<0.45$ or $T<5\,$GK for $Y_e>0.45$, we assume the composition to be inert and r-process heating to be irrelevant, i.e. all eight rates $R_k=0$.

This concludes the description of the RHINE scheme. The scheme will be validated by comparing the results from simulations using RHINE with full nuclear network results in Sect.~\ref{sec:applications}.

\section{Validation and Applications}\label{sec:applications}

We now discuss hydrodynamic simulations utilizing the RHINE scheme. A suite of spherically symmetric wind models is considered in Sect.~\ref{sec:spher-symm-winds} and subsequently in Sect.~\ref{sec:end-end-merger} the scheme is applied to long-term NSM models. The intention of these applications is twofold: First, we aim to test the ability of the ML-based RHINE scheme to reproduce full nuclear networks by comparing the RHINE simulations with the corresponding post-processing results from nuclear-network calculations. After validating the scheme, we study the impact of r-process heating and its dependence on the ejecta properties by comparing the RHINE models with corresponding models not using RHINE. None of the simulations discussed in this section have previously been used for training the ML models, i.e. we are applying the ML models to unseen data. The simulations are conducted with the finite-volume neutrino-hydrodynamics code Aenus-ALCAR adopting spherical polar coordinates. For more details about this code we refer to previous papers, e.g., \citet{Just2015b, Just2022b}.

\subsection{Spherically symmetric winds}\label{sec:spher-symm-winds}

In this first application of RHINE we consider relatively simple spherically symmetric, steady-state wind solutions. Although steady-state winds are not necessarily representative of (all) ejecta components encountered in NSMs, these tests -- in contrast to full-fledged merger simulations -- are, first, straightforward to reproduce and therefore suited as comparison tests for other schemes describing r-process heating, and second, they permit a systematic exploration of the dependence on the primarily relevant nucleosynthesis parameters electron fraction, expansion velocity, and entropy. For the current tests, we ignore all neutrino interactions apart from r-process related $\beta$-decay by setting $\srcnuye = \srcnumom= \srcnutau=0$ in Eqs.~(\ref{eq:hdevo}).

Winds are outflow solutions of the hydrodynamic equations that are characterized by the existence of a sonic point, i.e. a radius below (above) which the expansion velocity is smaller (greater) than the sound speed (see, e.g., Ref.~\cite{Frank2002b}). We assume Newtonian gravity with the mass of the central object being $M_c=2.5\,M_\odot$. For a given gravitational potential, a wind solution is defined by fixing the final outflow velocity, $v_{\infty}=v(r\rightarrow\infty)$, the electron fraction before onset of r-process nucleosynthesis, $Y_{e,0}$, and the entropy before onset of r-process nucleosynthesis\footnote{Note that in the case of ignoring r-process related source terms $Y_e=Y_{e,0}$ and $s=s_0$ in the entire wind.}, $s_0$. We first define a set of seven baseline models not using RHINE, labeled by ``noRHINE'' hereafter. Motivated by typical properties found in NSM models we choose the fiducial model to have $Y_{e,0}=0.25, v_{\infty}/c=0.1$,~and $s_0/k_B=15$ and vary these parameters within $v_{\infty}/c=0.03,0.3$,  $Y_{e,0}=0.1, 0.4$, and $s_0/k_B=8, 40$; see Table~\ref{tab:wind}. In the noRHINE models, all RHINE-related source terms are turned off during the evolution (i.e. $\srcrhinerest$, $\srcrhinexn$, $\srcrhinexp$, $\srcrhinexa$, $\srcrhinexh$, $\srcrhineah$, $\srcrhineye$, and $\srcrhinetau$; cf. Sect.~\ref{sec:evolution-equations}) and NSE is assumed down to temperatures of $T=5\,$GK, below which the composition is just advected but not changed.

The wind simulations are conducted by injecting material at some radius $r_{\rm min}$ into the computational domain and following the evolution until reaching a stationary state within some large radius $r_{\rm max}$ at which the velocity has saturated. In order to ensure the final wind velocity $v(r_{\rm max})$ to be equal to a desired $v_{\infty}$, we first have to find the correct boundary conditions, i.e. the values of $D, \tau, S^i$ etc. at $r_{\rm min}$. More details about this non-trivial step, which involves finding the aforementioned sonic point, are provided in Appendix~\ref{sec:constr-wind-solut}. We adopt a radial grid with 640 cells extending from $r_{\rm min}=100\,$km to $r_{\rm max}=3\times 10^7\,$km with cell widths increasing by about 2\,\% per cell. Once having constructed the noRHINE models, the corresponding RHINE simulations are then conducted using the same boundary conditions at $r_{\rm min}$ but switching on the RHINE source terms (i.e. activating $\srcrhinerest$, $\srcrhinexn$, $\srcrhinexp$, $\srcrhinexa$, $\srcrhinexh$, $\srcrhineah$, $\srcrhineye$, and $\srcrhinetau$ at temperatures below $T_{\rm NSE}=7$\,GK). In order to test the accuracy of the RHINE scheme, we post-process each RHINE model by running a full nuclear-network calculation on the corresponding outflow trajectory using the same nuclear-network setup as in \citet{just2023end} (called ``network B'' therein), namely the HFB21 mass model from \citet{goriely2010further} and $\beta$-decay rates from \citet{marketin2016largescale}. The post-processing results will be denoted as ``pp'' hereafter.

\begingroup
\begin{table*}[t]
\begin{ruledtabular}
\centering
\caption{\label{tab:wind} Parameters and results for the spherically symmetric wind models. Each wind model is run without (``noRHINE'') and with (``RHINE'') r-process heating, and the RHINE versions are additionally post-processed (``pp'') with a detailed nuclear network. The first three columns provide the defining parameters of all wind models, namely the electron fraction at the base of the wind (before onset of nucleosynthesis), the final velocity if no r-process heating is included, and the entropy at the base. The remaining columns provide the final wind velocity with r-process heating, the rest-mass energy per baryon released below $T=5\,$GK (cf. Eq.~(\ref{eq:deltilm})) as well as the energy per baryon emitted in neutrinos (Eq.~(\ref{eq:enu})) in a Lagrangian fluid element, and the relative difference of the net heating energy (Eq.~(\ref{eq:deleheat})) between the RHINE and pp results. In columns showing two numbers the first number provides the RHINE result and the second the pp result.}
\begin{tabular}{cccccccc}
  wind model & $Y_{e,0}$ & $v^{\rm noRHINE}_{\infty}$ & $s_0$   & $v_{\infty}^{\rm RHINE}$ & $\Delta\tilm^{\rm RHINE/pp}$ & $E_{\nu,\rm \infty}^{\rm RHINE/pp}$ & ($\Delta E_{\rm heat}^{\rm RHINE}/\Delta E_{\rm heat}^{\rm pp})-1$ \\
             &           & [$c$]                      & [$k_B$] & [$c$]                    & [\MeVc]                        & [MeV]                               & [\%]                                \\\colrule
  1          & 0.25      & 0.1                        & 15      & 0.127                    & 3.13/3.12                    & 0.53/0.73                           & $+9.0$                              \\
  2          & 0.1       & 0.1                        & 15      & 0.145                    & 6.29/6.03                    & 1.27/0.90                           & $-2.2$                              \\
  3          & 0.4       & 0.1                        & 15      & 0.102                    & 0.16/0.22                    & 0.03/0.03                           & $-29$                               \\
  4          & 0.25      & 0.03                       & 15      & 0.084                    & 3.12/3.03                    & 0.62/0.46                           & $-2.7$                              \\
  5          & 0.25      & 0.3                        & 15      & 0.309                    & 3.14/3.13                    & 0.58/0.55                           & $-0.7$                              \\
  6          & 0.25      & 0.1                        & 8       & 0.127                    & 2.93/2.85                    & 0.15/0.03                           & $-1.1$                              \\
  7          & 0.25      & 0.1                        & 40      & 0.124                    & 3.14/3.16                    & 0.54/0.58                           & $+0.7$                              \\
\end{tabular}
\end{ruledtabular}
\end{table*}
\endgroup

\begin{figure*}
    \centering
    \includegraphics[width=0.999\textwidth]{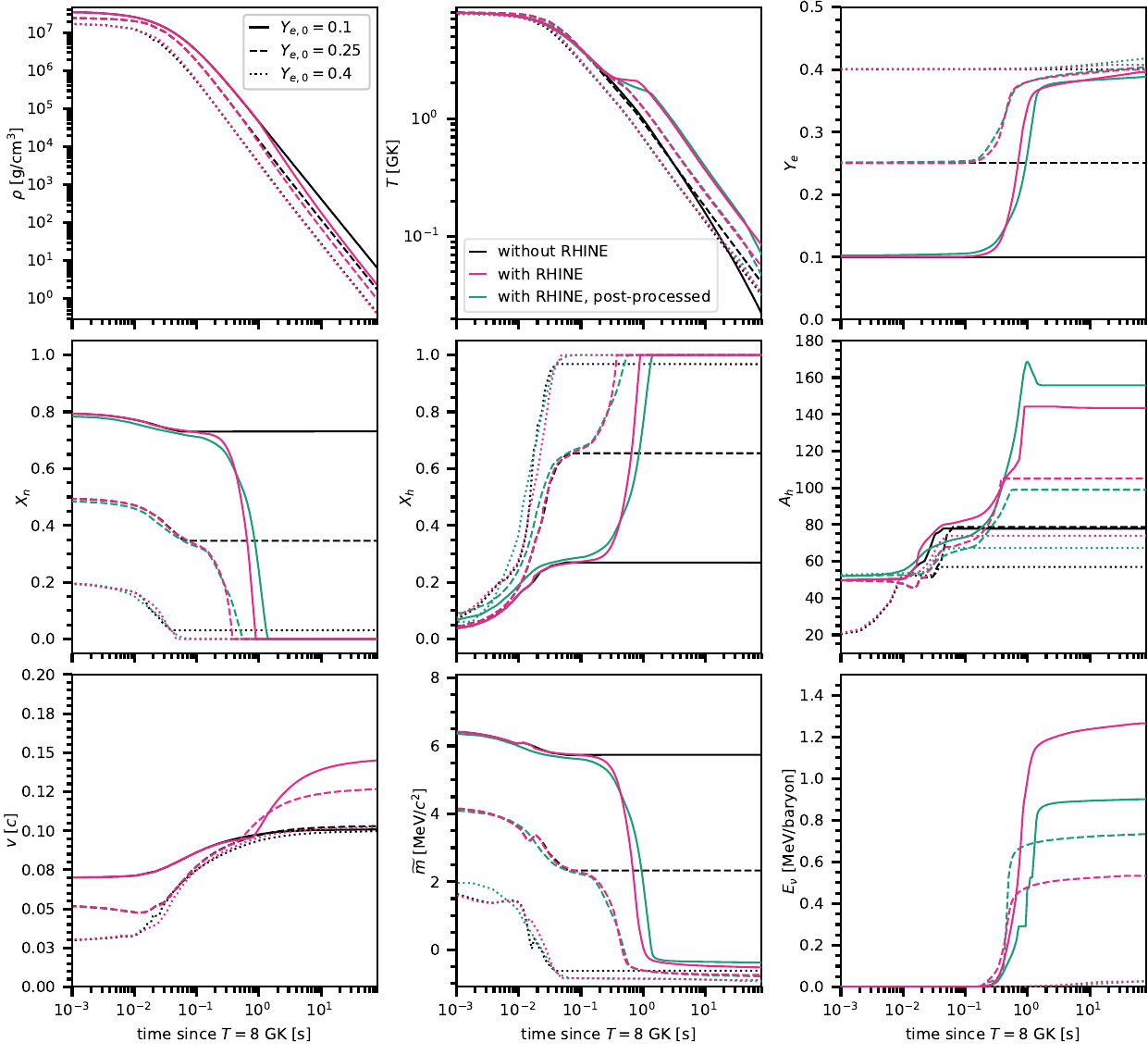}
    \caption{\label{fig:wind1} Global properties of wind models 1{--}3 with different initial electron fractions, $Y_{e,0}$, as functions of the time elapsed since the material reached a temperature of $T=8\,$GK, namely the density, $\rho$, temperature, $T$, electron fraction, $Y_e$, neutron mass fraction, $X_n$, heavy-nuclei mass fraction, $X_h$, average mass number of heavy nuclei, $A_h$, velocity, $v$, average mass excess per baryon, $\tilm$, and integrated energy per baryon lost to neutrinos in $\beta$-decays, $E_\nu$. Black lines denote hydrodynamic models without r-process heating, magenta lines hydrodynamic models with r-process heating, and cyan lines nuclear-network results obtained from post-processing the simulations with r-process heating.}
\end{figure*}
\begin{figure*}
    \centering
    \includegraphics[width=0.999\textwidth]{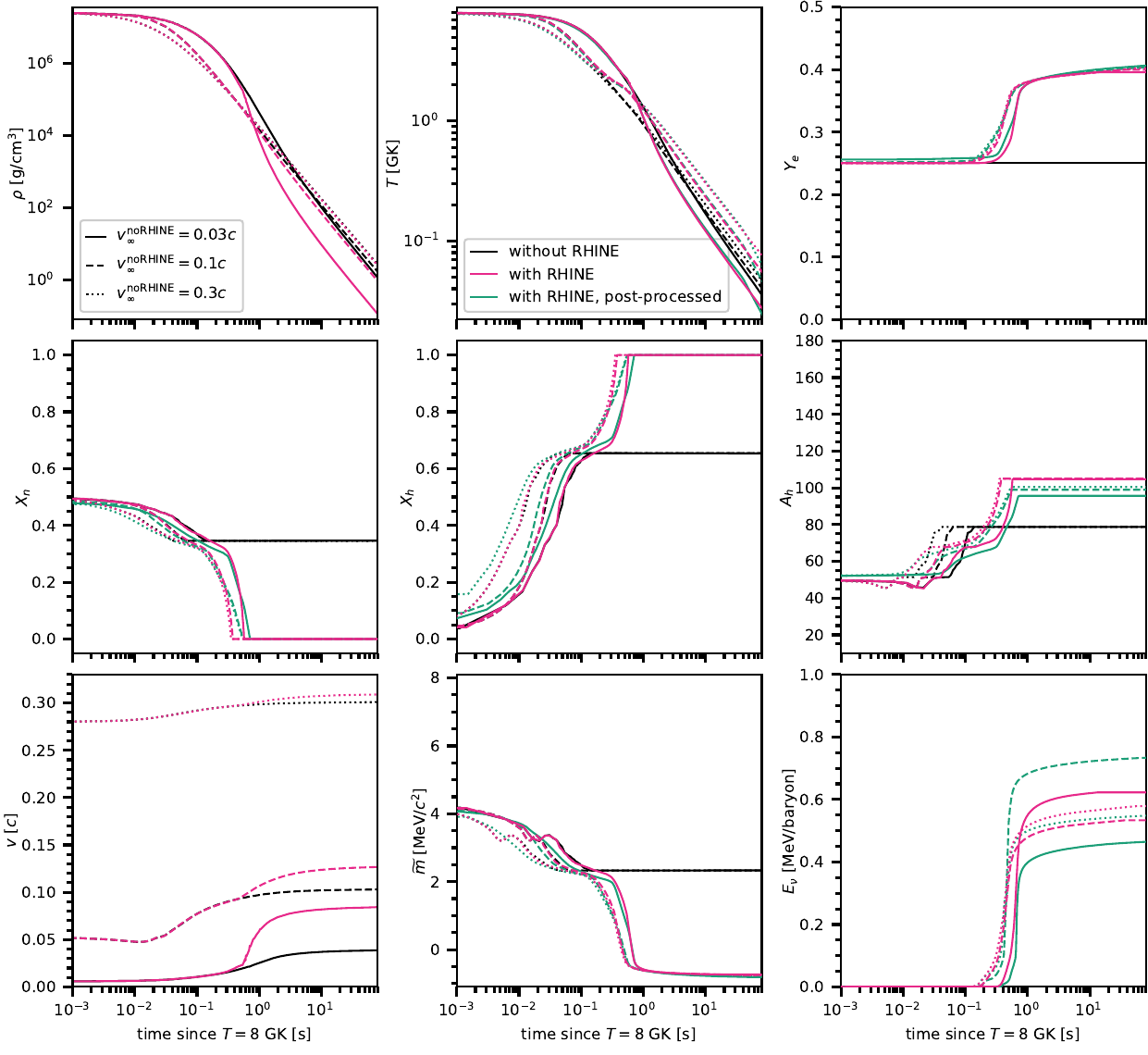}
    \caption{\label{fig:wind2} Same as Fig.~\ref{fig:wind1} but for wind models $4,1$,~and~$5$ that differ by the asymptotic velocity of the wind without r-process heating, $v_\infty^{\rm noRHINE}$.}
\end{figure*}
\begin{figure*}
    \centering
    \includegraphics[width=0.999\textwidth]{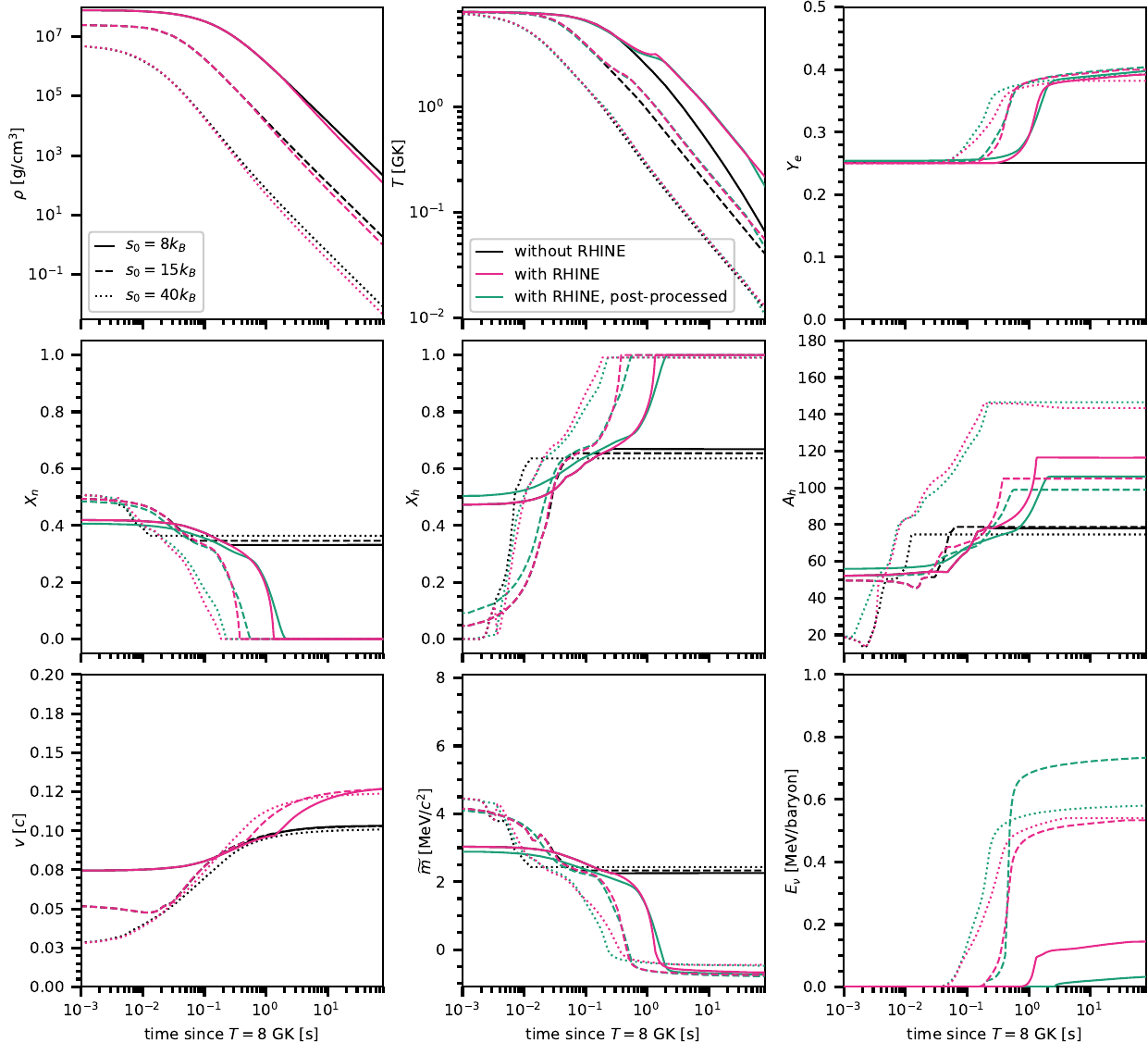}
    \caption{\label{fig:wind3} Same as Fig.~\ref{fig:wind1} but for wind models $6,1$,~and~$7$ that differ by the initial entropy per baryon of the wind, $s_0$.}
\end{figure*}

Table~\ref{tab:wind} summarizes the model parameters and main results for all investigated cases, and Figs.~\ref{fig:wind1},~\ref{fig:wind2},~and~\ref{fig:wind3} compare the time evolution of various quantities for the series of models varying the electron fraction, velocity, and entropy. Consistent with the post-processing results, all RHINE models show a transition of $Y_e$ towards $\sim$0.4, $X_n\rightarrow 0$, $X_h\rightarrow 1$, increase of $A_h$, reduction of $\tilm$ towards 0\,\MeVc or lower, and loss of neutrino energy between 0 and 1\,MeV within the characteristic r-process timescale of $\sim$1\,s. The agreement between the RHINE models (magenta lines in Figs.~\ref{fig:wind1}--\ref{fig:wind3}) and the post-processing results (cyan lines) is overall very good. This demonstrates the robustness and accuracy of the ML-based RHINE approach, particularly in view of the fact that the type of wind solutions considered in this section were not used for training the ML models. Most importantly, the net amount of heating energy per baryon, $\Delta E_{\rm heat}$ dumped into the material below $T=5\,$GK agrees to within less than $10\,\%$ (cf. Table~\ref{tab:wind}), except in wind model 3 where, however, the amount of heating energy is only $\sim$\,0.2\,MeV. We define the net heating energy as:
\begin{align}\label{eq:deleheat}
  \Delta E_{\rm heat}= \Delta \tilm c^2 - E_{\nu,\infty}  \, ,
\end{align}
i.e. as the difference between the rest-mass energy liberated below 5\,GK,
\begin{align}\label{eq:deltilm}
  \Delta \tilm c^2 = \tilm(T=5\,\mathrm{GK})c^2 - \tilm(t\rightarrow \infty)c^2 \, ,
\end{align}
and the cumulative neutrino energy lost in a Lagrangian fluid element during its expansion,
\begin{align}\label{eq:enu}
  E_{\nu}(t) = \int_{T=5\,\rm GK}^{t} \frac{\srcrhinetau(\tilde{t}) m_u}{D(\tilde{t})}\dd \tilde{t} \, ,
\end{align}
where $E_{\nu,\infty}=E_{\nu}(t\rightarrow\infty)$. The good agreement of the heating rates also explains the excellent match of the temperature evolution between RHINE and the post-processed models (cf. top middle panel in Figs.~\ref{fig:wind1}--\ref{fig:wind3}).

We also observe some differences between both sets of models. However, given the approximate nature of the underlying ML models, a perfect agreement is not expected. The most noticeable discrepancy is obtained for the neutrino energy losses, $E_{\nu}$ (bottom right panel), where the final values differ by tens of percent for several models. As mentioned in Sect.~\ref{sec:infer-source-terms}, the neutrino losses are particularly difficult to predict within the ML framework. Fortunately, the neutrino energy loss represents only a small contribution to the net heating energy, i.e. the corresponding error in $\Delta E_{\rm heat}$ remains small. For the very neutron-rich wind model 2 (case $Y_e=0.1$ shown in Fig.~\ref{fig:wind1}) we also notice a discrepancy in the evolution of $A_h$: In the post-processed data $A_h$ climbs to higher values and shows a drop after reaching a peak at around 1\,s, while in the RHINE model $A_h$ reaches slightly lower values at the end of the r-process and does not show a pronounced peak. The drop in $A_h$ is related to fission, which is known to be relevant only for very neutron-rich ejecta with $Y_e\lesssim 0.15$~\cite{Goriely:2015,Cowan2021origin}. In our training data used for building the ML models, such conditions are reached mainly in the dynamical ejecta which, however, are dominated by trajectories with higher outflow velocities (cf. Sect.~\ref{sec:end-end-merger}) compared to the current ones with $0.1\,c$. We therefore suspect that the discrepancy seen for the wind model with $Y_{e,0}=0.1$ is, at least partially, explained by the circumstance that the physical regime is not well sampled by the adopted set of training data (reflected by the small number of data points that is used to train ML model 16 adopted for predicting negative $\dot{A}_h$). Extending the set of training data could reduce the disagreement for $A_h$. However, the otherwise good results suggest that the impact of this limitation on $\Delta\tilm$ seems to be small.

We now take a look at the impact of r-process heating by comparing the RHINE models with the models not using RHINE (black lines in Figs.~\ref{fig:wind1}--\ref{fig:wind3}). As expected from the behavior of $\tilm_{\rm NSE}$ (see right plot of Fig.~\ref{fig:massex}) the total amount of rest-mass energy released below 5\,GK, $\Delta\tilm$ in Table~\ref{tab:wind}, is most sensitive to the choice of the initial $Y_e$, decreasing from about 6.3\,MeV to 0.2\,MeV for initial $Y_e$ values increasing from 0.1 to 0.4. Only very little variation of $\Delta\tilm$ is observed when changing the wind velocity and entropy. 

On the other hand, the impact on the velocity depends sensitively on the velocity regime (cf. Fig.~\ref{fig:wind2}): For similar values of $\Delta\tilm\approx 3\,$\MeVc, the final velocity $v_{\infty}$ is nearly tripled after switching on RHINE for an otherwise slow trajectory with $v^{\rm noRHINE}_\infty=0.03\,c$, while in the case of the trajectory originally reaching 0.3\,$c$ the RHINE-related velocity change is only 3\,\%. This strong sensitivity to the initial velocity is the expected result of the approximately quadratic dependence of the kinetic energy on the velocity (cf. Fig.~\ref{fig:vboost}). As can be seen particularly well for the trajectory with $v^{\rm noRHINE}=0.03\,c$, the r-process related acceleration can also significantly speed up the temporal decline of the density (top left panel of Fig.~\ref{fig:wind2}) and, by that, counteract the tendency of increasing the temperature relative to the noRHINE models.  This explains why the temperatures (top middle panel) drop faster, instead of more slowly, in the RHINE version compared to the noRHINE version of this model.

The value of the initial ejecta entropy $s_0$ (cf. Fig.~\ref{fig:wind3}) affects neither $\Delta\tilm$ nor the final velocity boost, $v_\infty^{\rm RHINE}-v_\infty^{\rm noRHINE}$, significantly (cf. Table~\ref{tab:wind}), i.e. seems to have little relevance for the dynamical impact of r-process heating. We notice some differences in the time when r-process acceleration kicks in (earlier for higher entropy). However, these timescale differences are likely just a ramification of the fact that the expansion timescales, $\rho/\dot{\rho}$, are shorter for the models with higher entropies, which in turn is related to the particular way of constructing the current trajectories that assume the same, spatially constant Bernoulli parameter; cf. Appendix~\ref{sec:constr-wind-solut}). Although not relevant for the dynamics, we observe a significantly stronger impact of heating on the temperature for initially low entropies, while the temperature evolution is nearly unaffected for the trajectory with $s_0=40\,k_B$, well in agreement with the nuclear-network post-processing calculation.

For all investigated cases we notice a significant delay between the release of rest-mass energy (which is typically finished within $\lesssim$\,1{--}2\,s) and the corresponding shift of the velocity to its new final value in response to heating (which may take a few seconds or tens of seconds). An analytical estimate of the acceleration timescale, $\tau_{\rm acc}$, on which thermal energy is converted to kinetic energy through pressure forces can be found from the first law of thermodynamics. Assuming that the expansion at late times is adiabatic and rest-mass energies do not vary anymore, the change of the thermal energy per baryon, $\epsilon=e_{\rm therm}/n_B$, in a fluid element of fixed baryon number is given by
\begin{align}\label{eq:deps}
 \dd \epsilon = \frac{P}{n_B}\frac{\dd \rho}{\rho} \, .
\end{align}
Using the continuity equation and ignoring relativistic corrections, this leads to
\begin{align}\label{eq:depsdt}
 \frac{\dd \epsilon}{\dd t} = - \frac{P}{n_B}\nabla_jv^j \simeq - \frac{\epsilon}{3}\nabla_jv^j \simeq - \epsilon\frac{v_r}{r} \, ,
\end{align}
where in the second step we assumed a radiation-dominated medium, $P\approx e_{\rm therm}/3$, and in the last step a homologous velocity field, $v_r\propto r$. It follows that
\begin{align}\label{eq:tauacc}
 \tau_{\rm acc} \approx \frac{\epsilon}{\left|\dd \epsilon/\dd t\right|} \simeq \frac{r}{v_r} \simeq t \, ,
\end{align}
i.e. thermal acceleration essentially takes place on the expansion timescale, $r/v_r$. This means that, once heating energy is released, the subsequent acceleration takes roughly the time required for the cloud to double its radial size.
\begin{figure*}
    \centering
  \includegraphics[trim= 0 0 10 0,clip,height=0.17\textheight]{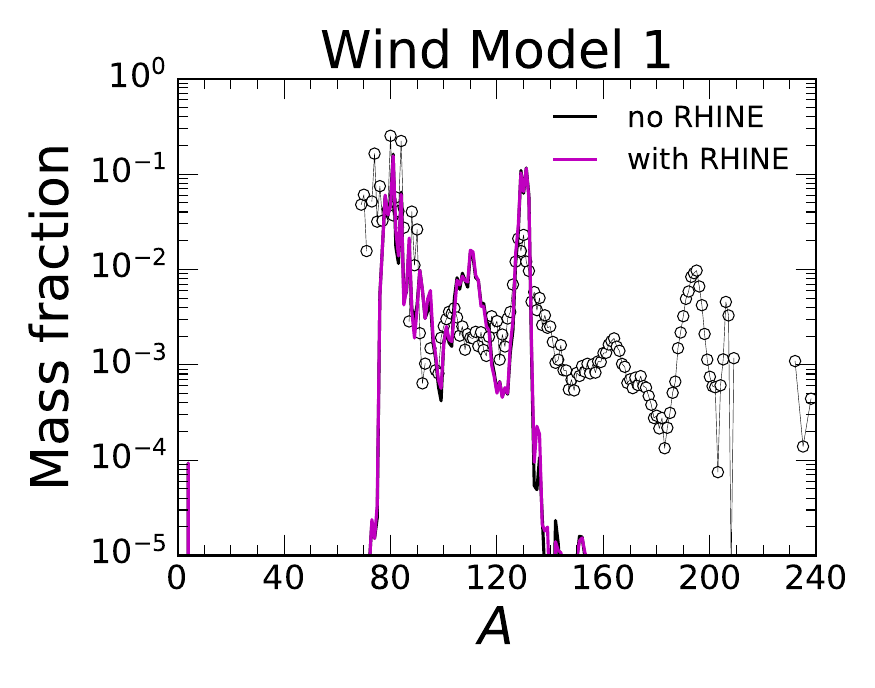}
  \includegraphics[trim=80 0 10 0,clip,height=0.17\textheight]{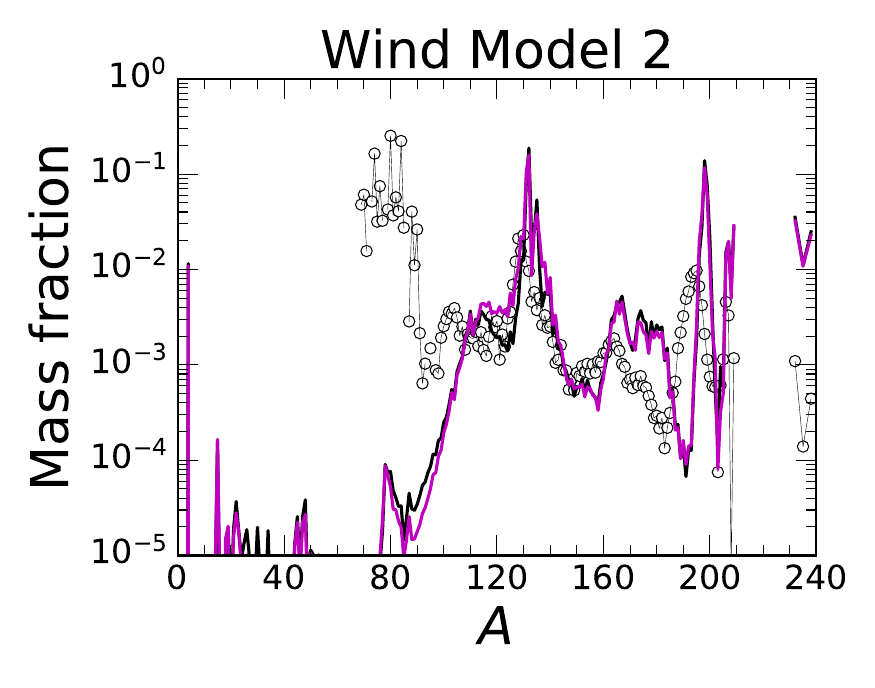}
  \includegraphics[trim=80 0 10 0,clip,height=0.17\textheight]{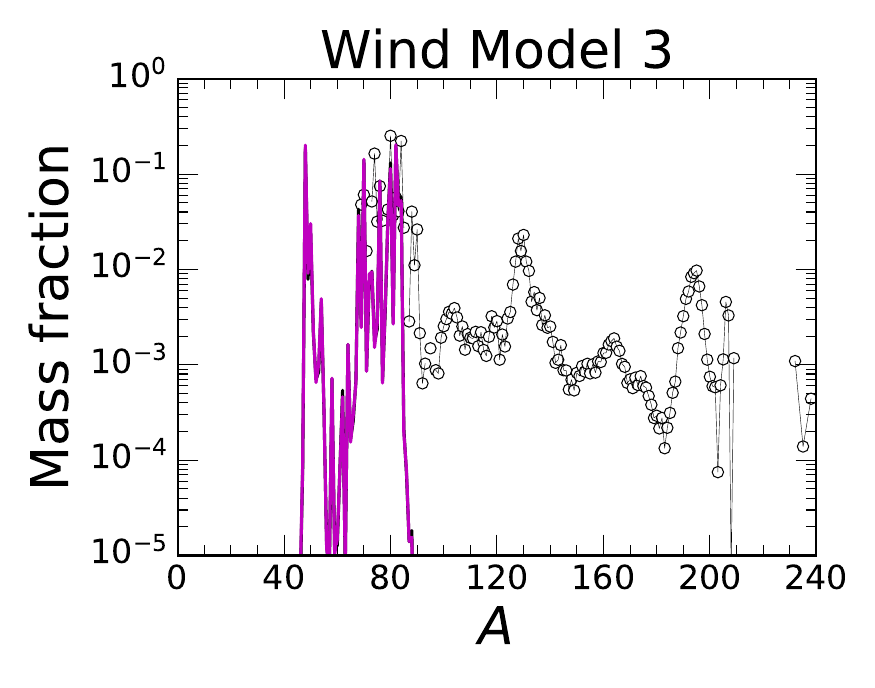}
  \includegraphics[trim=80 0 10 0,clip,height=0.17\textheight]{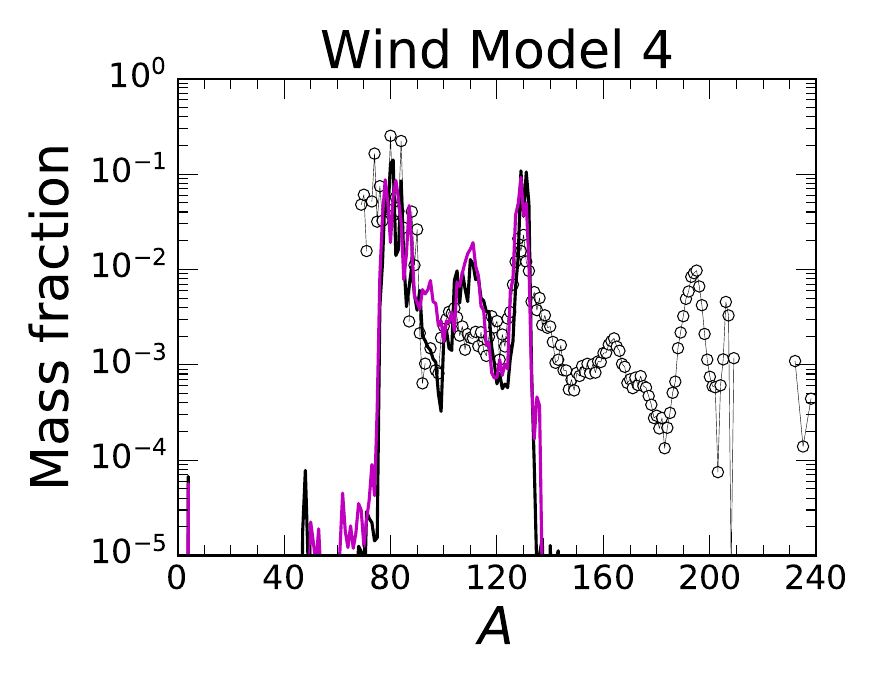}
  \includegraphics[trim= 0 0 10 0,clip,height=0.17\textheight]{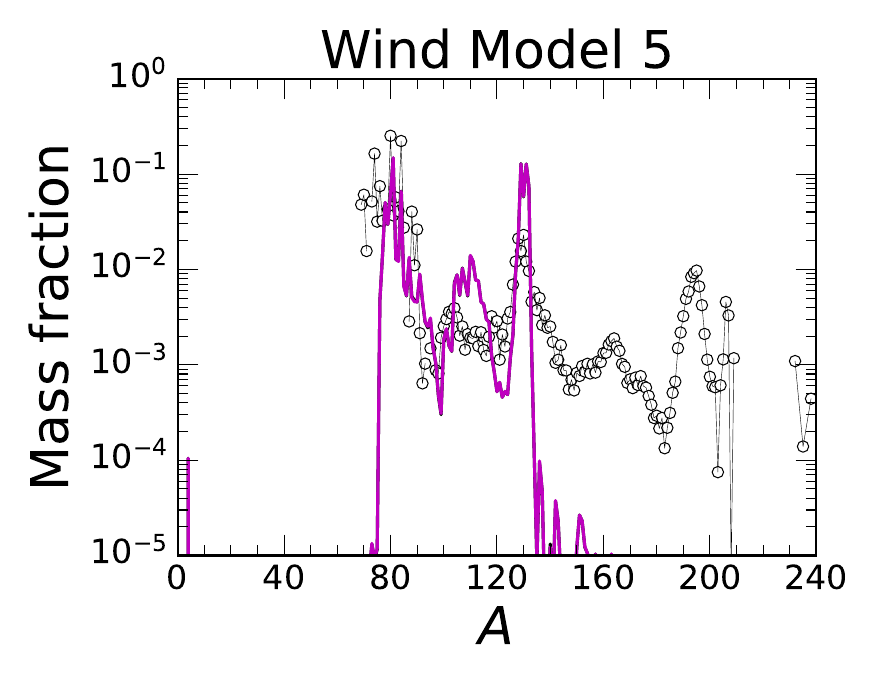}
  \includegraphics[trim=80 0 10 0,clip,height=0.17\textheight]{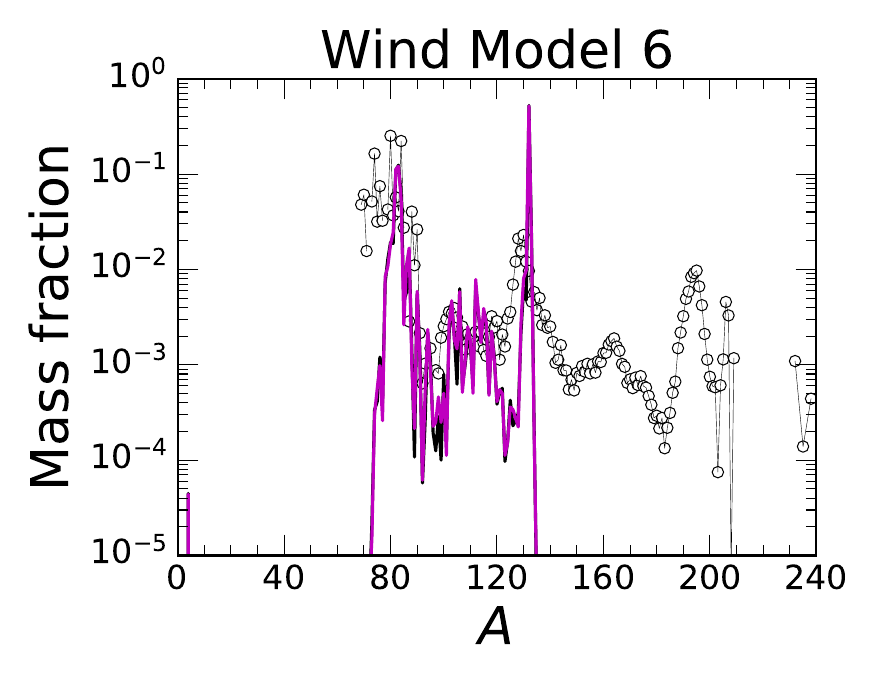}
  \includegraphics[trim=80 0 10 0,clip,height=0.17\textheight]{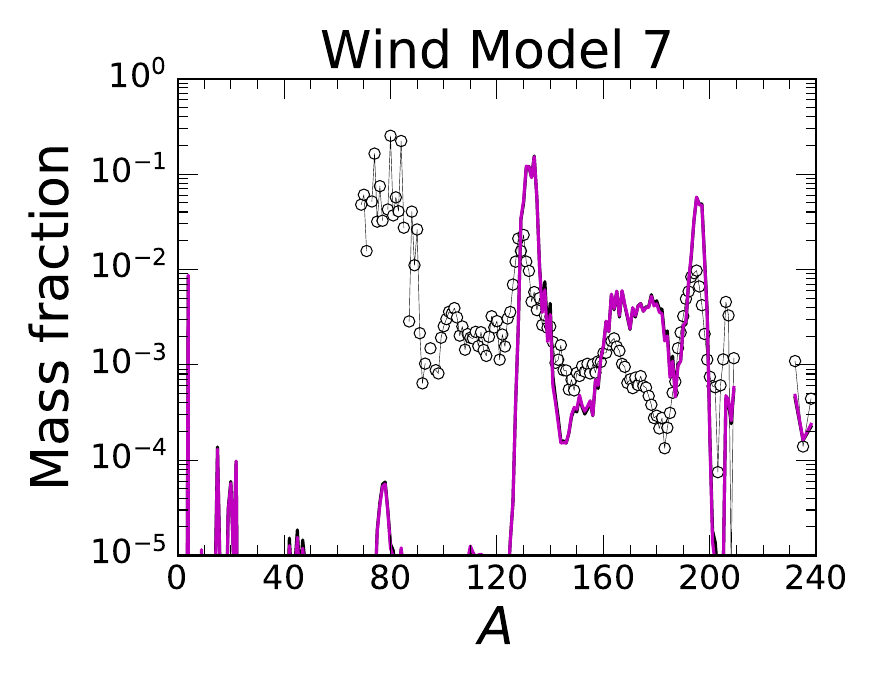}
  \caption{\label{fig:xwind} Distributions of mass fraction as function of mass number obtained from post-processing the steady-state wind models 1{--}7 with a nuclear network. Black lines denote the wind models without r-process heating and magenta lines the models with r-process heating. As a guidance, the (arbitrarily scaled) solar r-process distribution is shown with black circles.}
\end{figure*}

We finally examine the impact of r-process heating on the nucleosynthesis yields (measured at 1\,Gyr) by comparing in Fig.~\ref{fig:xwind} the post-processed noRHINE models (black lines) with the post-processed RHINE models (purple lines). None of the considered models leads to a qualitative change of the abundance pattern in the sense that an entire group of elements is produced in one case but not in the respective other. Most differences seen in the mass fractions are only a few (tens of) percent for individual nuclei. As anticipated, the two wind models 2 ($Y_{e,0}=0.1$) and 4 ($v_{\infty}^{\rm noRHINE}=0.03\,c$) with the strongest velocity boost, $v_\infty^{\rm RHINE}-v_\infty^{\rm noRHINE}$, also show the most notable differences in the abundance pattern. In both models, the region between the 1st and 2nd r-process peak seems to be affected most, with some elements being produced a factor of about two more efficiently with heating compared to without.

\subsection{Neutron-star mergers}\label{sec:end-end-merger}

Next we examine the impact of r-process heating in the scenario that RHINE was primarily developed for, namely long-term evolution models of binary NSMs. The merger model we consider is very similar, though not exactly identical\footnote{The noteworthy differences are: A refined treatment of the smoothing length $h_{\rm SPH}$ in the 3D merger model \citep{Ardevol-Pulpillo2019a} avoiding a limitation of $h_{\rm SPH}$ for ejecta at large radii; a new method to map the temperature distribution from the 3D merger to the 2D post-merger simulation assuming hydrodynamic rotational equilibrium, which reduces initial oscillations triggered by the mapping and leads to an overall more stable NS remnant surviving longer before BH-formation; inclusion of neutrino-electron scattering in the post-merger simulation (using the explicit treatment of Ref.~\cite{oconnor2015open}); a consistent transition from the high-density EOS to the low-density EOS regime retaining the correct values of $\tilm$ and $A_h$ instead of switching to a 4-species NSE composition (without advection of $\tilm$ and $A_h$) as done previously.}, to model sym-n1-a6 discussed in \citet{just2023end}. The model consists of a 3D simulation describing the early phase until a post-merger time of $t=10\,$ms, a 2D simulation following the subsequent evolution of the NS remnant, its collapse to a BH, and the viscous disintegration of the BH-torus system until $t=10$\,s, and a final simulation following the expansion of just the ejected material until $t_{\rm fin}=100$\,s. In the considered model, BH formation commences at $t=t_{\rm BH}\approx 214\,$ms. In each modeling step, the simulation domain is large enough to contain all of the ejected material, i.e. no material is lost through crossing the outer boundary. We again conduct two simulations of this model, one with RHINE and one without, in which all RHINE-related source terms are turned off and NSE is assumed above $T=5\,$GK. We then post-process both simulations with a full nuclear network (the same as used in Sect.~\ref{sec:spher-symm-winds}). The outflow trajectories, i.e. the Lagrangian properties of outflow mass elements, are extracted from both models by integrating the paths $\vec{r}_l(t)$ (with $l=1,\ldots,N_{\rm traj}$) of $N_{\rm traj}\sim 2000$ tracers per model sampling the ejecta backwards in time using the previously written output data. The initial points $\vec{r}_l(t_{\rm fin})$ for this path integration are given by a regular 2D mesh spanned by a radial grid and a polar-angle grid, and the masses associated with each trajectory are $m_l^{\rm ej}=D(\vec{r}_l(t_{\rm fin}),t_{\rm fin})\Delta V_l$ with the conserved density $D=\rho W$ and the volume element of the initialization mesh $\Delta V_l$. We ensured that the spatial sampling and temporal resolution used is fine enough to not affect the final results significantly.  For more details regarding the modeling techniques and physics properties of these models, see \citet{just2023end}.

For each model the set of trajectories is divided into three groups representing different ejecta components using the same criteria as in \citet{just2023end}: Trajectories with $|\vec{r}_l(t=10\,\mathrm{ms})|>250\,$km are classified as dynamical ejecta, those with $|\vec{r}_l(t=t_{\rm BH})|>1000\,$km as NS-torus ejecta, and the remaining trajectories as BH-torus ejecta. The mapping from 3D to 2D at $t=10\,$ms introduces a significant smearing of the $Y_e$ distribution in the dynamical ejecta. This smearing can be corrected for in the trajectory data by combining the dynamical-ejecta trajectories from the 3D and 2D simulations at $t=10\,$ms in the way outlined in Appendix~D of \citet{just2023end}. However, this procedure can only correct the output data used for post-processing but not the data ``seen'' by RHINE during the simulation. Hence, RHINE will be operating on a somewhat smeared $Y_e$ distribution of the dynamical ejecta, and we therefore anticipate a certain discrepancy between the RHINE simulation and the post-processing results for the dynamical-ejecta component. Nevertheless, due to their already high velocities, dynamical ejecta are expected to be less dramatically affected by r-process heating (cf. Fig.~\ref{fig:vboost}). We note for clarity that this smearing effect is not intrinsic to RHINE, only to the currently considered merger models.

\begingroup
\begin{table*}[t]
\begin{ruledtabular}
\centering
\caption{\label{tab:mergermodels} Masses, average electron fractions (measured at $T=5\,$GK), average final velocities of the ejecta in the neutron-star merger simulations without and with RHINE as well as the average r-process heating energy per baryon as obtained during the RHINE simulation and as resulting in the post-processing nuclear-network calculations of the trajectories extracted from the two simulations. Each column provides three numbers, which refer to the dynamical ejecta, NS-torus ejecta, and BH-torus ejecta in that order.}
\begin{tabular}{lcccc}
  merger model          & $M_{\rm ej}^{\rm dyn/NS/BH}$ & $\langle Y_{\rm 5\,\rm GK}\rangle ^{\rm dyn/NS/BH}$ & $\langle v_{\rm ej}\rangle ^{\rm dyn/NS/BH}$ & $\langle \Delta E_{\rm heat}\rangle ^{\rm dyn/NS/BH}$ \\
                        & [$10^{-2}M_\odot$]           &                                                     & [$c$]                                        & [MeV/baryon]                                          \\\colrule
  without RHINE              & 0.510/4.148/4.929            & 0.257/0.390/0.297                                   & 0.277/0.143/0.050                            & 0/0/0                                                  \\
  without RHINE, post-processed &                              &                                                     &                                              & 2.258/0.809/1.870 \\                  
   with RHINE                 & 0.514/4.158/6.000            & 0.258/0.390/0.279                                   & 0.283/0.157/0.073                            & 2.301/0.705/2.102                                     \\
 with RHINE, post-processed &                              &                                                     &                                              & 2.240/0.815/2.178                                     
\end{tabular}
\end{ruledtabular}
\end{table*}
\endgroup
\begin{figure*}
    \centering
     \includegraphics[width=\textwidth]{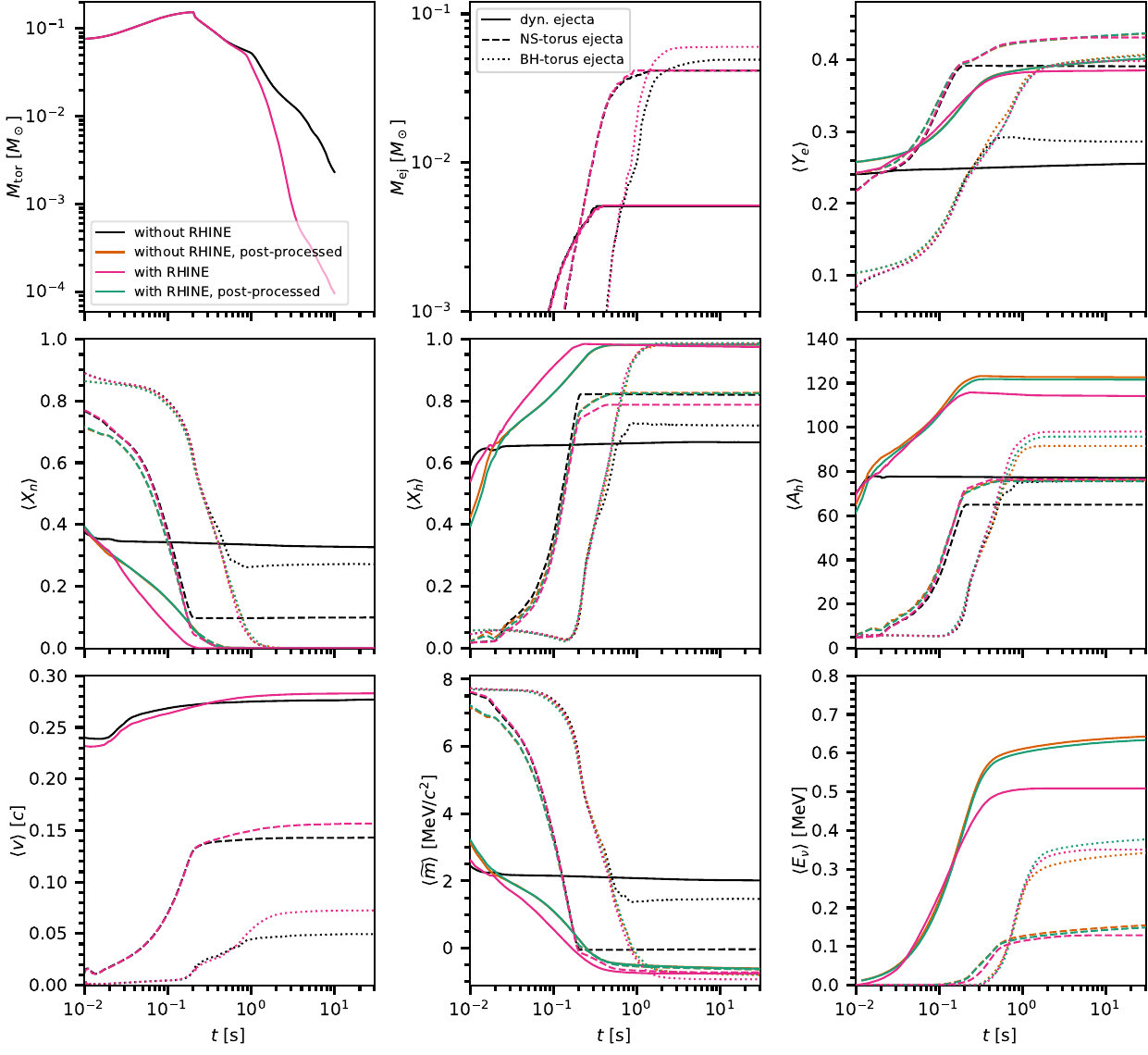}
    \caption{\label{fig:hmns} Global properties of the neutron-star merger simulations and post-processing calculations as functions of time post merger, namely the torus mass $M_{\rm tor}$ (cf. Eq.~(\ref{eq:mtor})) and ejecta mass $M_{\rm ej}$ (Eq.~(\ref{eq:mej})), as well as averages over the ejecta (computed using Eq.~(\ref{eq:trajavg})) of the electron fraction $Y_e$, neutron mass fraction $X_n$, heavy-nuclei mass fraction $X_h$, average mass number of nuclei $A_h$, velocity $v$, average mass excess per baryon $\tilm$, and integrated energy loss by $\beta$-decay neutrinos $\Delta E_\nu$. In all panels except the one for $M_{\rm tor}$ solid lines refer to dynamical ejecta, dashed lines to NS-torus ejecta, and dotted lines to BH-torus ejecta. In the panels for $M_{\rm tor}, M_{\rm ej},$~and~$\langle v\rangle$ the post-processing results are identical to the corresponding simulation data, because the post-processing does not change velocities or masses of tracer particles. In the other panels orange lines are, if invisible, overlaid by the cyan lines.}
\end{figure*}

Table~\ref{tab:mergermodels} provides the ejecta masses, velocities, and net deposited energies per baryon for the three ejecta components, and Fig.~\ref{fig:hmns} shows the time evolution of the torus mass,
\begin{align}\label{eq:mtor}
 M_{\rm tor}(t) = \int_{\rm torus } D(\vec{r},t) \dd V \, ,
\end{align}
with the integration considering only material inside a radius of $r=10^4\,$km and with density $D<10^{12}\,$g\,cm$^{-3}$, the cumulative mass of material ejected beyond $10^4$\,km,
\begin{align}\label{eq:mej}
 M_{\rm ej}(t) = \sum_{|\vec{r}_l(t)|>10^4\,\rm km}  m_l^{\rm ej} \, ,
\end{align}
and averages of various quantities $\mathcal{A}$, computed as
\begin{align}\label{eq:trajavg}
 \langle \mathcal{A}\rangle(t) = \frac{\sum_l \mathcal{A}_l(\vec{r}_l(t),t) m_l^{\rm ej}}{\sum_l m_l^{\rm ej}} \, .
\end{align}
As for the accuracy of RHINE in reproducing full nuclear-network results, the comparison between the purple and cyan lines in Fig.~\ref{fig:hmns} reveals very good agreement for the averaged characteristic quantities $Y_e, X_n, X_h, A_h, \tilm$,~and even $E_\nu$. For the dynamical ejecta the detailed time dependence is not recovered as accurately as for the other ejecta components, most likely due to the mapping-related smearing of the $Y_e$ distribution mentioned previously. However, the final amount of energy released per baryon, $\langle \Delta E_{\rm heat}\rangle$, agrees very well between RHINE and the corresponding full network results for all three ejecta components (cf. Table~\ref{tab:mergermodels}). The dynamical ejecta reach the largest value ($\langle \Delta E_{\rm heat}\rangle\approx 2.3\,$MeV), followed by the BH-torus ejecta (2.1\,MeV), and the NS-torus ejecta (0.7\,MeV). The amount of released energy per baryon is broadly consistent (modulo neutrino losses) with the energy reservoir that is left when departing from NSE conditions, i.e. with the difference between the freeze-out values of $\tilm_{\rm NSE}(Y_e,T\rightarrow 0)$ (cf. right panel of Fig.~\ref{fig:massex} evaluated at $Y_e\approx\langle Y_e\rangle$ for each ejecta component) and the asymptotic value $\tilm\approx -1\,$\MeVc typically reached at the end of r-process activity. Concerning the energetics, our results thus lend support to the treatment of \citet{Haddadi2023a}, who parametrize the temporal increase of $Y_e$ towards $Y_e\approx 0.4$ while keeping the composition in NSE. In this way, \citet{Haddadi2023a} likewise achieve a final state with $\tilm_{\rm NSE}(Y_e\rightarrow 0.4,T\rightarrow 0)\approx -1\,$\MeVc. However, their scheme is unable to capture the time dependence of $\langle Y_e\rangle(t)$ consistently, because they assume $Y_e$ to increase on the same global timescale everywhere, whereas we see different r-process timescales for each ejecta component\footnote{Curiously, in \citet{Haddadi2023a}, the simulations ``Nuc'' (with a similar NSE-EOS treatment as in our model without RHINE) and ``rproc'' (same as Nuc but with r-process heating) suggest that r-process heating effectively reduces the mass and velocity of the disk ejecta, in contrast to the expected impact and to our results. The authors argue, however, that this tendency is likely an artificial result of including the $\beta$-decay related source terms for energy and electron fraction in regions of high temperature and high density, in which in reality the r-process would not operate.}.

For completeness we also provide in Table~\ref{tab:mergermodels} and Fig.~\ref{fig:hmns} the results obtained by post-processing the simulation not using RHINE. This case represents the conventional post-processing treatment used in many previous studies not accounting for the dynamical impact of r-process heating. The good agreement between the two post-processing results with and without RHINE (cp. orange with cyan lines in Fig.~\ref{fig:hmns}) reflects the fact that the impact of r-process heating on the nucleosynthesis results is overall mild (see below for a discussion of the abundance patterns). Moreover, it justifies the use of ML models trained with data from simulations not including r-process heating. If the agreement had turned out to be poor, we would likely have to re-train the ML models using trajectories from the RHINE simulations and repeat this process until reaching convergence. The only noteworthy discrepancy between the two post-processing results is seen for $\langle \Delta E_{\rm heat}\rangle^{\rm BH}$ in Table~\ref{tab:mergermodels}, which is likely explained by the tendency that the additional BH-torus ejecta generated through r-process heating are slightly more neutron rich than the average BH-torus ejecta launched in the model without RHINE (see next two paragraphs).

Consistent with the expectations from energetic considerations (cf. Fig.~\ref{fig:vboost}), the most substantial dynamical feedback from r-process heating is observed for the slowest ejecta component, namely the BH-torus ejecta, of which the average velocity is boosted by  $\sim$\,40\,\%. In addition to the acceleration, however, this ejecta component is also $\sim$\,20\,\% more massive in the RHINE model, suggesting that r-process heating amplifies the, otherwise mainly viscously driven, matter ejection process during the BH-torus evolution. This interpretation is supported by the behavior of $M_{\rm tor}$ and $M_{\rm ej}$ in Fig.~\ref{fig:hmns}, showing that the torus disintegrates (and drives outflows) considerably faster in the RHINE model compared to the reference model. The outcome that only the ejecta mass of the BH-torus component, and barely that of the other components, is affected is likely explained by two circumstances: First, most of the r-process energy is released on timescales of tens to hundreds of milliseconds, which is too late to get a significant effect on the launching processes of the dynamical and NS-torus ejecta. Second, due to viscous spreading the average radius of the torus increases \citep[e.g.][]{Metzger2009b} and, as a consequence, the (absolute value of the) gravitational binding energy decreases with time, reaching values comparable to the released r-process energy of a few MeV per baryon. Compared to more strongly bound material located at smaller radii, loosely bound disk material is thus affected more by r-process heating. A qualitatively similar increase in the mass and velocity of disk ejecta was reported by \cite{Wu2016a}, who parametrized r-process heating based on the ejecta temperature, and by \cite{Kawaguchi2024c}, who followed the long-term evolution of disk- and dynamical-ejecta material beyond $r=3000\,$km from NS-BH mergers previously post-processed with nucleosynthesis calculations. We note that hydrodynamical models that simulate only the ejected material by starting at large radii are, by design, less likely to capture this feedback effect of r-process heating responsible for enhancing the disk-ejecta masses \citep[e.g.][]{Klion2022a}.

\begin{figure*}
    \centering
    \includegraphics[width=\textwidth]{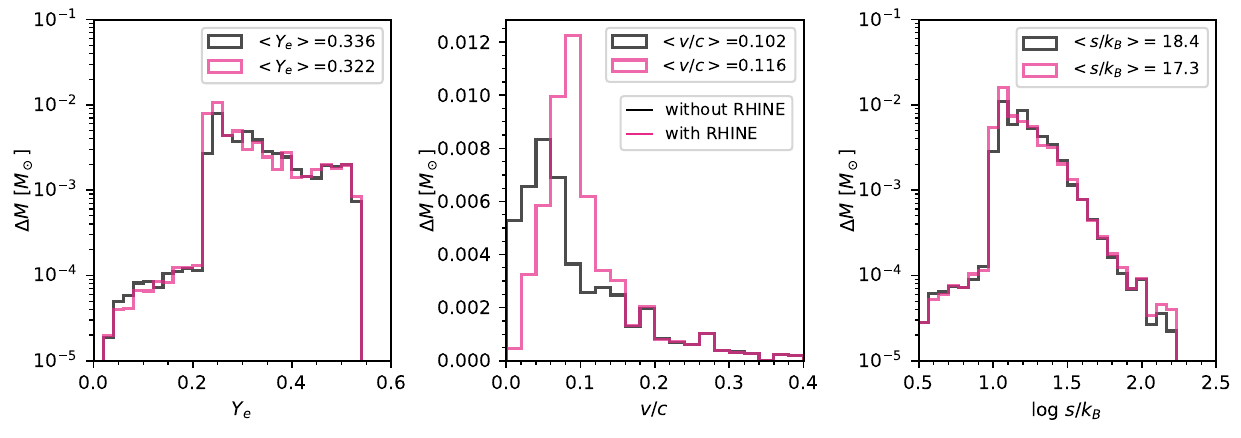}
    \caption{\label{fig:hist} Histograms depicting the mass distribution in $Y_e$ at $T=5\,$GK, final velocity, and entropy per baryon at $T=5\,$GK of the material ejected in the neutron-star merger simulations without (black lines) and with (magenta lines) r-process heating.}
\end{figure*}
\begin{figure*}
    \centering
    \includegraphics[width=\textwidth]{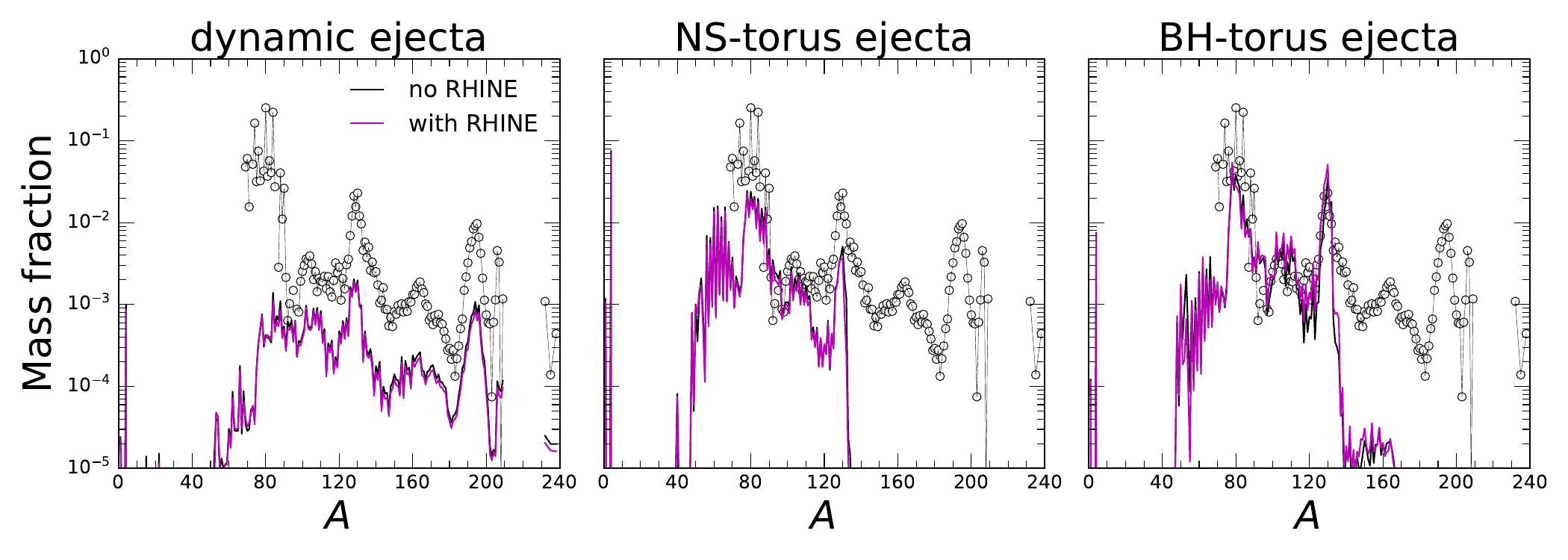}
    \caption{\label{fig:abundance} Nucleosynthesis yields as functions of mass number in the material ejected in the neutron-star merger simulations without (black lines) and with (magenta lines) r-process heating individually for each ejecta component. Black circles show the (arbitrarily scaled) solar r-process distribution.}
\end{figure*}

The mass-velocity histogram in Fig.~\ref{fig:hist} confirms that the final velocity distribution is mainly affected in the low-velocity regime, $v/c\lesssim 0.1$, while the distribution at $v/c\gtrsim 0.2$ remains nearly unchanged. Thus, despite a greater amount of injected energy compared to BH-torus ejecta, the dynamical ejecta component is only little affected by r-process heating. This result may seem contradictory to \citet{Rosswog2014}, who report a relatively strong impact on the dynamical ejecta in their NSM simulations. However, the dynamical ejecta in their Newtonian models are significantly slower compared to more realistic general relativistic results \citep[e.g.][]{Bauswein2013, Hotokezaka2013b} and, hence, the relative impact of heating on the velocity distribution is likely overestimated in these models. Our results are also not in conflict with \citet{Kawaguchi2024c}, who find a relatively strong impact of r-process heating in their dynamical ejecta from NS-BH mergers. In these events, in contrast to NS-NS mergers, the majority of the dynamical ejecta is expelled in the form of cold tidal tails unaffected by shocks or weak interactions. The electron fraction in NS-BH dynamical ejecta is therefore lower and, correspondingly (cf. Fig.~\ref{fig:massex}), the amount of r-process heating higher.

Looking again at Fig.~\ref{fig:hist}, the distributions of $Y_e$ and $s$ in the ejecta measured at $T=5\,$GK seem to be barely affected by r-process heating, which is plausible because it mainly acts below this temperature. Not surprisingly, therefore, we see only small differences in the global nucleosynthesis yields of the three ejecta components, shown in Fig.~\ref{fig:abundance}. The BH-torus ejecta again exhibit the most noticeable differences, although the discrepancy between the two models reaches only up to a factor of $\sim$\,2 for individual nuclei (at least when considering just the nuclei that are produced by a significant number). Just like in the case of the wind trajectories (cf. Sect.~\ref{sec:spher-symm-winds}), r-process heating thus only has a small impact on the final abundance pattern in NSM ejecta, consistent with previous studies using more simplified treatments \citep[e.g.][]{Rosswog2014,Just2015a,Wu2016a}. It is worth pointing out that this conclusion is not in tension with the recent study by \citet{Magistrelli2024b}, who performed a set of Lagrangian hydrodynamics simulations including full nuclear-network calculations in each grid cell. The (compared to our results) slightly more significant differences seen in their comparison of abundances (see, e.g., Fig.~2 of that study) are mainly attributed by the authors to the circumstance that their reference models, called ``coupled homologous'' and ``post-process'', assume a parametrized density evolution of the ejecta (after 30\,ms post merger or along the entire expansion in the two respective models). Our models, in contrast, follow the hydrodynamic evolution much longer, namely until $100$\,s. Thus, while their study highlights the importance of evolving the ejecta for a sufficiently long time (as opposed to parameterizing the expansion), it does not allow to draw conclusions on just the dynamical impact of nuclear energy release, which is the purpose of our study.

\begin{figure*}
    \centering
    \includegraphics[width=0.495\textwidth]{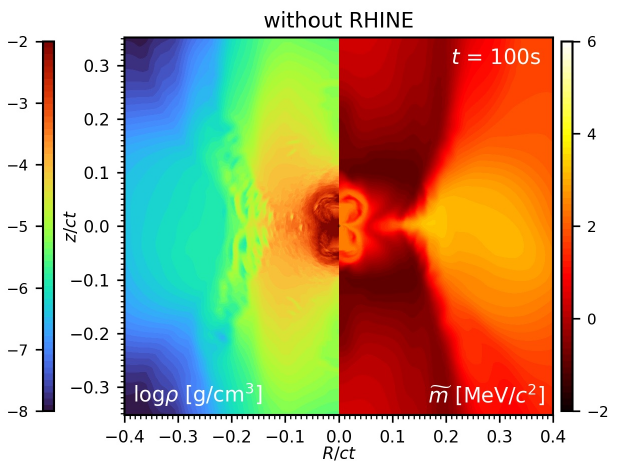}
    \includegraphics[width=0.495\textwidth]{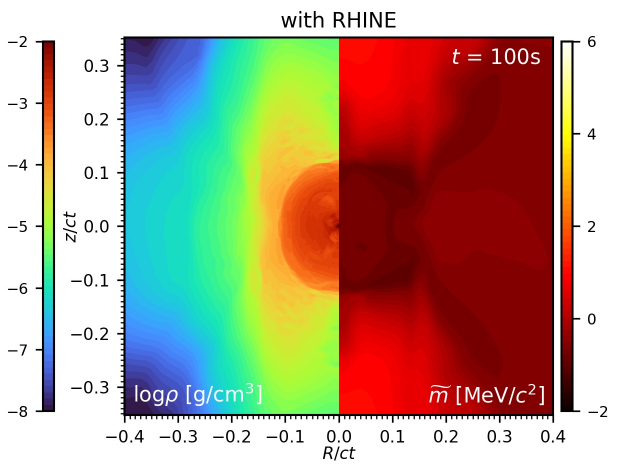}
    \caption{\label{fig:morphology} Snapshots showing the distribution of density (left sides) and average mass excess per baryon (right sides) of the material ejected in the models without r-process heating (left panel) and with r-process heating (right panel) at $t=100\,$s. R-process heating smoothens small-scale variations and causes the low-velocity BH-torus ejecta to inflate and attain a more spherical geometry.}
\end{figure*}

In Fig.~\ref{fig:morphology} snapshots are compared of $\rho$ and $\tilm$ at $t=100\,$s for the two hydrodynamic models. Apart from accelerating the BH-torus ejecta and enhancing their mass, r-process heating also has a bearing on the ejecta geometry. The thermal expansion fueled by r-process heating inflates the material into all directions, not just the radial direction, effectively smoothening the ejecta structure and creating a more spherical distribution. This tendency can be seen clearly for the innermost BH-torus ejecta. In the RHINE case they become significantly more spherical as they expand into the surrounding NS-torus ejecta that are not subject to substantial r-process heating. The tendency of r-process heating to drive the ejecta towards sphericity has been reported also in previous studies using simulations with a more simplified heating treatment \citep[e.g.][]{Rosswog2014, Fernandez2015c, Klion2022a, Sneppen2023b} or considering NS-BH instead of NS-NS mergers \citep[e.g.][]{Darbha2021a, Kawaguchi2024c}. However, given the relatively small amount of injected r-process energy (of $\sim$\,2{--}3\, MeV/baryon on average), the geometry of material at higher velocities ($v/c\gtrsim 0.1$) is only marginally affected. R-process heating (alone) is thus probably not a viable mechanism to create sphericity at velocities of $v/c\sim 0.2\ldots 0.3$. The spherical geometry suggested by the early kilonova spectrum of AT2017gfo \citep{Sneppen2023b} therefore remains puzzling. 

\begin{figure}
    \centering
    \includegraphics[width=0.99\columnwidth]{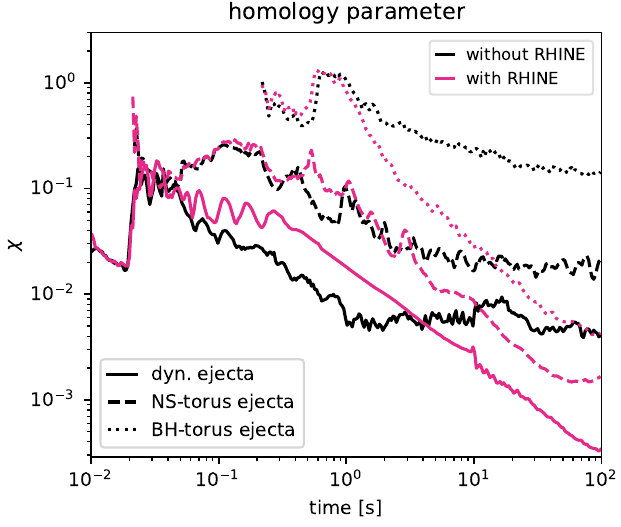}
    \caption{\label{fig:homol} Homology parameter as function time measuring the average degree by which fluid elements in the three ejecta components are frozen out in velocity space. With r-process heating (magenta lines) the ejecta are less homologous during the early phase of r-process acceleration but subsequently become homologous faster than ejecta without r-process heating (black lines).}
\end{figure}

Another aspect related to the density structure is connected to the question whether r-process heating can possibly delay or shorten the time for the distribution to become homologous, i.e. to freeze out in velocity space such that $r\propto v$. A suitable measure for the degree of homology is the ratio between the current time and the characteristic timescale of velocity changes \citep[e.g.][]{Rosswog2014, Neuweiler2023a}. Various possibilities exist and have been used in the literature for computing a homology parameter $\chi$. Here we are primarily interested in assessing the overall trend of $\chi$ in our models and ejecta components and not in a quantitative comparison with previous studies. In order to compute $\chi$ we use the Lagrangian velocities, $v_l(t)=|\vec{v}_l(t)|$, recorded along the set of $N_{\rm traj}$ ejecta trajectories that we extracted a posteriori from our simulations. We first define a homology parameter for each trajectory as
\begin{align}\label{eq:hom1}
 \chi_l(t) =  \frac{t}{v_l(t)}\frac{\dd v_l(t)}{\dd t} 
\end{align}
and from this compute a global homology parameter by mass-weighted averaging,
\begin{align}\label{eq:hom2}
 \chi(t) =  \frac{\sum_l \chi_l(t)m^{\rm ej}_l}{\sum_l m^{\rm ej}_l} \, ,
\end{align}
including in the sums only trajectories of one of the three ejecta components and ignoring trajectories with $r(t)<1000\,$km and $v(t)<0.003\,c$. The resulting time evolution of $\chi(t)$ is provided\footnote{It is possible that $\chi$ is affected by numerical errors associated with the path integration of the trajectories via post processing. However, considering that the same path-integration time steps are used for both simulations and all ejecta components, the relative tendencies between the different models and ejecta components indicated by Fig.~\ref{fig:homol} should be robust.} in Fig.~\ref{fig:homol}. In both hydrodynamic models $\chi$ decreases the fastest for the dynamical ejecta and the slowest for the BH-torus ejecta, which is the expected behavior considering the order of expansion speeds for the three ejecta components (see Table~\ref{tab:mergermodels}). Switching on r-process heating leads initially, i.e. roughly within the first second of evolution, to higher values of $\chi$ as a result of the enhanced thermal acceleration. At later times, however, $\chi$ drops significantly faster for each component in the RHINE model compared to the reference model. This can be understood from the tendency of heating to obliterate local extrema in the pressure distribution (e.g. clumps and gaps; cf. Fig.~\ref{fig:morphology}) that otherwise could induce velocity changes at later times. Overall, we thus find that r-process heating expedites the transition into homologous expansion.

\begin{figure*}
    \centering
    \includegraphics[width=0.98\textwidth]{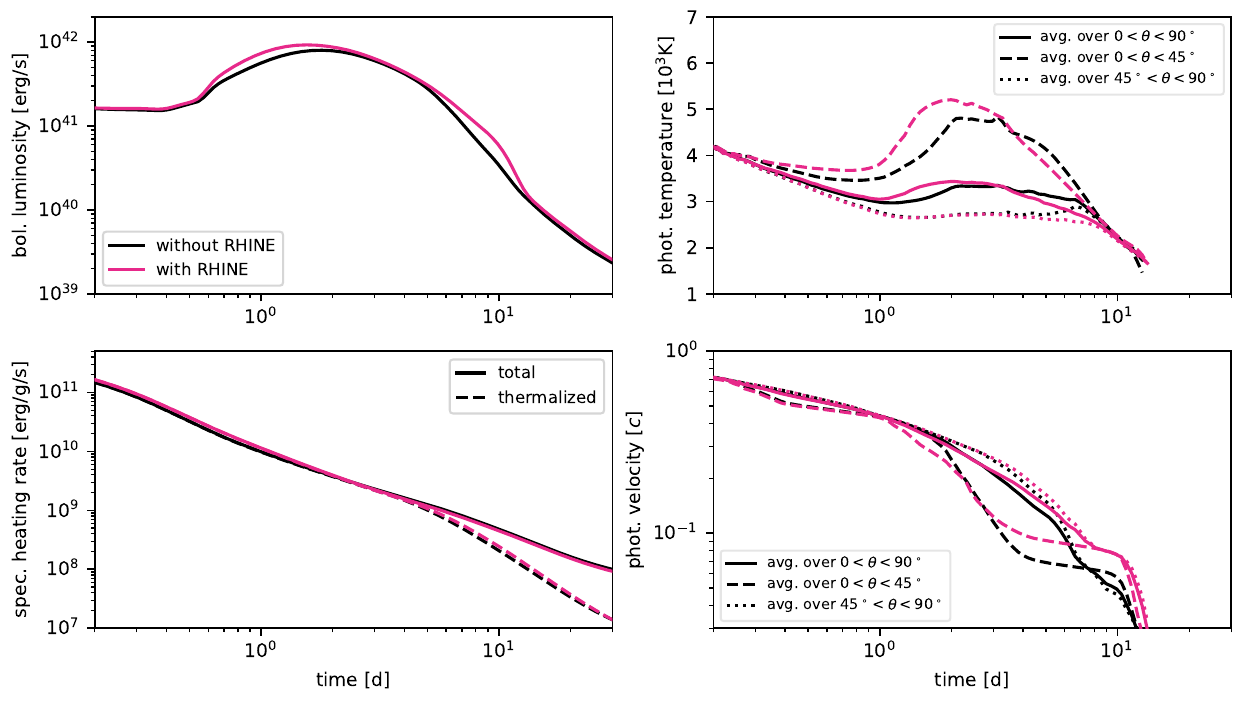}
    \caption{\label{fig:kilonova} Properties of the kilonova signal for the neutron-star merger models without r-process heating (black lines) and with r-process heating (magenta lines). The left panels show the bolometric luminosity integrated over all directions and the specific rate by which radioactive energy is released (solid lines) and deposited (dashed lines), not including neutrino losses. The right panels show the gas temperature and velocity measured at and averaged over the photosphere, where the average includes either the entire northern hemisphere (solid lines), or just the regions extending $45^\circ$ from the north pole (dashed lines) or from the equator (dotted lines).}
\end{figure*}

The final aspect that we briefly consider is the impact of r-process heating on the kilonova light curve. To this end, we use the extracted outflow trajectories and corresponding nucleosynthesis data and adopt the same methods as in \citet{just2023end} (originally developed in \cite{just2022dynamical}) to estimate the kilonova signal using an approximate M1 radiative-transfer scheme. The resulting bolometric light curve, mass-averaged heating rate, photospheric temperature, and photospheric velocity are plotted in Fig.~\ref{fig:kilonova}. The main impact of r-process heating is a luminosity excess at late times, around $\sim$\,10\,d, which is roughly when the photosphere reaches the BH-torus ejecta. One reason for the excess is the velocity boost from r-process heating, which allows the ejecta to release a larger fraction of photons at earlier times when the ejecta are still hotter. A similar effect was also witnessed in \citet{Klion2022a}, where r-process heating was included parametrically in simulations of ejecta from BH-torus systems. The second reason for the excess is the greater mass of the BH-torus ejecta in the RHINE model, which not only leads to a higher nuclear energy-deposition rate but also better thermalization efficiency \citep[e.g.][]{Barnes2016a}, with the result being an almost twice as powerful net heating rate at $t\approx 10\,$d. In contrast to the late-time bump, the reason for the early luminosity excess at around $t\sim 1\,$d is rather puzzling, because the light curve is produced almost entirely by very fast ($v\gtrsim 0.4\,c$) material at these times, for which according to our estimates in Fig.~\ref{fig:vboost} a sizable impact of r-process heating is not expected. Nevertheless, the \emph{lateral} velocity component of the dynamical ejecta pushed away by the polar neutrino-driven ejecta is significantly smaller than the total velocity. We suspect that the (if only very subtle) lateral expansion of those polar ejecta induced by r-process heating may explain the early bump. However, tracing back the origin of the radiation produced at very early times (when emission is released only from a very small amount of mass) is not straightforward, especially given the approximate nature of our kilonova scheme based on a two-moment local-closure transport method. We postpone a detailed investigation of this aspect, as well as on other aspects related to the kilonova signal of these models, to a future study using more elaborate radiative-transfer methods.

\section{Summary and conclusions}\label{sec:summary-conclusions}

In this paper we presented a novel method of incorporating r-process heating in multi-dimensional hydrodynamical simulations, which was developed to avoid the prohibitively large computational demands required for evolving the abundances of thousands of nuclei with a detailed reaction network. The approach of our scheme, called RHINE, is to advect only a small number of composition-related quantities in addition to the conventional hydrodynamic variables and to obtain the source terms needed for their evolution using fits provided by machine-learning (ML) models. The ML models have been trained by a large set of full nuclear-network calculations from representative outflow trajectories and provide on-the-fly approximations of r-process related rates of change as functions of the evolved variables.

The defining features of the scheme are that it is 1) consistent with full nuclear networks regarding the released amount of energy, 2) consistent regarding the timescale of energy release, in particular avoiding any explicit dependence of the heating rate on the evolution time, 3) self-sufficient, in the sense that no previous post-processing steps are necessary, 4) numerically simple, e.g. avoiding non-local operations, implicit time stepping, or online tracers, and 5) computationally efficient. 

In order to incorporate RHINE into an existing hydrodynamics solver, six additional quantities need to be advected: The mass fractions of neutrons, protons, $\alpha$-particles, and heavy nuclei, $X_i$, the average mass number of heavy nuclei, $A_h$, and the average mass excess per baryon, $\tilm$. The source terms representing the microphysical rates of change of these quantities are, at each location and time step, inferred from ML models (namely multilayer perceptron neural networks) as functions of the evolved variables and integrated explicitly in time.

We validated the scheme using a suite of spherically symmetric wind outflows as well as state-of-the-art simulations of binary NSMs. In both cases we compared the outflow trajectories of the simulations using RHINE with the post-processing results from full nuclear-network calculations and found overall very good agreement. In most cases the time dependence of the r-process related change of the composition is reproduced well and the total amount of heat released from nuclear reactions is accurate to within $\lesssim 10\,\%$. The fraction of energy lost to $\beta$-decay neutrinos exhibits relatively large uncertainties which, however, are only of little relevance because of the overall relatively small contribution of neutrino losses. Good agreement is found also for the evolution of the temperature, which can be affected significantly by r-process heating.

The second purpose of our application tests was to quantify the dynamical impact of r-process heating for the various ejecta components emerging from NSMs. In all investigated cases the amount of released heat per baryon tends to be (modulo $\mathcal{O}(10\,\%)$ neutrino losses) well in agreement with the rest-energy reservoir at NSE, i.e. with the difference between the mass excess at NSE freeze-out, $\tilm=\tilm_{\rm NSE}(Y_e)$, and the asymptotic value reached for a composition consisting only of strongly bound nuclei, $\tilm\approx -1\,$\MeVc (cf. Fig.~\ref{fig:massex}). This tendency was pointed out already by \citet{Foucart2021a} and used to develop a quasi-NSE scheme of r-process heating based on a parametric increase of $Y_e$. Furthermore, the impact on the velocity corresponding to the amount of heat per baryon was in all cases well in line with the velocity boost $\Delta v$ expected from energy conservation, in particular the tendency that $\Delta v$ decreases for higher initial velocities (cf. Fig.~\ref{fig:vboost}) because of the non-linear dependence of the kinetic energy on velocity. Consistent with these tendencies, we found in the considered long-term simulations of NSMs the relatively strongest impact for the slowest ejecta component, namely the BH-torus ejecta, where the average heating energy per baryon of about $2.1$\,MeV leads to a velocity boost by about $40\,\%$. Moreover, for this component we also found a significant increase of the ejecta mass. Since NS-torus ejecta consist of matter with on average high $Y_e$, they are subject only to a small amount of r-process heat of about $0.7$\,MeV. The dynamical ejecta, which are on average the most neutron-rich but also fastest ejecta component, experience the largest energetic gain from r-process heating (about $2.3\,$MeV) but only a small velocity boost because of their already high kinetic energies. The impact on the global nucleosynthesis yields was found to be relatively small. R-process heating essentially retains the global abundance pattern, although it can induce changes up to a factor of several for individual elements. The kilonova becomes significantly brighter with r-process heating, by about a factor of two, mainly because of the boosted and more massive BH-torus ejecta.

In conclusion, our applications of RHINE suggest that r-process heating is an important effect that should be taken into account for accurate kilonova modeling, particularly for the slow ejecta components. We note, however, that although it gives rise to quantitative corrections, r-process heating does not seem to induce qualitatively different ejecta properties or kilonova signals and is thus not a leading-order physics ingredient such as, for instance, neutrino absorption or turbulent angular momentum transport (which are still being neglected occasionally in hydrodynamical models of NSMs).

RHINE adopts a novel approach for incorporating nuclear physics into multi-dimensional hydrodynamics by emulating nuclear networks with ML models. By demonstrating the viability of coupling hydrodynamics with pre-trained ML models, the scheme motivates future efforts exploring how ML models can be used for describing other nuclear-physics regimes \citep[e.g.][]{Grichener2025a} or other physics ingredients, such as neutrino-interaction rates or neutrino flavor oscillations \citep[e.g.][]{abbar2024physics,richers2024asymptoticstate} or conservative-to-primitive recovery \citep{Dieselhorst2021c}, in hydrodynamic simulations.

The source code of RHINE including the trained ML models is publicly available from Zenodo \cite{Just2025a} and the git repository of GSI \cite{rhine_git}.

\begin{acknowledgments}
We thank Andreas Bauswein, Rodrigo Fernandez, Stephane Goriely, Li-Ting Ma, and Meng-Ru Wu for insightful discussions as well as the referee for constructive comments. The authors acknowledge support of the European Research Council (ERC) under the European Union's Horizon 2020 research and innovation programme (KILONOVA No.~885281, HeavyMetal No.~101071865), by the Deutsche Forschungsgemeinschaft (DFG, German Research Foundation) through Project - ID 279384907 – SFB 1245 and MA 4248/3-1, and by the State of Hesse within the Cluster Project ELEMENTS. ZX acknowledges support by the ERC through grant NeuTrAE (No.~101165138). The work is partially funded by the European Union. Views and opinions expressed are however those of the authors only and do not necessarily reflect those of the European Union or the European Research Council Executive Agency. Neither the European Union nor the granting authority can be held responsible for them. This research was supported in part by the cluster computing resource provided by the IT Department at the GSI Helmholtzzentrum f\"ur Schwerionenforschung, Darmstadt, Germany. We acknowledge the following software: 
Numpy~\citep{numpy},  
Matplotlib~\citep{matplotlib}, 
Scipy~\citep{scipy},
Pandas~\citep{pandas}, and
PyTorch~\citep{pytorch}.
\end{acknowledgments}




\begin{thebibliography}{113}%
\makeatletter
\providecommand \@ifxundefined [1]{%
 \@ifx{#1\undefined}
}%
\providecommand \@ifnum [1]{%
 \ifnum #1\expandafter \@firstoftwo
 \else \expandafter \@secondoftwo
 \fi
}%
\providecommand \@ifx [1]{%
 \ifx #1\expandafter \@firstoftwo
 \else \expandafter \@secondoftwo
 \fi
}%
\providecommand \natexlab [1]{#1}%
\providecommand \enquote  [1]{``#1''}%
\providecommand \bibnamefont  [1]{#1}%
\providecommand \bibfnamefont [1]{#1}%
\providecommand \citenamefont [1]{#1}%
\providecommand \href@noop [0]{\@secondoftwo}%
\providecommand \href [0]{\begingroup \@sanitize@url \@href}%
\providecommand \@href[1]{\@@startlink{#1}\@@href}%
\providecommand \@@href[1]{\endgroup#1\@@endlink}%
\providecommand \@sanitize@url [0]{\catcode `\\12\catcode `\$12\catcode
  `\&12\catcode `\#12\catcode `\^12\catcode `\_12\catcode `\%12\relax}%
\providecommand \@@startlink[1]{}%
\providecommand \@@endlink[0]{}%
\providecommand \url  [0]{\begingroup\@sanitize@url \@url }%
\providecommand \@url [1]{\endgroup\@href {#1}{\urlprefix }}%
\providecommand \urlprefix  [0]{URL }%
\providecommand \Eprint [0]{\href }%
\providecommand \doibase [0]{https://doi.org/}%
\providecommand \selectlanguage [0]{\@gobble}%
\providecommand \bibinfo  [0]{\@secondoftwo}%
\providecommand \bibfield  [0]{\@secondoftwo}%
\providecommand \translation [1]{[#1]}%
\providecommand \BibitemOpen [0]{}%
\providecommand \bibitemStop [0]{}%
\providecommand \bibitemNoStop [0]{.\EOS\space}%
\providecommand \EOS [0]{\spacefactor3000\relax}%
\providecommand \BibitemShut  [1]{\csname bibitem#1\endcsname}%
\let\auto@bib@innerbib\@empty
\bibitem [{\citenamefont {{Abbott}}\ \emph {et~al.}(2017)\citenamefont
  {{Abbott}}, \citenamefont {{Abbott}}, \citenamefont {{Abbott}}, \citenamefont
  {{Acernese}}, \citenamefont {{Ackley}}, \citenamefont {{Adams}},
  \citenamefont {{Adams}}, \citenamefont {{Addesso}}, \citenamefont
  {{Adhikari}}, \citenamefont {{Adya}},\ and\ \citenamefont
  {et~al.}}]{Abbott2017b}%
  \BibitemOpen
  \bibfield  {author} {\bibinfo {author} {\bibfnamefont {B.~P.}\ \bibnamefont
  {{Abbott}}}, \bibinfo {author} {\bibfnamefont {R.}~\bibnamefont {{Abbott}}},
  \bibinfo {author} {\bibfnamefont {T.~D.}\ \bibnamefont {{Abbott}}}, \bibinfo
  {author} {\bibfnamefont {F.}~\bibnamefont {{Acernese}}}, \bibinfo {author}
  {\bibfnamefont {K.}~\bibnamefont {{Ackley}}}, \bibinfo {author}
  {\bibfnamefont {C.}~\bibnamefont {{Adams}}}, \bibinfo {author} {\bibfnamefont
  {T.}~\bibnamefont {{Adams}}}, \bibinfo {author} {\bibfnamefont
  {P.}~\bibnamefont {{Addesso}}}, \bibinfo {author} {\bibfnamefont {R.~X.}\
  \bibnamefont {{Adhikari}}}, \bibinfo {author} {\bibfnamefont {V.~B.}\
  \bibnamefont {{Adya}}},\ and\ \bibinfo {author} {\bibnamefont {et~al.}},\
  }\href@noop {} {\bibfield  {journal} {\bibinfo  {journal} {{Astrophys. J.
  Lett.}}\ }\textbf {\bibinfo {volume} {848}},\ \bibinfo {pages} {L12}
  (\bibinfo {year} {2017})}\BibitemShut {NoStop}%
\bibitem [{\citenamefont {{Smartt}}\ \emph {et~al.}(2017)\citenamefont
  {{Smartt}}, \citenamefont {{Chen}}, \citenamefont {{Jerkstrand}},
  \citenamefont {{Coughlin}}, \citenamefont {{Kankare}}, \citenamefont {{Sim}},
  \citenamefont {{Fraser}}, \citenamefont {{Inserra}}, \citenamefont
  {{Maguire}}, \citenamefont {{Chambers}}, \citenamefont {{Huber}},
  \citenamefont {{Kr{\"u}hler}}, \citenamefont {{Leloudas}}, \citenamefont
  {{Magee}}, \citenamefont {{Shingles}}, \citenamefont {{Smith}}, \citenamefont
  {{Young}}, \citenamefont {{Tonry}}, \citenamefont {{Kotak}}, \citenamefont
  {{Gal-Yam}}, \citenamefont {{Lyman}}, \citenamefont {{Homan}}, \citenamefont
  {{Agliozzo}}, \citenamefont {{Anderson}}, \citenamefont {{Angus}},
  \citenamefont {{Ashall}}, \citenamefont {{Barbarino}}, \citenamefont
  {{Bauer}}, \citenamefont {{Berton}}, \citenamefont {{Botticella}},
  \citenamefont {{Bulla}}, \citenamefont {{Bulger}}, \citenamefont
  {{Cannizzaro}}, \citenamefont {{Cano}}, \citenamefont {{Cartier}},
  \citenamefont {{Cikota}}, \citenamefont {{Clark}}, \citenamefont {{De Cia}},
  \citenamefont {{Della Valle}}, \citenamefont {{Denneau}}, \citenamefont
  {{Dennefeld}}, \citenamefont {{Dessart}}, \citenamefont {{Dimitriadis}},
  \citenamefont {{Elias-Rosa}}, \citenamefont {{Firth}}, \citenamefont
  {{Flewelling}}, \citenamefont {{Fl{\"o}rs}}, \citenamefont {{Franckowiak}},
  \citenamefont {{Frohmaier}}, \citenamefont {{Galbany}}, \citenamefont
  {{Gonz{\'a}lez-Gait{\'a}n}}, \citenamefont {{Greiner}}, \citenamefont
  {{Gromadzki}}, \citenamefont {{Guelbenzu}}, \citenamefont {{Guti{\'e}rrez}},
  \citenamefont {{Hamanowicz}}, \citenamefont {{Hanlon}}, \citenamefont
  {{Harmanen}}, \citenamefont {{Heintz}}, \citenamefont {{Heinze}},
  \citenamefont {{Hernandez}}, \citenamefont {{Hodgkin}}, \citenamefont
  {{Hook}}, \citenamefont {{Izzo}}, \citenamefont {{James}}, \citenamefont
  {{Jonker}}, \citenamefont {{Kerzendorf}}, \citenamefont {{Klose}},
  \citenamefont {{Kostrzewa-Rutkowska}}, \citenamefont {{Kowalski}},
  \citenamefont {{Kromer}}, \citenamefont {{Kuncarayakti}}, \citenamefont
  {{Lawrence}}, \citenamefont {{Lowe}}, \citenamefont {{Magnier}},
  \citenamefont {{Manulis}}, \citenamefont {{Martin-Carrillo}}, \citenamefont
  {{Mattila}}, \citenamefont {{McBrien}}, \citenamefont {{M{\"u}ller}},
  \citenamefont {{Nordin}}, \citenamefont {{O'Neill}}, \citenamefont {{Onori}},
  \citenamefont {{Palmerio}}, \citenamefont {{Pastorello}}, \citenamefont
  {{Patat}}, \citenamefont {{Pignata}}, \citenamefont {{Podsiadlowski}},
  \citenamefont {{Pumo}}, \citenamefont {{Prentice}}, \citenamefont {{Rau}},
  \citenamefont {{Razza}}, \citenamefont {{Rest}}, \citenamefont {{Reynolds}},
  \citenamefont {{Roy}}, \citenamefont {{Ruiter}}, \citenamefont {{Rybicki}},
  \citenamefont {{Salmon}}, \citenamefont {{Schady}}, \citenamefont
  {{Schultz}}, \citenamefont {{Schweyer}}, \citenamefont {{Seitenzahl}},
  \citenamefont {{Smith}}, \citenamefont {{Sollerman}}, \citenamefont
  {{Stalder}}, \citenamefont {{Stubbs}}, \citenamefont {{Sullivan}},
  \citenamefont {{Szegedi}}, \citenamefont {{Taddia}}, \citenamefont
  {{Taubenberger}}, \citenamefont {{Terreran}}, \citenamefont {{van Soelen}},
  \citenamefont {{Vos}}, \citenamefont {{Wainscoat}}, \citenamefont {{Walton}},
  \citenamefont {{Waters}}, \citenamefont {{Weiland}}, \citenamefont
  {{Willman}}, \citenamefont {{Wiseman}}, \citenamefont {{Wright}},
  \citenamefont {{Wyrzykowski}},\ and\ \citenamefont {{Yaron}}}]{Smartt2017s}%
  \BibitemOpen
  \bibfield  {author} {\bibinfo {author} {\bibfnamefont {S.~J.}\ \bibnamefont
  {{Smartt}}}, \bibinfo {author} {\bibfnamefont {T.~W.}\ \bibnamefont
  {{Chen}}}, \bibinfo {author} {\bibfnamefont {A.}~\bibnamefont
  {{Jerkstrand}}}, \bibinfo {author} {\bibfnamefont {M.}~\bibnamefont
  {{Coughlin}}}, \bibinfo {author} {\bibfnamefont {E.}~\bibnamefont
  {{Kankare}}}, \bibinfo {author} {\bibfnamefont {S.~A.}\ \bibnamefont
  {{Sim}}}, \bibinfo {author} {\bibfnamefont {M.}~\bibnamefont {{Fraser}}},
  \bibinfo {author} {\bibfnamefont {C.}~\bibnamefont {{Inserra}}}, \bibinfo
  {author} {\bibfnamefont {K.}~\bibnamefont {{Maguire}}}, \bibinfo {author}
  {\bibfnamefont {K.~C.}\ \bibnamefont {{Chambers}}}, \bibinfo {author}
  {\bibfnamefont {M.~E.}\ \bibnamefont {{Huber}}}, \bibinfo {author}
  {\bibfnamefont {T.}~\bibnamefont {{Kr{\"u}hler}}}, \bibinfo {author}
  {\bibfnamefont {G.}~\bibnamefont {{Leloudas}}}, \bibinfo {author}
  {\bibfnamefont {M.}~\bibnamefont {{Magee}}}, \bibinfo {author} {\bibfnamefont
  {L.~J.}\ \bibnamefont {{Shingles}}}, \bibinfo {author} {\bibfnamefont
  {K.~W.}\ \bibnamefont {{Smith}}}, \bibinfo {author} {\bibfnamefont {D.~R.}\
  \bibnamefont {{Young}}}, \bibinfo {author} {\bibfnamefont {J.}~\bibnamefont
  {{Tonry}}}, \bibinfo {author} {\bibfnamefont {R.}~\bibnamefont {{Kotak}}},
  \bibinfo {author} {\bibfnamefont {A.}~\bibnamefont {{Gal-Yam}}}, \bibinfo
  {author} {\bibfnamefont {J.~D.}\ \bibnamefont {{Lyman}}}, \bibinfo {author}
  {\bibfnamefont {D.~S.}\ \bibnamefont {{Homan}}}, \bibinfo {author}
  {\bibfnamefont {C.}~\bibnamefont {{Agliozzo}}}, \bibinfo {author}
  {\bibfnamefont {J.~P.}\ \bibnamefont {{Anderson}}}, \bibinfo {author}
  {\bibfnamefont {C.~R.}\ \bibnamefont {{Angus}}}, \bibinfo {author}
  {\bibfnamefont {C.}~\bibnamefont {{Ashall}}}, \bibinfo {author}
  {\bibfnamefont {C.}~\bibnamefont {{Barbarino}}}, \bibinfo {author}
  {\bibfnamefont {F.~E.}\ \bibnamefont {{Bauer}}}, \bibinfo {author}
  {\bibfnamefont {M.}~\bibnamefont {{Berton}}}, \bibinfo {author}
  {\bibfnamefont {M.~T.}\ \bibnamefont {{Botticella}}}, \bibinfo {author}
  {\bibfnamefont {M.}~\bibnamefont {{Bulla}}}, \bibinfo {author} {\bibfnamefont
  {J.}~\bibnamefont {{Bulger}}}, \bibinfo {author} {\bibfnamefont
  {G.}~\bibnamefont {{Cannizzaro}}}, \bibinfo {author} {\bibfnamefont
  {Z.}~\bibnamefont {{Cano}}}, \bibinfo {author} {\bibfnamefont
  {R.}~\bibnamefont {{Cartier}}}, \bibinfo {author} {\bibfnamefont
  {A.}~\bibnamefont {{Cikota}}}, \bibinfo {author} {\bibfnamefont
  {P.}~\bibnamefont {{Clark}}}, \bibinfo {author} {\bibfnamefont
  {A.}~\bibnamefont {{De Cia}}}, \bibinfo {author} {\bibfnamefont
  {M.}~\bibnamefont {{Della Valle}}}, \bibinfo {author} {\bibfnamefont
  {L.}~\bibnamefont {{Denneau}}}, \bibinfo {author} {\bibfnamefont
  {M.}~\bibnamefont {{Dennefeld}}}, \bibinfo {author} {\bibfnamefont
  {L.}~\bibnamefont {{Dessart}}}, \bibinfo {author} {\bibfnamefont
  {G.}~\bibnamefont {{Dimitriadis}}}, \bibinfo {author} {\bibfnamefont
  {N.}~\bibnamefont {{Elias-Rosa}}}, \bibinfo {author} {\bibfnamefont {R.~E.}\
  \bibnamefont {{Firth}}}, \bibinfo {author} {\bibfnamefont {H.}~\bibnamefont
  {{Flewelling}}}, \bibinfo {author} {\bibfnamefont {A.}~\bibnamefont
  {{Fl{\"o}rs}}}, \bibinfo {author} {\bibfnamefont {A.}~\bibnamefont
  {{Franckowiak}}}, \bibinfo {author} {\bibfnamefont {C.}~\bibnamefont
  {{Frohmaier}}}, \bibinfo {author} {\bibfnamefont {L.}~\bibnamefont
  {{Galbany}}}, \bibinfo {author} {\bibfnamefont {S.}~\bibnamefont
  {{Gonz{\'a}lez-Gait{\'a}n}}}, \bibinfo {author} {\bibfnamefont
  {J.}~\bibnamefont {{Greiner}}}, \bibinfo {author} {\bibfnamefont
  {M.}~\bibnamefont {{Gromadzki}}}, \bibinfo {author} {\bibfnamefont {A.~N.}\
  \bibnamefont {{Guelbenzu}}}, \bibinfo {author} {\bibfnamefont {C.~P.}\
  \bibnamefont {{Guti{\'e}rrez}}}, \bibinfo {author} {\bibfnamefont
  {A.}~\bibnamefont {{Hamanowicz}}}, \bibinfo {author} {\bibfnamefont
  {L.}~\bibnamefont {{Hanlon}}}, \bibinfo {author} {\bibfnamefont
  {J.}~\bibnamefont {{Harmanen}}}, \bibinfo {author} {\bibfnamefont {K.~E.}\
  \bibnamefont {{Heintz}}}, \bibinfo {author} {\bibfnamefont {A.}~\bibnamefont
  {{Heinze}}}, \bibinfo {author} {\bibfnamefont {M.~S.}\ \bibnamefont
  {{Hernandez}}}, \bibinfo {author} {\bibfnamefont {S.~T.}\ \bibnamefont
  {{Hodgkin}}}, \bibinfo {author} {\bibfnamefont {I.~M.}\ \bibnamefont
  {{Hook}}}, \bibinfo {author} {\bibfnamefont {L.}~\bibnamefont {{Izzo}}},
  \bibinfo {author} {\bibfnamefont {P.~A.}\ \bibnamefont {{James}}}, \bibinfo
  {author} {\bibfnamefont {P.~G.}\ \bibnamefont {{Jonker}}}, \bibinfo {author}
  {\bibfnamefont {W.~E.}\ \bibnamefont {{Kerzendorf}}}, \bibinfo {author}
  {\bibfnamefont {S.}~\bibnamefont {{Klose}}}, \bibinfo {author} {\bibfnamefont
  {Z.}~\bibnamefont {{Kostrzewa-Rutkowska}}}, \bibinfo {author} {\bibfnamefont
  {M.}~\bibnamefont {{Kowalski}}}, \bibinfo {author} {\bibfnamefont
  {M.}~\bibnamefont {{Kromer}}}, \bibinfo {author} {\bibfnamefont
  {H.}~\bibnamefont {{Kuncarayakti}}}, \bibinfo {author} {\bibfnamefont
  {A.}~\bibnamefont {{Lawrence}}}, \bibinfo {author} {\bibfnamefont {T.~B.}\
  \bibnamefont {{Lowe}}}, \bibinfo {author} {\bibfnamefont {E.~A.}\
  \bibnamefont {{Magnier}}}, \bibinfo {author} {\bibfnamefont {I.}~\bibnamefont
  {{Manulis}}}, \bibinfo {author} {\bibfnamefont {A.}~\bibnamefont
  {{Martin-Carrillo}}}, \bibinfo {author} {\bibfnamefont {S.}~\bibnamefont
  {{Mattila}}}, \bibinfo {author} {\bibfnamefont {O.}~\bibnamefont
  {{McBrien}}}, \bibinfo {author} {\bibfnamefont {A.}~\bibnamefont
  {{M{\"u}ller}}}, \bibinfo {author} {\bibfnamefont {J.}~\bibnamefont
  {{Nordin}}}, \bibinfo {author} {\bibfnamefont {D.}~\bibnamefont {{O'Neill}}},
  \bibinfo {author} {\bibfnamefont {F.}~\bibnamefont {{Onori}}}, \bibinfo
  {author} {\bibfnamefont {J.~T.}\ \bibnamefont {{Palmerio}}}, \bibinfo
  {author} {\bibfnamefont {A.}~\bibnamefont {{Pastorello}}}, \bibinfo {author}
  {\bibfnamefont {F.}~\bibnamefont {{Patat}}}, \bibinfo {author} {\bibfnamefont
  {G.}~\bibnamefont {{Pignata}}}, \bibinfo {author} {\bibfnamefont
  {P.}~\bibnamefont {{Podsiadlowski}}}, \bibinfo {author} {\bibfnamefont
  {M.~L.}\ \bibnamefont {{Pumo}}}, \bibinfo {author} {\bibfnamefont {S.~J.}\
  \bibnamefont {{Prentice}}}, \bibinfo {author} {\bibfnamefont
  {A.}~\bibnamefont {{Rau}}}, \bibinfo {author} {\bibfnamefont
  {A.}~\bibnamefont {{Razza}}}, \bibinfo {author} {\bibfnamefont
  {A.}~\bibnamefont {{Rest}}}, \bibinfo {author} {\bibfnamefont
  {T.}~\bibnamefont {{Reynolds}}}, \bibinfo {author} {\bibfnamefont
  {R.}~\bibnamefont {{Roy}}}, \bibinfo {author} {\bibfnamefont {A.~J.}\
  \bibnamefont {{Ruiter}}}, \bibinfo {author} {\bibfnamefont {K.~A.}\
  \bibnamefont {{Rybicki}}}, \bibinfo {author} {\bibfnamefont {L.}~\bibnamefont
  {{Salmon}}}, \bibinfo {author} {\bibfnamefont {P.}~\bibnamefont {{Schady}}},
  \bibinfo {author} {\bibfnamefont {A.~S.~B.}\ \bibnamefont {{Schultz}}},
  \bibinfo {author} {\bibfnamefont {T.}~\bibnamefont {{Schweyer}}}, \bibinfo
  {author} {\bibfnamefont {I.~R.}\ \bibnamefont {{Seitenzahl}}}, \bibinfo
  {author} {\bibfnamefont {M.}~\bibnamefont {{Smith}}}, \bibinfo {author}
  {\bibfnamefont {J.}~\bibnamefont {{Sollerman}}}, \bibinfo {author}
  {\bibfnamefont {B.}~\bibnamefont {{Stalder}}}, \bibinfo {author}
  {\bibfnamefont {C.~W.}\ \bibnamefont {{Stubbs}}}, \bibinfo {author}
  {\bibfnamefont {M.}~\bibnamefont {{Sullivan}}}, \bibinfo {author}
  {\bibfnamefont {H.}~\bibnamefont {{Szegedi}}}, \bibinfo {author}
  {\bibfnamefont {F.}~\bibnamefont {{Taddia}}}, \bibinfo {author}
  {\bibfnamefont {S.}~\bibnamefont {{Taubenberger}}}, \bibinfo {author}
  {\bibfnamefont {G.}~\bibnamefont {{Terreran}}}, \bibinfo {author}
  {\bibfnamefont {B.}~\bibnamefont {{van Soelen}}}, \bibinfo {author}
  {\bibfnamefont {J.}~\bibnamefont {{Vos}}}, \bibinfo {author} {\bibfnamefont
  {R.~J.}\ \bibnamefont {{Wainscoat}}}, \bibinfo {author} {\bibfnamefont
  {N.~A.}\ \bibnamefont {{Walton}}}, \bibinfo {author} {\bibfnamefont
  {C.}~\bibnamefont {{Waters}}}, \bibinfo {author} {\bibfnamefont
  {H.}~\bibnamefont {{Weiland}}}, \bibinfo {author} {\bibfnamefont
  {M.}~\bibnamefont {{Willman}}}, \bibinfo {author} {\bibfnamefont
  {P.}~\bibnamefont {{Wiseman}}}, \bibinfo {author} {\bibfnamefont {D.~E.}\
  \bibnamefont {{Wright}}}, \bibinfo {author} {\bibfnamefont
  {{\L}.}~\bibnamefont {{Wyrzykowski}}},\ and\ \bibinfo {author} {\bibfnamefont
  {O.}~\bibnamefont {{Yaron}}},\ }\href@noop {} {\bibfield  {journal} {\bibinfo
   {journal} {\nat}\ }\textbf {\bibinfo {volume} {551}},\ \bibinfo {pages} {75}
  (\bibinfo {year} {2017})}\BibitemShut {NoStop}%
\bibitem [{\citenamefont {{Tanaka}}\ \emph {et~al.}(2017)\citenamefont
  {{Tanaka}}, \citenamefont {{Utsumi}}, \citenamefont {{Mazzali}},
  \citenamefont {{Tominaga}}, \citenamefont {{Yoshida}}, \citenamefont
  {{Sekiguchi}}, \citenamefont {{Morokuma}}, \citenamefont {{Motohara}},
  \citenamefont {{Ohta}}, \citenamefont {{Kawabata}}, \citenamefont {{Abe}},
  \citenamefont {{Aoki}}, \citenamefont {{Asakura}}, \citenamefont {{Baar}},
  \citenamefont {{Barway}}, \citenamefont {{Bond}}, \citenamefont {{Doi}},
  \citenamefont {{Fujiyoshi}}, \citenamefont {{Furusawa}}, \citenamefont
  {{Honda}}, \citenamefont {{Itoh}}, \citenamefont {{Kawabata}}, \citenamefont
  {{Kawai}}, \citenamefont {{Kim}}, \citenamefont {{Lee}}, \citenamefont
  {{Miyazaki}}, \citenamefont {{Morihana}}, \citenamefont {{Nagashima}},
  \citenamefont {{Nagayama}}, \citenamefont {{Nakaoka}}, \citenamefont
  {{Nakata}}, \citenamefont {{Ohsawa}}, \citenamefont {{Ohshima}},
  \citenamefont {{Okita}}, \citenamefont {{Saito}}, \citenamefont {{Sumi}},
  \citenamefont {{Tajitsu}}, \citenamefont {{Takahashi}}, \citenamefont
  {{Takayama}}, \citenamefont {{Tamura}}, \citenamefont {{Tanaka}},
  \citenamefont {{Terai}}, \citenamefont {{Tristram}}, \citenamefont
  {{Yasuda}},\ and\ \citenamefont {{Zenko}}}]{Tanaka2017t}%
  \BibitemOpen
  \bibfield  {author} {\bibinfo {author} {\bibfnamefont {M.}~\bibnamefont
  {{Tanaka}}}, \bibinfo {author} {\bibfnamefont {Y.}~\bibnamefont {{Utsumi}}},
  \bibinfo {author} {\bibfnamefont {P.~A.}\ \bibnamefont {{Mazzali}}}, \bibinfo
  {author} {\bibfnamefont {N.}~\bibnamefont {{Tominaga}}}, \bibinfo {author}
  {\bibfnamefont {M.}~\bibnamefont {{Yoshida}}}, \bibinfo {author}
  {\bibfnamefont {Y.}~\bibnamefont {{Sekiguchi}}}, \bibinfo {author}
  {\bibfnamefont {T.}~\bibnamefont {{Morokuma}}}, \bibinfo {author}
  {\bibfnamefont {K.}~\bibnamefont {{Motohara}}}, \bibinfo {author}
  {\bibfnamefont {K.}~\bibnamefont {{Ohta}}}, \bibinfo {author} {\bibfnamefont
  {K.~S.}\ \bibnamefont {{Kawabata}}}, \bibinfo {author} {\bibfnamefont
  {F.}~\bibnamefont {{Abe}}}, \bibinfo {author} {\bibfnamefont
  {K.}~\bibnamefont {{Aoki}}}, \bibinfo {author} {\bibfnamefont
  {Y.}~\bibnamefont {{Asakura}}}, \bibinfo {author} {\bibfnamefont
  {S.}~\bibnamefont {{Baar}}}, \bibinfo {author} {\bibfnamefont
  {S.}~\bibnamefont {{Barway}}}, \bibinfo {author} {\bibfnamefont {I.~A.}\
  \bibnamefont {{Bond}}}, \bibinfo {author} {\bibfnamefont {M.}~\bibnamefont
  {{Doi}}}, \bibinfo {author} {\bibfnamefont {T.}~\bibnamefont {{Fujiyoshi}}},
  \bibinfo {author} {\bibfnamefont {H.}~\bibnamefont {{Furusawa}}}, \bibinfo
  {author} {\bibfnamefont {S.}~\bibnamefont {{Honda}}}, \bibinfo {author}
  {\bibfnamefont {Y.}~\bibnamefont {{Itoh}}}, \bibinfo {author} {\bibfnamefont
  {M.}~\bibnamefont {{Kawabata}}}, \bibinfo {author} {\bibfnamefont
  {N.}~\bibnamefont {{Kawai}}}, \bibinfo {author} {\bibfnamefont {J.~H.}\
  \bibnamefont {{Kim}}}, \bibinfo {author} {\bibfnamefont {C.-H.}\ \bibnamefont
  {{Lee}}}, \bibinfo {author} {\bibfnamefont {S.}~\bibnamefont {{Miyazaki}}},
  \bibinfo {author} {\bibfnamefont {K.}~\bibnamefont {{Morihana}}}, \bibinfo
  {author} {\bibfnamefont {H.}~\bibnamefont {{Nagashima}}}, \bibinfo {author}
  {\bibfnamefont {T.}~\bibnamefont {{Nagayama}}}, \bibinfo {author}
  {\bibfnamefont {T.}~\bibnamefont {{Nakaoka}}}, \bibinfo {author}
  {\bibfnamefont {F.}~\bibnamefont {{Nakata}}}, \bibinfo {author}
  {\bibfnamefont {R.}~\bibnamefont {{Ohsawa}}}, \bibinfo {author}
  {\bibfnamefont {T.}~\bibnamefont {{Ohshima}}}, \bibinfo {author}
  {\bibfnamefont {H.}~\bibnamefont {{Okita}}}, \bibinfo {author} {\bibfnamefont
  {T.}~\bibnamefont {{Saito}}}, \bibinfo {author} {\bibfnamefont
  {T.}~\bibnamefont {{Sumi}}}, \bibinfo {author} {\bibfnamefont
  {A.}~\bibnamefont {{Tajitsu}}}, \bibinfo {author} {\bibfnamefont
  {J.}~\bibnamefont {{Takahashi}}}, \bibinfo {author} {\bibfnamefont
  {M.}~\bibnamefont {{Takayama}}}, \bibinfo {author} {\bibfnamefont
  {Y.}~\bibnamefont {{Tamura}}}, \bibinfo {author} {\bibfnamefont
  {I.}~\bibnamefont {{Tanaka}}}, \bibinfo {author} {\bibfnamefont
  {T.}~\bibnamefont {{Terai}}}, \bibinfo {author} {\bibfnamefont {P.~J.}\
  \bibnamefont {{Tristram}}}, \bibinfo {author} {\bibfnamefont
  {N.}~\bibnamefont {{Yasuda}}},\ and\ \bibinfo {author} {\bibfnamefont
  {T.}~\bibnamefont {{Zenko}}},\ }\href@noop {} {\bibfield  {journal} {\bibinfo
   {journal} {{Publications of the Astronomical Society of Japan}}\ }\textbf
  {\bibinfo {volume} {69}},\ \bibinfo {pages} {102} (\bibinfo {year}
  {2017})}\BibitemShut {NoStop}%
\bibitem [{\citenamefont {{Frebel}}(2018)}]{Frebel2018a}%
  \BibitemOpen
  \bibfield  {author} {\bibinfo {author} {\bibfnamefont {A.}~\bibnamefont
  {{Frebel}}},\ }\href@noop {} {\bibfield  {journal} {\bibinfo  {journal}
  {Annual Review of Nuclear and Particle Science}\ }\textbf {\bibinfo {volume}
  {68}},\ \bibinfo {pages} {237} (\bibinfo {year} {2018})}\BibitemShut
  {NoStop}%
\bibitem [{\citenamefont {{Lombardo}}\ \emph {et~al.}(2025)\citenamefont
  {{Lombardo}}, \citenamefont {{Hansen}}, \citenamefont {{Rizzuti}},
  \citenamefont {{Cescutti}}, \citenamefont {{Mashonkina}}, \citenamefont
  {{Fran{\c{c}}ois}}, \citenamefont {{Bonifacio}}, \citenamefont {{Caffau}},
  \citenamefont {{Alencastro Puls}}, \citenamefont {{Fernandes de Melo}},
  \citenamefont {{Gallagher}}, \citenamefont {{Sk{\'u}lad{\'o}ttir}},
  \citenamefont {{Koch-Hansen}},\ and\ \citenamefont
  {{Sbordone}}}]{Lombardo2025a}%
  \BibitemOpen
  \bibfield  {author} {\bibinfo {author} {\bibfnamefont {L.}~\bibnamefont
  {{Lombardo}}}, \bibinfo {author} {\bibfnamefont {C.~J.}\ \bibnamefont
  {{Hansen}}}, \bibinfo {author} {\bibfnamefont {F.}~\bibnamefont {{Rizzuti}}},
  \bibinfo {author} {\bibfnamefont {G.}~\bibnamefont {{Cescutti}}}, \bibinfo
  {author} {\bibfnamefont {L.~I.}\ \bibnamefont {{Mashonkina}}}, \bibinfo
  {author} {\bibfnamefont {P.}~\bibnamefont {{Fran{\c{c}}ois}}}, \bibinfo
  {author} {\bibfnamefont {P.}~\bibnamefont {{Bonifacio}}}, \bibinfo {author}
  {\bibfnamefont {E.}~\bibnamefont {{Caffau}}}, \bibinfo {author}
  {\bibfnamefont {A.}~\bibnamefont {{Alencastro Puls}}}, \bibinfo {author}
  {\bibfnamefont {R.}~\bibnamefont {{Fernandes de Melo}}}, \bibinfo {author}
  {\bibfnamefont {A.~J.}\ \bibnamefont {{Gallagher}}}, \bibinfo {author}
  {\bibfnamefont {{\'A}.}~\bibnamefont {{Sk{\'u}lad{\'o}ttir}}}, \bibinfo
  {author} {\bibfnamefont {A.~J.}\ \bibnamefont {{Koch-Hansen}}},\ and\
  \bibinfo {author} {\bibfnamefont {L.}~\bibnamefont {{Sbordone}}},\
  }\href@noop {} {\bibfield  {journal} {\bibinfo  {journal} {\aap}\ }\textbf
  {\bibinfo {volume} {693}},\ \bibinfo {pages} {A293} (\bibinfo {year}
  {2025})}\BibitemShut {NoStop}%
\bibitem [{\citenamefont {{Shen}}\ \emph {et~al.}(2015)\citenamefont {{Shen}},
  \citenamefont {{Cooke}}, \citenamefont {{Ramirez-Ruiz}}, \citenamefont
  {{Madau}}, \citenamefont {{Mayer}},\ and\ \citenamefont
  {{Guedes}}}]{Shen2015}%
  \BibitemOpen
  \bibfield  {author} {\bibinfo {author} {\bibfnamefont {S.}~\bibnamefont
  {{Shen}}}, \bibinfo {author} {\bibfnamefont {R.~J.}\ \bibnamefont {{Cooke}}},
  \bibinfo {author} {\bibfnamefont {E.}~\bibnamefont {{Ramirez-Ruiz}}},
  \bibinfo {author} {\bibfnamefont {P.}~\bibnamefont {{Madau}}}, \bibinfo
  {author} {\bibfnamefont {L.}~\bibnamefont {{Mayer}}},\ and\ \bibinfo {author}
  {\bibfnamefont {J.}~\bibnamefont {{Guedes}}},\ }\href@noop {} {\bibfield
  {journal} {\bibinfo  {journal} {\apj}\ }\textbf {\bibinfo {volume} {807}},\
  \bibinfo {pages} {115} (\bibinfo {year} {2015})}\BibitemShut {NoStop}%
\bibitem [{\citenamefont {{van de Voort}}\ \emph {et~al.}(2015)\citenamefont
  {{van de Voort}}, \citenamefont {{Quataert}}, \citenamefont {{Hopkins}},
  \citenamefont {{Kere{\v s}}},\ and\ \citenamefont
  {{Faucher-Gigu{\`e}re}}}]{vandeVoort2015}%
  \BibitemOpen
  \bibfield  {author} {\bibinfo {author} {\bibfnamefont {F.}~\bibnamefont {{van
  de Voort}}}, \bibinfo {author} {\bibfnamefont {E.}~\bibnamefont
  {{Quataert}}}, \bibinfo {author} {\bibfnamefont {P.~F.}\ \bibnamefont
  {{Hopkins}}}, \bibinfo {author} {\bibfnamefont {D.}~\bibnamefont {{Kere{\v
  s}}}},\ and\ \bibinfo {author} {\bibfnamefont {C.-A.}\ \bibnamefont
  {{Faucher-Gigu{\`e}re}}},\ }\href@noop {} {\bibfield  {journal} {\bibinfo
  {journal} {\mnras}\ }\textbf {\bibinfo {volume} {447}},\ \bibinfo {pages}
  {140} (\bibinfo {year} {2015})}\BibitemShut {NoStop}%
\bibitem [{\citenamefont {{Kobayashi}}\ \emph {et~al.}(2020)\citenamefont
  {{Kobayashi}}, \citenamefont {{Karakas}},\ and\ \citenamefont
  {{Lugaro}}}]{Kobayashi2020a}%
  \BibitemOpen
  \bibfield  {author} {\bibinfo {author} {\bibfnamefont {C.}~\bibnamefont
  {{Kobayashi}}}, \bibinfo {author} {\bibfnamefont {A.~I.}\ \bibnamefont
  {{Karakas}}},\ and\ \bibinfo {author} {\bibfnamefont {M.}~\bibnamefont
  {{Lugaro}}},\ }\href@noop {} {\bibfield  {journal} {\bibinfo  {journal}
  {\apj}\ }\textbf {\bibinfo {volume} {900}},\ \bibinfo {pages} {179} (\bibinfo
  {year} {2020})}\BibitemShut {NoStop}%
\bibitem [{\citenamefont {{Baiotti}}\ and\ \citenamefont
  {{Rezzolla}}(2017)}]{Baiotti2017a}%
  \BibitemOpen
  \bibfield  {author} {\bibinfo {author} {\bibfnamefont {L.}~\bibnamefont
  {{Baiotti}}}\ and\ \bibinfo {author} {\bibfnamefont {L.}~\bibnamefont
  {{Rezzolla}}},\ }\href@noop {} {\bibfield  {journal} {\bibinfo  {journal}
  {Reports on Progress in Physics}\ }\textbf {\bibinfo {volume} {80}},\
  \bibinfo {pages} {096901} (\bibinfo {year} {2017})}\BibitemShut {NoStop}%
\bibitem [{\citenamefont {Duez}\ and\ \citenamefont
  {Zlochower}(2018)}]{Duez2018y}%
  \BibitemOpen
  \bibfield  {author} {\bibinfo {author} {\bibfnamefont {M.~D.}\ \bibnamefont
  {Duez}}\ and\ \bibinfo {author} {\bibfnamefont {Y.}~\bibnamefont
  {Zlochower}},\ }\href@noop {} {\bibfield  {journal} {\bibinfo  {journal}
  {Reports on Progress in Physics}\ }\textbf {\bibinfo {volume} {82}},\
  \bibinfo {pages} {016902} (\bibinfo {year} {2018})}\BibitemShut {NoStop}%
\bibitem [{\citenamefont {Shibata}\ and\ \citenamefont
  {Hotokezaka}(2019)}]{Shibata2019c}%
  \BibitemOpen
  \bibfield  {author} {\bibinfo {author} {\bibfnamefont {M.}~\bibnamefont
  {Shibata}}\ and\ \bibinfo {author} {\bibfnamefont {K.}~\bibnamefont
  {Hotokezaka}},\ }\href@noop {} {\bibfield  {journal} {\bibinfo  {journal}
  {Annual Review of Nuclear and Particle Science}\ }\textbf {\bibinfo {volume}
  {69}},\ \bibinfo {pages} {41} (\bibinfo {year} {2019})}\BibitemShut {NoStop}%
\bibitem [{\citenamefont {{Bernuzzi}}(2020)}]{Bernuzzi2020b}%
  \BibitemOpen
  \bibfield  {author} {\bibinfo {author} {\bibfnamefont {S.}~\bibnamefont
  {{Bernuzzi}}},\ }\href@noop {} {\bibfield  {journal} {\bibinfo  {journal}
  {General Relativity and Gravitation}\ }\textbf {\bibinfo {volume} {52}},\
  \bibinfo {pages} {108} (\bibinfo {year} {2020})}\BibitemShut {NoStop}%
\bibitem [{\citenamefont {{Janka}}\ and\ \citenamefont
  {{Bauswein}}(2022)}]{Janka2022a}%
  \BibitemOpen
  \bibfield  {author} {\bibinfo {author} {\bibfnamefont {H.~T.}\ \bibnamefont
  {{Janka}}}\ and\ \bibinfo {author} {\bibfnamefont {A.}~\bibnamefont
  {{Bauswein}}},\ }\href@noop {} {\bibfield  {journal} {\bibinfo  {journal}
  {arXiv e-prints}\ ,\ \bibinfo {pages} {arXiv:2212.07498}} (\bibinfo {year}
  {2022})}\BibitemShut {NoStop}%
\bibitem [{\citenamefont {{Foucart}}(2023)}]{Foucart2023b}%
  \BibitemOpen
  \bibfield  {author} {\bibinfo {author} {\bibfnamefont {F.}~\bibnamefont
  {{Foucart}}},\ }\href@noop {} {\bibfield  {journal} {\bibinfo  {journal}
  {Living Reviews in Computational Astrophysics}\ }\textbf {\bibinfo {volume}
  {9}},\ \bibinfo {pages} {1} (\bibinfo {year} {2023})}\BibitemShut {NoStop}%
\bibitem [{\citenamefont {{Siegel}}(2022)}]{Siegel2022k}%
  \BibitemOpen
  \bibfield  {author} {\bibinfo {author} {\bibfnamefont {D.~M.}\ \bibnamefont
  {{Siegel}}},\ }\href@noop {} {\bibfield  {journal} {\bibinfo  {journal}
  {Nature Reviews Physics}\ }\textbf {\bibinfo {volume} {4}},\ \bibinfo {pages}
  {306} (\bibinfo {year} {2022})}\BibitemShut {NoStop}%
\bibitem [{\citenamefont {{Rosswog}}\ and\ \citenamefont
  {{Korobkin}}(2024)}]{Rosswog2024b}%
  \BibitemOpen
  \bibfield  {author} {\bibinfo {author} {\bibfnamefont {S.}~\bibnamefont
  {{Rosswog}}}\ and\ \bibinfo {author} {\bibfnamefont {O.}~\bibnamefont
  {{Korobkin}}},\ }\href@noop {} {\bibfield  {journal} {\bibinfo  {journal}
  {Annalen der Physik}\ }\textbf {\bibinfo {volume} {536}},\ \bibinfo {pages}
  {2200306} (\bibinfo {year} {2024})}\BibitemShut {NoStop}%
\bibitem [{\citenamefont {{Horowitz}}\ \emph {et~al.}(2018)\citenamefont
  {{Horowitz}}, \citenamefont {{Arcones}}, \citenamefont {{C{\^o}t{\'e}}},
  \citenamefont {{Dillmann}}, \citenamefont {{Nazarewicz}}, \citenamefont
  {{Roederer}}, \citenamefont {{Schatz}}, \citenamefont {{Aprahamian}},
  \citenamefont {{Atanasov}}, \citenamefont {{Bauswein}}, \citenamefont
  {{Bliss}}, \citenamefont {{Brodeur}}, \citenamefont {{Clark}}, \citenamefont
  {{Frebel}}, \citenamefont {{Foucart}}, \citenamefont {{Hansen}},
  \citenamefont {{Just}}, \citenamefont {{Kankainen}}, \citenamefont
  {{McLaughlin}}, \citenamefont {{Kelly}}, \citenamefont {{Liddick}},
  \citenamefont {{Lee}}, \citenamefont {{Lippuner}}, \citenamefont {{Martin}},
  \citenamefont {{Mendoza-Temis}}, \citenamefont {{Metzger}}, \citenamefont
  {{Mumpower}}, \citenamefont {{Perdikakis}}, \citenamefont {{Pereira}},
  \citenamefont {{O'Shea}}, \citenamefont {{Reifarth}}, \citenamefont
  {{Rogers}}, \citenamefont {{Siegel}}, \citenamefont {{Spyrou}}, \citenamefont
  {{Surman}}, \citenamefont {{Tang}}, \citenamefont {{Uesaka}},\ and\
  \citenamefont {{Wang}}}]{Horowitz2018a}%
  \BibitemOpen
  \bibfield  {author} {\bibinfo {author} {\bibfnamefont {C.~J.}\ \bibnamefont
  {{Horowitz}}}, \bibinfo {author} {\bibfnamefont {A.}~\bibnamefont
  {{Arcones}}}, \bibinfo {author} {\bibfnamefont {B.}~\bibnamefont
  {{C{\^o}t{\'e}}}}, \bibinfo {author} {\bibfnamefont {I.}~\bibnamefont
  {{Dillmann}}}, \bibinfo {author} {\bibfnamefont {W.}~\bibnamefont
  {{Nazarewicz}}}, \bibinfo {author} {\bibfnamefont {I.~U.}\ \bibnamefont
  {{Roederer}}}, \bibinfo {author} {\bibfnamefont {H.}~\bibnamefont
  {{Schatz}}}, \bibinfo {author} {\bibfnamefont {A.}~\bibnamefont
  {{Aprahamian}}}, \bibinfo {author} {\bibfnamefont {D.}~\bibnamefont
  {{Atanasov}}}, \bibinfo {author} {\bibfnamefont {A.}~\bibnamefont
  {{Bauswein}}}, \bibinfo {author} {\bibfnamefont {J.}~\bibnamefont {{Bliss}}},
  \bibinfo {author} {\bibfnamefont {M.}~\bibnamefont {{Brodeur}}}, \bibinfo
  {author} {\bibfnamefont {J.~A.}\ \bibnamefont {{Clark}}}, \bibinfo {author}
  {\bibfnamefont {A.}~\bibnamefont {{Frebel}}}, \bibinfo {author}
  {\bibfnamefont {F.}~\bibnamefont {{Foucart}}}, \bibinfo {author}
  {\bibfnamefont {C.~J.}\ \bibnamefont {{Hansen}}}, \bibinfo {author}
  {\bibfnamefont {O.}~\bibnamefont {{Just}}}, \bibinfo {author} {\bibfnamefont
  {A.}~\bibnamefont {{Kankainen}}}, \bibinfo {author} {\bibfnamefont {G.~C.}\
  \bibnamefont {{McLaughlin}}}, \bibinfo {author} {\bibfnamefont {J.~M.}\
  \bibnamefont {{Kelly}}}, \bibinfo {author} {\bibfnamefont {S.~N.}\
  \bibnamefont {{Liddick}}}, \bibinfo {author} {\bibfnamefont {D.~M.}\
  \bibnamefont {{Lee}}}, \bibinfo {author} {\bibfnamefont {J.}~\bibnamefont
  {{Lippuner}}}, \bibinfo {author} {\bibfnamefont {D.}~\bibnamefont
  {{Martin}}}, \bibinfo {author} {\bibfnamefont {J.}~\bibnamefont
  {{Mendoza-Temis}}}, \bibinfo {author} {\bibfnamefont {B.~D.}\ \bibnamefont
  {{Metzger}}}, \bibinfo {author} {\bibfnamefont {M.~R.}\ \bibnamefont
  {{Mumpower}}}, \bibinfo {author} {\bibfnamefont {G.}~\bibnamefont
  {{Perdikakis}}}, \bibinfo {author} {\bibfnamefont {J.}~\bibnamefont
  {{Pereira}}}, \bibinfo {author} {\bibfnamefont {B.~W.}\ \bibnamefont
  {{O'Shea}}}, \bibinfo {author} {\bibfnamefont {R.}~\bibnamefont
  {{Reifarth}}}, \bibinfo {author} {\bibfnamefont {A.~M.}\ \bibnamefont
  {{Rogers}}}, \bibinfo {author} {\bibfnamefont {D.~M.}\ \bibnamefont
  {{Siegel}}}, \bibinfo {author} {\bibfnamefont {A.}~\bibnamefont {{Spyrou}}},
  \bibinfo {author} {\bibfnamefont {R.}~\bibnamefont {{Surman}}}, \bibinfo
  {author} {\bibfnamefont {X.}~\bibnamefont {{Tang}}}, \bibinfo {author}
  {\bibfnamefont {T.}~\bibnamefont {{Uesaka}}},\ and\ \bibinfo {author}
  {\bibfnamefont {M.}~\bibnamefont {{Wang}}},\ }\href@noop {} {\bibfield
  {journal} {\bibinfo  {journal} {ArXiv e-prints}\ } (\bibinfo {year}
  {2018})}\BibitemShut {NoStop}%
\bibitem [{\citenamefont {{Arnould}}\ and\ \citenamefont
  {{Goriely}}(2020)}]{Arnould2020f}%
  \BibitemOpen
  \bibfield  {author} {\bibinfo {author} {\bibfnamefont {M.}~\bibnamefont
  {{Arnould}}}\ and\ \bibinfo {author} {\bibfnamefont {S.}~\bibnamefont
  {{Goriely}}},\ }\href@noop {} {\bibfield  {journal} {\bibinfo  {journal}
  {Progress in Particle and Nuclear Physics}\ }\textbf {\bibinfo {volume}
  {112}},\ \bibinfo {pages} {103766} (\bibinfo {year} {2020})}\BibitemShut
  {NoStop}%
\bibitem [{\citenamefont {Cowan}\ \emph {et~al.}(2021)\citenamefont {Cowan},
  \citenamefont {Sneden}, \citenamefont {Lawler}, \citenamefont {Aprahamian},
  \citenamefont {Wiescher}, \citenamefont {Langanke}, \citenamefont
  {Mart{\'i}nez-Pinedo},\ and\ \citenamefont {Thielemann}}]{Cowan2021origin}%
  \BibitemOpen
  \bibfield  {author} {\bibinfo {author} {\bibfnamefont {J.~J.}\ \bibnamefont
  {Cowan}}, \bibinfo {author} {\bibfnamefont {C.}~\bibnamefont {Sneden}},
  \bibinfo {author} {\bibfnamefont {J.~E.}\ \bibnamefont {Lawler}}, \bibinfo
  {author} {\bibfnamefont {A.}~\bibnamefont {Aprahamian}}, \bibinfo {author}
  {\bibfnamefont {M.}~\bibnamefont {Wiescher}}, \bibinfo {author}
  {\bibfnamefont {K.}~\bibnamefont {Langanke}}, \bibinfo {author}
  {\bibfnamefont {G.}~\bibnamefont {Mart{\'i}nez-Pinedo}},\ and\ \bibinfo
  {author} {\bibfnamefont {F.~K.}\ \bibnamefont {Thielemann}},\ }\href
  {https://doi.org/10.1103/RevModPhys.93.015002} {\bibfield  {journal}
  {\bibinfo  {journal} {Rev. Mod. Phys.}\ }\textbf {\bibinfo {volume} {93}},\
  \bibinfo {pages} {015002} (\bibinfo {year} {2021})},\ \Eprint
  {https://arxiv.org/abs/1901.01410} {arXiv:1901.01410} \BibitemShut {NoStop}%
\bibitem [{\citenamefont {Holmbeck}\ \emph {et~al.}(2023)\citenamefont
  {Holmbeck}, \citenamefont {Sprouse},\ and\ \citenamefont
  {Mumpower}}]{Holmbeck2023c}%
  \BibitemOpen
  \bibfield  {author} {\bibinfo {author} {\bibfnamefont {E.~M.}\ \bibnamefont
  {Holmbeck}}, \bibinfo {author} {\bibfnamefont {T.~M.}\ \bibnamefont
  {Sprouse}},\ and\ \bibinfo {author} {\bibfnamefont {M.~R.}\ \bibnamefont
  {Mumpower}},\ }\href@noop {} {\bibfield  {journal} {\bibinfo  {journal} {The
  European Physical Journal A}\ }\textbf {\bibinfo {volume} {59}},\ \bibinfo
  {pages} {28} (\bibinfo {year} {2023})}\BibitemShut {NoStop}%
\bibitem [{\citenamefont {Metzger}\ \emph {et~al.}(2010)\citenamefont
  {Metzger}, \citenamefont {Mart{\'{i}}nez-Pinedo}, \citenamefont {Darbha},
  \citenamefont {Quataert}, \citenamefont {Arcones}, \citenamefont {Kasen},
  \citenamefont {Thomas}, \citenamefont {Nugent}, \citenamefont {Panov},\ and\
  \citenamefont {Zinner}}]{metzger2010electromagnetic}%
  \BibitemOpen
  \bibfield  {author} {\bibinfo {author} {\bibfnamefont {B.~D.}\ \bibnamefont
  {Metzger}}, \bibinfo {author} {\bibfnamefont {G.}~\bibnamefont
  {Mart{\'{i}}nez-Pinedo}}, \bibinfo {author} {\bibfnamefont {S.}~\bibnamefont
  {Darbha}}, \bibinfo {author} {\bibfnamefont {E.}~\bibnamefont {Quataert}},
  \bibinfo {author} {\bibfnamefont {A.}~\bibnamefont {Arcones}}, \bibinfo
  {author} {\bibfnamefont {D.}~\bibnamefont {Kasen}}, \bibinfo {author}
  {\bibfnamefont {R.}~\bibnamefont {Thomas}}, \bibinfo {author} {\bibfnamefont
  {P.}~\bibnamefont {Nugent}}, \bibinfo {author} {\bibfnamefont {I.~V.}\
  \bibnamefont {Panov}},\ and\ \bibinfo {author} {\bibfnamefont {N.~T.}\
  \bibnamefont {Zinner}},\ }\href
  {https://doi.org/10.1111/j.1365-2966.2010.16864.x} {\bibfield  {journal}
  {\bibinfo  {journal} {Mon. Not. R. Astron. Soc.}\ }\textbf {\bibinfo {volume}
  {406}},\ \bibinfo {pages} {2650} (\bibinfo {year} {2010})},\ \Eprint
  {https://arxiv.org/abs/1001.5029} {arXiv:1001.5029} \BibitemShut {NoStop}%
\bibitem [{\citenamefont {{Roberts}}\ \emph {et~al.}(2011)\citenamefont
  {{Roberts}}, \citenamefont {{Kasen}}, \citenamefont {{Lee}},\ and\
  \citenamefont {{Ramirez-Ruiz}}}]{Roberts2011}%
  \BibitemOpen
  \bibfield  {author} {\bibinfo {author} {\bibfnamefont {L.~F.}\ \bibnamefont
  {{Roberts}}}, \bibinfo {author} {\bibfnamefont {D.}~\bibnamefont {{Kasen}}},
  \bibinfo {author} {\bibfnamefont {W.~H.}\ \bibnamefont {{Lee}}},\ and\
  \bibinfo {author} {\bibfnamefont {E.}~\bibnamefont {{Ramirez-Ruiz}}},\
  }\href@noop {} {\bibfield  {journal} {\bibinfo  {journal} {{Astrophys. J.
  Lett.}}\ }\textbf {\bibinfo {volume} {736}},\ \bibinfo {pages} {L21+}
  (\bibinfo {year} {2011})}\BibitemShut {NoStop}%
\bibitem [{\citenamefont {{Metzger}}(2019)}]{Metzger2019a}%
  \BibitemOpen
  \bibfield  {author} {\bibinfo {author} {\bibfnamefont {B.~D.}\ \bibnamefont
  {{Metzger}}},\ }\href@noop {} {\bibfield  {journal} {\bibinfo  {journal}
  {Living Reviews in Relativity}\ }\textbf {\bibinfo {volume} {23}},\ \bibinfo
  {pages} {1} (\bibinfo {year} {2019})}\BibitemShut {NoStop}%
\bibitem [{\citenamefont {{Burns}}(2020)}]{Burns2020a}%
  \BibitemOpen
  \bibfield  {author} {\bibinfo {author} {\bibfnamefont {E.}~\bibnamefont
  {{Burns}}},\ }\href@noop {} {\bibfield  {journal} {\bibinfo  {journal}
  {Living Reviews in Relativity}\ }\textbf {\bibinfo {volume} {23}},\ \bibinfo
  {pages} {4} (\bibinfo {year} {2020})}\BibitemShut {NoStop}%
\bibitem [{\citenamefont {{Kasen}}\ \emph {et~al.}(2015)\citenamefont
  {{Kasen}}, \citenamefont {{Fern{\'a}ndez}},\ and\ \citenamefont
  {{Metzger}}}]{Kasen2015}%
  \BibitemOpen
  \bibfield  {author} {\bibinfo {author} {\bibfnamefont {D.}~\bibnamefont
  {{Kasen}}}, \bibinfo {author} {\bibfnamefont {R.}~\bibnamefont
  {{Fern{\'a}ndez}}},\ and\ \bibinfo {author} {\bibfnamefont {B.~D.}\
  \bibnamefont {{Metzger}}},\ }\href@noop {} {\bibfield  {journal} {\bibinfo
  {journal} {{Mon. Not. R. Astron. Soc.}}\ }\textbf {\bibinfo {volume} {450}},\
  \bibinfo {pages} {1777} (\bibinfo {year} {2015})}\BibitemShut {NoStop}%
\bibitem [{\citenamefont {{Miller}}\ \emph {et~al.}(2019)\citenamefont
  {{Miller}}, \citenamefont {{Ryan}}, \citenamefont {{Dolence}}, \citenamefont
  {{Burrows}}, \citenamefont {{Fontes}}, \citenamefont {{Fryer}}, \citenamefont
  {{Korobkin}}, \citenamefont {{Lippuner}}, \citenamefont {{Mumpower}},\ and\
  \citenamefont {{Wollaeger}}}]{Miller2019a}%
  \BibitemOpen
  \bibfield  {author} {\bibinfo {author} {\bibfnamefont {J.~M.}\ \bibnamefont
  {{Miller}}}, \bibinfo {author} {\bibfnamefont {B.~R.}\ \bibnamefont
  {{Ryan}}}, \bibinfo {author} {\bibfnamefont {J.~C.}\ \bibnamefont
  {{Dolence}}}, \bibinfo {author} {\bibfnamefont {A.}~\bibnamefont
  {{Burrows}}}, \bibinfo {author} {\bibfnamefont {C.~J.}\ \bibnamefont
  {{Fontes}}}, \bibinfo {author} {\bibfnamefont {C.~L.}\ \bibnamefont
  {{Fryer}}}, \bibinfo {author} {\bibfnamefont {O.}~\bibnamefont {{Korobkin}}},
  \bibinfo {author} {\bibfnamefont {J.}~\bibnamefont {{Lippuner}}}, \bibinfo
  {author} {\bibfnamefont {M.~R.}\ \bibnamefont {{Mumpower}}},\ and\ \bibinfo
  {author} {\bibfnamefont {R.~T.}\ \bibnamefont {{Wollaeger}}},\ }\href@noop {}
  {\bibfield  {journal} {\bibinfo  {journal} {\prd}\ }\textbf {\bibinfo
  {volume} {100}},\ \bibinfo {pages} {023008} (\bibinfo {year}
  {2019})}\BibitemShut {NoStop}%
\bibitem [{\citenamefont {{Kawaguchi}}\ \emph {et~al.}(2021)\citenamefont
  {{Kawaguchi}}, \citenamefont {{Fujibayashi}}, \citenamefont {{Shibata}},
  \citenamefont {{Tanaka}},\ and\ \citenamefont {{Wanajo}}}]{Kawaguchi2021a}%
  \BibitemOpen
  \bibfield  {author} {\bibinfo {author} {\bibfnamefont {K.}~\bibnamefont
  {{Kawaguchi}}}, \bibinfo {author} {\bibfnamefont {S.}~\bibnamefont
  {{Fujibayashi}}}, \bibinfo {author} {\bibfnamefont {M.}~\bibnamefont
  {{Shibata}}}, \bibinfo {author} {\bibfnamefont {M.}~\bibnamefont
  {{Tanaka}}},\ and\ \bibinfo {author} {\bibfnamefont {S.}~\bibnamefont
  {{Wanajo}}},\ }\href@noop {} {\bibfield  {journal} {\bibinfo  {journal}
  {\apj}\ }\textbf {\bibinfo {volume} {913}},\ \bibinfo {pages} {100} (\bibinfo
  {year} {2021})}\BibitemShut {NoStop}%
\bibitem [{\citenamefont {Just}\ \emph {et~al.}(2023)\citenamefont {Just},
  \citenamefont {Vijayan}, \citenamefont {Xiong}, \citenamefont {Goriely},
  \citenamefont {Soultanis}, \citenamefont {Bauswein}, \citenamefont {Guilet},
  \citenamefont {Janka},\ and\ \citenamefont
  {Mart{\'{i}}nez-Pinedo}}]{just2023end}%
  \BibitemOpen
  \bibfield  {author} {\bibinfo {author} {\bibfnamefont {O.}~\bibnamefont
  {Just}}, \bibinfo {author} {\bibfnamefont {V.}~\bibnamefont {Vijayan}},
  \bibinfo {author} {\bibfnamefont {Z.}~\bibnamefont {Xiong}}, \bibinfo
  {author} {\bibfnamefont {S.}~\bibnamefont {Goriely}}, \bibinfo {author}
  {\bibfnamefont {T.}~\bibnamefont {Soultanis}}, \bibinfo {author}
  {\bibfnamefont {A.}~\bibnamefont {Bauswein}}, \bibinfo {author}
  {\bibfnamefont {J.}~\bibnamefont {Guilet}}, \bibinfo {author} {\bibfnamefont
  {H.-T.}\ \bibnamefont {Janka}},\ and\ \bibinfo {author} {\bibfnamefont
  {G.}~\bibnamefont {Mart{\'{i}}nez-Pinedo}},\ }\href
  {https://doi.org/10.3847/2041-8213/acdad2} {\bibfield  {journal} {\bibinfo
  {journal} {Astrophys. J. Lett.}\ }\textbf {\bibinfo {volume} {951}},\
  \bibinfo {pages} {L12} (\bibinfo {year} {2023})},\ \Eprint
  {https://arxiv.org/abs/2302.10928} {arXiv:2302.10928} \BibitemShut {NoStop}%
\bibitem [{\citenamefont {{Klion}}\ \emph {et~al.}(2022)\citenamefont
  {{Klion}}, \citenamefont {{Tchekhovskoy}}, \citenamefont {{Kasen}},
  \citenamefont {{Kathirgamaraju}}, \citenamefont {{Quataert}},\ and\
  \citenamefont {{Fern{\'a}ndez}}}]{Klion2022a}%
  \BibitemOpen
  \bibfield  {author} {\bibinfo {author} {\bibfnamefont {H.}~\bibnamefont
  {{Klion}}}, \bibinfo {author} {\bibfnamefont {A.}~\bibnamefont
  {{Tchekhovskoy}}}, \bibinfo {author} {\bibfnamefont {D.}~\bibnamefont
  {{Kasen}}}, \bibinfo {author} {\bibfnamefont {A.}~\bibnamefont
  {{Kathirgamaraju}}}, \bibinfo {author} {\bibfnamefont {E.}~\bibnamefont
  {{Quataert}}},\ and\ \bibinfo {author} {\bibfnamefont {R.}~\bibnamefont
  {{Fern{\'a}ndez}}},\ }\href@noop {} {\bibfield  {journal} {\bibinfo
  {journal} {{Mon. Not. R. Astron. Soc.}}\ }\textbf {\bibinfo {volume} {510}},\
  \bibinfo {pages} {2968} (\bibinfo {year} {2022})}\BibitemShut {NoStop}%
\bibitem [{\citenamefont {Shingles}\ \emph {et~al.}(2023)\citenamefont
  {Shingles}, \citenamefont {Collins}, \citenamefont {Vijayan}, \citenamefont
  {Fl{\"{o}}rs}, \citenamefont {Just}, \citenamefont {Leck}, \citenamefont
  {Xiong}, \citenamefont {Bauswein}, \citenamefont {Mart{\'{i}}nez-Pinedo},\
  and\ \citenamefont {Sim}}]{shingles2023self}%
  \BibitemOpen
  \bibfield  {author} {\bibinfo {author} {\bibfnamefont {L.~J.}\ \bibnamefont
  {Shingles}}, \bibinfo {author} {\bibfnamefont {C.~E.}\ \bibnamefont
  {Collins}}, \bibinfo {author} {\bibfnamefont {V.}~\bibnamefont {Vijayan}},
  \bibinfo {author} {\bibfnamefont {A.}~\bibnamefont {Fl{\"{o}}rs}}, \bibinfo
  {author} {\bibfnamefont {O.}~\bibnamefont {Just}}, \bibinfo {author}
  {\bibfnamefont {G.}~\bibnamefont {Leck}}, \bibinfo {author} {\bibfnamefont
  {Z.}~\bibnamefont {Xiong}}, \bibinfo {author} {\bibfnamefont
  {A.}~\bibnamefont {Bauswein}}, \bibinfo {author} {\bibfnamefont
  {G.}~\bibnamefont {Mart{\'{i}}nez-Pinedo}},\ and\ \bibinfo {author}
  {\bibfnamefont {S.~A.}\ \bibnamefont {Sim}},\ }\href@noop {} {\bibfield
  {journal} {\bibinfo  {journal} {Astrophys. J. Lett.}\ }\textbf {\bibinfo
  {volume} {954}},\ \bibinfo {pages} {L41} (\bibinfo {year} {2023})},\ \Eprint
  {https://arxiv.org/abs/2306.17612} {arXiv:2306.17612} \BibitemShut {NoStop}%
\bibitem [{\citenamefont {{Combi}}\ and\ \citenamefont
  {{Siegel}}(2023)}]{Combi2023b}%
  \BibitemOpen
  \bibfield  {author} {\bibinfo {author} {\bibfnamefont {L.}~\bibnamefont
  {{Combi}}}\ and\ \bibinfo {author} {\bibfnamefont {D.~M.}\ \bibnamefont
  {{Siegel}}},\ }\href@noop {} {\bibfield  {journal} {\bibinfo  {journal}
  {\prl}\ }\textbf {\bibinfo {volume} {131}},\ \bibinfo {pages} {231402}
  (\bibinfo {year} {2023})}\BibitemShut {NoStop}%
\bibitem [{\citenamefont {Curtis}\ \emph {et~al.}(2024)\citenamefont {Curtis},
  \citenamefont {Bosch}, \citenamefont {M{\"o}sta}, \citenamefont {Radice},
  \citenamefont {Bernuzzi}, \citenamefont {Perego}, \citenamefont {Haas},\ and\
  \citenamefont {Schnetter}}]{Curtis2024a}%
  \BibitemOpen
  \bibfield  {author} {\bibinfo {author} {\bibfnamefont {S.}~\bibnamefont
  {Curtis}}, \bibinfo {author} {\bibfnamefont {P.}~\bibnamefont {Bosch}},
  \bibinfo {author} {\bibfnamefont {P.}~\bibnamefont {M{\"o}sta}}, \bibinfo
  {author} {\bibfnamefont {D.}~\bibnamefont {Radice}}, \bibinfo {author}
  {\bibfnamefont {S.}~\bibnamefont {Bernuzzi}}, \bibinfo {author}
  {\bibfnamefont {A.}~\bibnamefont {Perego}}, \bibinfo {author} {\bibfnamefont
  {R.}~\bibnamefont {Haas}},\ and\ \bibinfo {author} {\bibfnamefont
  {E.}~\bibnamefont {Schnetter}},\ }\href@noop {} {\bibfield  {journal}
  {\bibinfo  {journal} {The Astrophysical Journal Letters}\ }\textbf {\bibinfo
  {volume} {961}},\ \bibinfo {pages} {L26} (\bibinfo {year}
  {2024})}\BibitemShut {NoStop}%
\bibitem [{\citenamefont {{Magistrelli}}\ \emph {et~al.}(2024)\citenamefont
  {{Magistrelli}}, \citenamefont {{Bernuzzi}}, \citenamefont {{Perego}},\ and\
  \citenamefont {{Radice}}}]{Magistrelli2024b}%
  \BibitemOpen
  \bibfield  {author} {\bibinfo {author} {\bibfnamefont {F.}~\bibnamefont
  {{Magistrelli}}}, \bibinfo {author} {\bibfnamefont {S.}~\bibnamefont
  {{Bernuzzi}}}, \bibinfo {author} {\bibfnamefont {A.}~\bibnamefont
  {{Perego}}},\ and\ \bibinfo {author} {\bibfnamefont {D.}~\bibnamefont
  {{Radice}}},\ }\href@noop {} {\bibfield  {journal} {\bibinfo  {journal}
  {{Astrophys. J. Lett.}}\ }\textbf {\bibinfo {volume} {974}},\ \bibinfo
  {pages} {L5} (\bibinfo {year} {2024})}\BibitemShut {NoStop}%
\bibitem [{\citenamefont {{Lee}}\ \emph {et~al.}(2005)\citenamefont {{Lee}},
  \citenamefont {{Ramirez-Ruiz}},\ and\ \citenamefont {{Page}}}]{Lee2005}%
  \BibitemOpen
  \bibfield  {author} {\bibinfo {author} {\bibfnamefont {W.~H.}\ \bibnamefont
  {{Lee}}}, \bibinfo {author} {\bibfnamefont {E.}~\bibnamefont
  {{Ramirez-Ruiz}}},\ and\ \bibinfo {author} {\bibfnamefont {D.}~\bibnamefont
  {{Page}}},\ }\href@noop {} {\bibfield  {journal} {\bibinfo  {journal} {ApJ}\
  }\textbf {\bibinfo {volume} {632}},\ \bibinfo {pages} {421} (\bibinfo {year}
  {2005})}\BibitemShut {NoStop}%
\bibitem [{\citenamefont {{Fern{\'a}ndez}}\ and\ \citenamefont
  {{Metzger}}(2013)}]{Fernandez2013b}%
  \BibitemOpen
  \bibfield  {author} {\bibinfo {author} {\bibfnamefont {R.}~\bibnamefont
  {{Fern{\'a}ndez}}}\ and\ \bibinfo {author} {\bibfnamefont {B.~D.}\
  \bibnamefont {{Metzger}}},\ }\href@noop {} {\bibfield  {journal} {\bibinfo
  {journal} {{Mon. Not. R. Astron. Soc.}}\ }\textbf {\bibinfo {volume} {435}},\
  \bibinfo {pages} {502} (\bibinfo {year} {2013})}\BibitemShut {NoStop}%
\bibitem [{\citenamefont {{Siegel}}\ and\ \citenamefont
  {{Metzger}}(2018)}]{Siegel2018c}%
  \BibitemOpen
  \bibfield  {author} {\bibinfo {author} {\bibfnamefont {D.~M.}\ \bibnamefont
  {{Siegel}}}\ and\ \bibinfo {author} {\bibfnamefont {B.~D.}\ \bibnamefont
  {{Metzger}}},\ }\href@noop {} {\bibfield  {journal} {\bibinfo  {journal}
  {\apj}\ }\textbf {\bibinfo {volume} {858}},\ \bibinfo {pages} {52} (\bibinfo
  {year} {2018})}\BibitemShut {NoStop}%
\bibitem [{\citenamefont {{Haddadi}}\ \emph {et~al.}(2023)\citenamefont
  {{Haddadi}}, \citenamefont {{Duez}}, \citenamefont {{Foucart}}, \citenamefont
  {{Ramirez}}, \citenamefont {{Fern{\'a}ndez}}, \citenamefont {{Knight}},
  \citenamefont {{Jesse}}, \citenamefont {{H{\'e}bert}}, \citenamefont
  {{Kidder}}, \citenamefont {{Pfeiffer}},\ and\ \citenamefont
  {{Scheel}}}]{Haddadi2023a}%
  \BibitemOpen
  \bibfield  {author} {\bibinfo {author} {\bibfnamefont {M.}~\bibnamefont
  {{Haddadi}}}, \bibinfo {author} {\bibfnamefont {M.~D.}\ \bibnamefont
  {{Duez}}}, \bibinfo {author} {\bibfnamefont {F.}~\bibnamefont {{Foucart}}},
  \bibinfo {author} {\bibfnamefont {T.}~\bibnamefont {{Ramirez}}}, \bibinfo
  {author} {\bibfnamefont {R.}~\bibnamefont {{Fern{\'a}ndez}}}, \bibinfo
  {author} {\bibfnamefont {A.~L.}\ \bibnamefont {{Knight}}}, \bibinfo {author}
  {\bibfnamefont {J.}~\bibnamefont {{Jesse}}}, \bibinfo {author} {\bibfnamefont
  {F.}~\bibnamefont {{H{\'e}bert}}}, \bibinfo {author} {\bibfnamefont {L.~E.}\
  \bibnamefont {{Kidder}}}, \bibinfo {author} {\bibfnamefont {H.~P.}\
  \bibnamefont {{Pfeiffer}}},\ and\ \bibinfo {author} {\bibfnamefont {M.~A.}\
  \bibnamefont {{Scheel}}},\ }\href@noop {} {\bibfield  {journal} {\bibinfo
  {journal} {Classical and Quantum Gravity}\ }\textbf {\bibinfo {volume}
  {40}},\ \bibinfo {pages} {085008} (\bibinfo {year} {2023})}\BibitemShut
  {NoStop}%
\bibitem [{\citenamefont {{Fern{\'a}ndez}}\ \emph {et~al.}(2024)\citenamefont
  {{Fern{\'a}ndez}}, \citenamefont {{Just}}, \citenamefont {{Xiong}},\ and\
  \citenamefont {{Mart{\'\i}nez-Pinedo}}}]{Fernandez2024a}%
  \BibitemOpen
  \bibfield  {author} {\bibinfo {author} {\bibfnamefont {R.}~\bibnamefont
  {{Fern{\'a}ndez}}}, \bibinfo {author} {\bibfnamefont {O.}~\bibnamefont
  {{Just}}}, \bibinfo {author} {\bibfnamefont {Z.}~\bibnamefont {{Xiong}}},\
  and\ \bibinfo {author} {\bibfnamefont {G.}~\bibnamefont
  {{Mart{\'\i}nez-Pinedo}}},\ }\href@noop {} {\bibfield  {journal} {\bibinfo
  {journal} {\prd}\ }\textbf {\bibinfo {volume} {110}},\ \bibinfo {pages}
  {023001} (\bibinfo {year} {2024})}\BibitemShut {NoStop}%
\bibitem [{\citenamefont {{Rosswog}}\ \emph {et~al.}(2014)\citenamefont
  {{Rosswog}}, \citenamefont {{Korobkin}}, \citenamefont {{Arcones}},
  \citenamefont {{Thielemann}},\ and\ \citenamefont {{Piran}}}]{Rosswog2014}%
  \BibitemOpen
  \bibfield  {author} {\bibinfo {author} {\bibfnamefont {S.}~\bibnamefont
  {{Rosswog}}}, \bibinfo {author} {\bibfnamefont {O.}~\bibnamefont
  {{Korobkin}}}, \bibinfo {author} {\bibfnamefont {A.}~\bibnamefont
  {{Arcones}}}, \bibinfo {author} {\bibfnamefont {F.-K.}\ \bibnamefont
  {{Thielemann}}},\ and\ \bibinfo {author} {\bibfnamefont {T.}~\bibnamefont
  {{Piran}}},\ }\href@noop {} {\bibfield  {journal} {\bibinfo  {journal} {{Mon.
  Not. R. Astron. Soc.}}\ }\textbf {\bibinfo {volume} {439}},\ \bibinfo {pages}
  {744} (\bibinfo {year} {2014})}\BibitemShut {NoStop}%
\bibitem [{\citenamefont {{Fern{\'a}ndez}}\ \emph {et~al.}(2015)\citenamefont
  {{Fern{\'a}ndez}}, \citenamefont {{Quataert}}, \citenamefont {{Schwab}},
  \citenamefont {{Kasen}},\ and\ \citenamefont {{Rosswog}}}]{Fernandez2015c}%
  \BibitemOpen
  \bibfield  {author} {\bibinfo {author} {\bibfnamefont {R.}~\bibnamefont
  {{Fern{\'a}ndez}}}, \bibinfo {author} {\bibfnamefont {E.}~\bibnamefont
  {{Quataert}}}, \bibinfo {author} {\bibfnamefont {J.}~\bibnamefont
  {{Schwab}}}, \bibinfo {author} {\bibfnamefont {D.}~\bibnamefont {{Kasen}}},\
  and\ \bibinfo {author} {\bibfnamefont {S.}~\bibnamefont {{Rosswog}}},\
  }\href@noop {} {\bibfield  {journal} {\bibinfo  {journal} {{Mon. Not. R.
  Astron. Soc.}}\ }\textbf {\bibinfo {volume} {449}},\ \bibinfo {pages} {390}
  (\bibinfo {year} {2015})}\BibitemShut {NoStop}%
\bibitem [{\citenamefont {{Just}}\ \emph
  {et~al.}(2015{\natexlab{a}})\citenamefont {{Just}}, \citenamefont
  {{Bauswein}}, \citenamefont {{Pulpillo}}, \citenamefont {{Goriely}},\ and\
  \citenamefont {{Janka}}}]{Just2015a}%
  \BibitemOpen
  \bibfield  {author} {\bibinfo {author} {\bibfnamefont {O.}~\bibnamefont
  {{Just}}}, \bibinfo {author} {\bibfnamefont {A.}~\bibnamefont {{Bauswein}}},
  \bibinfo {author} {\bibfnamefont {R.~A.}\ \bibnamefont {{Pulpillo}}},
  \bibinfo {author} {\bibfnamefont {S.}~\bibnamefont {{Goriely}}},\ and\
  \bibinfo {author} {\bibfnamefont {H.-T.}\ \bibnamefont {{Janka}}},\
  }\href@noop {} {\bibfield  {journal} {\bibinfo  {journal} {{Mon. Not. R.
  Astron. Soc.}}\ }\textbf {\bibinfo {volume} {448}},\ \bibinfo {pages} {541}
  (\bibinfo {year} {2015}{\natexlab{a}})}\BibitemShut {NoStop}%
\bibitem [{\citenamefont {{Wu}}\ \emph {et~al.}(2016)\citenamefont {{Wu}},
  \citenamefont {{Fern{\'a}ndez}}, \citenamefont {{Mart{\'{\i}}nez-Pinedo}},\
  and\ \citenamefont {{Metzger}}}]{Wu2016a}%
  \BibitemOpen
  \bibfield  {author} {\bibinfo {author} {\bibfnamefont {M.-R.}\ \bibnamefont
  {{Wu}}}, \bibinfo {author} {\bibfnamefont {R.}~\bibnamefont
  {{Fern{\'a}ndez}}}, \bibinfo {author} {\bibfnamefont {G.}~\bibnamefont
  {{Mart{\'{\i}}nez-Pinedo}}},\ and\ \bibinfo {author} {\bibfnamefont {B.~D.}\
  \bibnamefont {{Metzger}}},\ }\href@noop {} {\bibfield  {journal} {\bibinfo
  {journal} {{Mon. Not. R. Astron. Soc.}}\ }\textbf {\bibinfo {volume} {463}},\
  \bibinfo {pages} {2323} (\bibinfo {year} {2016})},\ \Eprint
  {https://arxiv.org/abs/1607.05290} {arXiv:1607.05290 [astro-ph.HE]}
  \BibitemShut {NoStop}%
\bibitem [{\citenamefont {{Desai}}\ \emph {et~al.}(2019)\citenamefont
  {{Desai}}, \citenamefont {{Metzger}},\ and\ \citenamefont
  {{Foucart}}}]{Desai2019a}%
  \BibitemOpen
  \bibfield  {author} {\bibinfo {author} {\bibfnamefont {D.}~\bibnamefont
  {{Desai}}}, \bibinfo {author} {\bibfnamefont {B.~D.}\ \bibnamefont
  {{Metzger}}},\ and\ \bibinfo {author} {\bibfnamefont {F.}~\bibnamefont
  {{Foucart}}},\ }\href@noop {} {\bibfield  {journal} {\bibinfo  {journal}
  {{Mon. Not. R. Astron. Soc.}}\ }\textbf {\bibinfo {volume} {485}},\ \bibinfo
  {pages} {4404} (\bibinfo {year} {2019})}\BibitemShut {NoStop}%
\bibitem [{\citenamefont {{Foucart}}\ \emph {et~al.}(2021)\citenamefont
  {{Foucart}}, \citenamefont {{M{\"o}sta}}, \citenamefont {{Ramirez}},
  \citenamefont {{Wright}}, \citenamefont {{Darbha}},\ and\ \citenamefont
  {{Kasen}}}]{Foucart2021a}%
  \BibitemOpen
  \bibfield  {author} {\bibinfo {author} {\bibfnamefont {F.}~\bibnamefont
  {{Foucart}}}, \bibinfo {author} {\bibfnamefont {P.}~\bibnamefont
  {{M{\"o}sta}}}, \bibinfo {author} {\bibfnamefont {T.}~\bibnamefont
  {{Ramirez}}}, \bibinfo {author} {\bibfnamefont {A.~J.}\ \bibnamefont
  {{Wright}}}, \bibinfo {author} {\bibfnamefont {S.}~\bibnamefont {{Darbha}}},\
  and\ \bibinfo {author} {\bibfnamefont {D.}~\bibnamefont {{Kasen}}},\
  }\href@noop {} {\bibfield  {journal} {\bibinfo  {journal} {\prd}\ }\textbf
  {\bibinfo {volume} {104}},\ \bibinfo {pages} {123010} (\bibinfo {year}
  {2021})}\BibitemShut {NoStop}%
\bibitem [{\citenamefont {{Darbha}}\ \emph {et~al.}(2021)\citenamefont
  {{Darbha}}, \citenamefont {{Kasen}}, \citenamefont {{Foucart}},\ and\
  \citenamefont {{Price}}}]{Darbha2021a}%
  \BibitemOpen
  \bibfield  {author} {\bibinfo {author} {\bibfnamefont {S.}~\bibnamefont
  {{Darbha}}}, \bibinfo {author} {\bibfnamefont {D.}~\bibnamefont {{Kasen}}},
  \bibinfo {author} {\bibfnamefont {F.}~\bibnamefont {{Foucart}}},\ and\
  \bibinfo {author} {\bibfnamefont {D.~J.}\ \bibnamefont {{Price}}},\
  }\href@noop {} {\bibfield  {journal} {\bibinfo  {journal} {\apj}\ }\textbf
  {\bibinfo {volume} {915}},\ \bibinfo {pages} {69} (\bibinfo {year}
  {2021})}\BibitemShut {NoStop}%
\bibitem [{\citenamefont {{Sneppen}}\ \emph {et~al.}(2023)\citenamefont
  {{Sneppen}}, \citenamefont {{Watson}}, \citenamefont {{Bauswein}},
  \citenamefont {{Just}}, \citenamefont {{Kotak}}, \citenamefont {{Nakar}},
  \citenamefont {{Poznanski}},\ and\ \citenamefont {{Sim}}}]{Sneppen2023b}%
  \BibitemOpen
  \bibfield  {author} {\bibinfo {author} {\bibfnamefont {A.}~\bibnamefont
  {{Sneppen}}}, \bibinfo {author} {\bibfnamefont {D.}~\bibnamefont {{Watson}}},
  \bibinfo {author} {\bibfnamefont {A.}~\bibnamefont {{Bauswein}}}, \bibinfo
  {author} {\bibfnamefont {O.}~\bibnamefont {{Just}}}, \bibinfo {author}
  {\bibfnamefont {R.}~\bibnamefont {{Kotak}}}, \bibinfo {author} {\bibfnamefont
  {E.}~\bibnamefont {{Nakar}}}, \bibinfo {author} {\bibfnamefont
  {D.}~\bibnamefont {{Poznanski}}},\ and\ \bibinfo {author} {\bibfnamefont
  {S.}~\bibnamefont {{Sim}}},\ }\href@noop {} {\bibfield  {journal} {\bibinfo
  {journal} {\nat}\ }\textbf {\bibinfo {volume} {614}},\ \bibinfo {pages} {436}
  (\bibinfo {year} {2023})}\BibitemShut {NoStop}%
\bibitem [{\citenamefont {{Rosswog}}(2007)}]{Rosswog2007c}%
  \BibitemOpen
  \bibfield  {author} {\bibinfo {author} {\bibfnamefont {S.}~\bibnamefont
  {{Rosswog}}},\ }\href@noop {} {\bibfield  {journal} {\bibinfo  {journal}
  {\mnras}\ }\textbf {\bibinfo {volume} {376}},\ \bibinfo {pages} {L48}
  (\bibinfo {year} {2007})}\BibitemShut {NoStop}%
\bibitem [{\citenamefont {{Metzger}}\ \emph {et~al.}(2010)\citenamefont
  {{Metzger}}, \citenamefont {{Arcones}}, \citenamefont {{Quataert}},\ and\
  \citenamefont {{Mart{\'{\i}}nez-Pinedo}}}]{Metzger2010}%
  \BibitemOpen
  \bibfield  {author} {\bibinfo {author} {\bibfnamefont {B.~D.}\ \bibnamefont
  {{Metzger}}}, \bibinfo {author} {\bibfnamefont {A.}~\bibnamefont
  {{Arcones}}}, \bibinfo {author} {\bibfnamefont {E.}~\bibnamefont
  {{Quataert}}},\ and\ \bibinfo {author} {\bibfnamefont {G.}~\bibnamefont
  {{Mart{\'{\i}}nez-Pinedo}}},\ }\href@noop {} {\bibfield  {journal} {\bibinfo
  {journal} {\mnras}\ }\textbf {\bibinfo {volume} {402}},\ \bibinfo {pages}
  {2771} (\bibinfo {year} {2010})},\ \Eprint {https://arxiv.org/abs/0908.0530}
  {arXiv:0908.0530 [astro-ph.HE]} \BibitemShut {NoStop}%
\bibitem [{\citenamefont {{Ishizaki}}\ \emph {et~al.}(2021)\citenamefont
  {{Ishizaki}}, \citenamefont {{Kiuchi}}, \citenamefont {{Ioka}},\ and\
  \citenamefont {{Wanajo}}}]{Ishizaki2021a}%
  \BibitemOpen
  \bibfield  {author} {\bibinfo {author} {\bibfnamefont {W.}~\bibnamefont
  {{Ishizaki}}}, \bibinfo {author} {\bibfnamefont {K.}~\bibnamefont
  {{Kiuchi}}}, \bibinfo {author} {\bibfnamefont {K.}~\bibnamefont {{Ioka}}},\
  and\ \bibinfo {author} {\bibfnamefont {S.}~\bibnamefont {{Wanajo}}},\
  }\href@noop {} {\bibfield  {journal} {\bibinfo  {journal} {\apj}\ }\textbf
  {\bibinfo {volume} {922}},\ \bibinfo {pages} {185} (\bibinfo {year}
  {2021})}\BibitemShut {NoStop}%
\bibitem [{\citenamefont {{Musolino}}\ \emph {et~al.}(2024)\citenamefont
  {{Musolino}}, \citenamefont {{Duqu{\'e}}},\ and\ \citenamefont
  {{Rezzolla}}}]{Musolino2024e}%
  \BibitemOpen
  \bibfield  {author} {\bibinfo {author} {\bibfnamefont {C.}~\bibnamefont
  {{Musolino}}}, \bibinfo {author} {\bibfnamefont {R.}~\bibnamefont
  {{Duqu{\'e}}}},\ and\ \bibinfo {author} {\bibfnamefont {L.}~\bibnamefont
  {{Rezzolla}}},\ }\href@noop {} {\bibfield  {journal} {\bibinfo  {journal}
  {\apjl}\ }\textbf {\bibinfo {volume} {966}},\ \bibinfo {pages} {L31}
  (\bibinfo {year} {2024})}\BibitemShut {NoStop}%
\bibitem [{\citenamefont {{Kawaguchi}}\ \emph {et~al.}(2024)\citenamefont
  {{Kawaguchi}}, \citenamefont {{Domoto}}, \citenamefont {{Fujibayashi}},
  \citenamefont {{Hamidani}}, \citenamefont {{Hayashi}}, \citenamefont
  {{Shibata}}, \citenamefont {{Tanaka}},\ and\ \citenamefont
  {{Wanajo}}}]{Kawaguchi2024c}%
  \BibitemOpen
  \bibfield  {author} {\bibinfo {author} {\bibfnamefont {K.}~\bibnamefont
  {{Kawaguchi}}}, \bibinfo {author} {\bibfnamefont {N.}~\bibnamefont
  {{Domoto}}}, \bibinfo {author} {\bibfnamefont {S.}~\bibnamefont
  {{Fujibayashi}}}, \bibinfo {author} {\bibfnamefont {H.}~\bibnamefont
  {{Hamidani}}}, \bibinfo {author} {\bibfnamefont {K.}~\bibnamefont
  {{Hayashi}}}, \bibinfo {author} {\bibfnamefont {M.}~\bibnamefont
  {{Shibata}}}, \bibinfo {author} {\bibfnamefont {M.}~\bibnamefont
  {{Tanaka}}},\ and\ \bibinfo {author} {\bibfnamefont {S.}~\bibnamefont
  {{Wanajo}}},\ }\href@noop {} {\bibfield  {journal} {\bibinfo  {journal}
  {{Mon. Not. R. Astron. Soc.}}\ }\textbf {\bibinfo {volume} {535}},\ \bibinfo
  {pages} {3711} (\bibinfo {year} {2024})}\BibitemShut {NoStop}%
\bibitem [{\citenamefont {{Ma}}\ \emph {et~al.}(2025)\citenamefont {{Ma}},
  \citenamefont {{Pan}}, \citenamefont {{Wu}},\ and\ \citenamefont
  {{Fern{\'a}ndez}}}]{Ma2025c}%
  \BibitemOpen
  \bibfield  {author} {\bibinfo {author} {\bibfnamefont {L.-T.}\ \bibnamefont
  {{Ma}}}, \bibinfo {author} {\bibfnamefont {K.-C.}\ \bibnamefont {{Pan}}},
  \bibinfo {author} {\bibfnamefont {M.-R.}\ \bibnamefont {{Wu}}},\ and\
  \bibinfo {author} {\bibfnamefont {R.}~\bibnamefont {{Fern{\'a}ndez}}},\
  }\href@noop {} {\bibfield  {journal} {\bibinfo  {journal} {arXiv e-prints}\ }
  (\bibinfo {year} {2025})},\ \Eprint {https://arxiv.org/abs/arXiv:2508.15288}
  {arXiv:2508.15288} \BibitemShut {NoStop}%
\bibitem [{\citenamefont {Rosenblatt}(1958)}]{Rosenblatt1958a}%
  \BibitemOpen
  \bibfield  {author} {\bibinfo {author} {\bibfnamefont {F.}~\bibnamefont
  {Rosenblatt}},\ }\href@noop {} {\bibfield  {journal} {\bibinfo  {journal}
  {Psychological Review}\ }\textbf {\bibinfo {volume} {65}},\ \bibinfo {pages}
  {386} (\bibinfo {year} {1958})}\BibitemShut {NoStop}%
\bibitem [{\citenamefont {LeCun}\ \emph {et~al.}(1989)\citenamefont {LeCun},
  \citenamefont {Boser}, \citenamefont {Denker}, \citenamefont {Henderson},
  \citenamefont {Howard}, \citenamefont {Hubbard},\ and\ \citenamefont
  {Jackel}}]{LeCun1989a}%
  \BibitemOpen
  \bibfield  {author} {\bibinfo {author} {\bibfnamefont {Y.}~\bibnamefont
  {LeCun}}, \bibinfo {author} {\bibfnamefont {B.}~\bibnamefont {Boser}},
  \bibinfo {author} {\bibfnamefont {J.~S.}\ \bibnamefont {Denker}}, \bibinfo
  {author} {\bibfnamefont {D.}~\bibnamefont {Henderson}}, \bibinfo {author}
  {\bibfnamefont {R.~E.}\ \bibnamefont {Howard}}, \bibinfo {author}
  {\bibfnamefont {W.}~\bibnamefont {Hubbard}},\ and\ \bibinfo {author}
  {\bibfnamefont {L.~D.}\ \bibnamefont {Jackel}},\ }\href@noop {} {\bibfield
  {journal} {\bibinfo  {journal} {Neural Computation}\ }\textbf {\bibinfo
  {volume} {1}},\ \bibinfo {pages} {541} (\bibinfo {year} {1989})}\BibitemShut
  {NoStop}%
\bibitem [{\citenamefont {Hornik}\ \emph {et~al.}(1989)\citenamefont {Hornik},
  \citenamefont {Stinchcombe},\ and\ \citenamefont {White}}]{Hornik1989a}%
  \BibitemOpen
  \bibfield  {author} {\bibinfo {author} {\bibfnamefont {K.}~\bibnamefont
  {Hornik}}, \bibinfo {author} {\bibfnamefont {M.}~\bibnamefont
  {Stinchcombe}},\ and\ \bibinfo {author} {\bibfnamefont {H.}~\bibnamefont
  {White}},\ }\href@noop {} {\bibfield  {journal} {\bibinfo  {journal} {Neural
  Networks}\ }\textbf {\bibinfo {volume} {2}},\ \bibinfo {pages} {359}
  (\bibinfo {year} {1989})}\BibitemShut {NoStop}%
\bibitem [{\citenamefont {Cybenko}(1989)}]{Cybenko1989a}%
  \BibitemOpen
  \bibfield  {author} {\bibinfo {author} {\bibfnamefont {G.}~\bibnamefont
  {Cybenko}},\ }\href@noop {} {\bibfield  {journal} {\bibinfo  {journal}
  {Mathematics of Control, Signals and Systems}\ }\textbf {\bibinfo {volume}
  {2}},\ \bibinfo {pages} {303} (\bibinfo {year} {1989})}\BibitemShut {NoStop}%
\bibitem [{\citenamefont {{Schmidhuber}}(2014)}]{Schmidhuber2014a}%
  \BibitemOpen
  \bibfield  {author} {\bibinfo {author} {\bibfnamefont {J.}~\bibnamefont
  {{Schmidhuber}}},\ }\href@noop {} {\bibfield  {journal} {\bibinfo  {journal}
  {arXiv e-prints}\ ,\ \bibinfo {pages} {arXiv:1404.7828}} (\bibinfo {year}
  {2014})}\BibitemShut {NoStop}%
\bibitem [{\citenamefont {Soultanis}\ \emph {et~al.}(2025)\citenamefont
  {Soultanis}, \citenamefont {Maltsev}, \citenamefont {Bauswein}, \citenamefont
  {Chatziioannou}, \citenamefont {R\"opke},\ and\ \citenamefont
  {Stergioulas}}]{Soultanis2025a}%
  \BibitemOpen
  \bibfield  {author} {\bibinfo {author} {\bibfnamefont {T.}~\bibnamefont
  {Soultanis}}, \bibinfo {author} {\bibfnamefont {K.}~\bibnamefont {Maltsev}},
  \bibinfo {author} {\bibfnamefont {A.}~\bibnamefont {Bauswein}}, \bibinfo
  {author} {\bibfnamefont {K.}~\bibnamefont {Chatziioannou}}, \bibinfo {author}
  {\bibfnamefont {F.~K.}\ \bibnamefont {R\"opke}},\ and\ \bibinfo {author}
  {\bibfnamefont {N.}~\bibnamefont {Stergioulas}},\ }\href@noop {} {\bibfield
  {journal} {\bibinfo  {journal} {Phys. Rev. D}\ }\textbf {\bibinfo {volume}
  {111}},\ \bibinfo {pages} {023002} (\bibinfo {year} {2025})}\BibitemShut
  {NoStop}%
\bibitem [{\citenamefont {Mirouh}\ \emph {et~al.}(2018)\citenamefont {Mirouh},
  \citenamefont {Angelou}, \citenamefont {Reese},\ and\ \citenamefont
  {Costa}}]{Mirouh2018a}%
  \BibitemOpen
  \bibfield  {author} {\bibinfo {author} {\bibfnamefont {G.~M.}\ \bibnamefont
  {Mirouh}}, \bibinfo {author} {\bibfnamefont {G.~C.}\ \bibnamefont {Angelou}},
  \bibinfo {author} {\bibfnamefont {D.~R.}\ \bibnamefont {Reese}},\ and\
  \bibinfo {author} {\bibfnamefont {G.}~\bibnamefont {Costa}},\ }\href@noop {}
  {\bibfield  {journal} {\bibinfo  {journal} {Monthly Notices of the Royal
  Astronomical Society: Letters}\ }\textbf {\bibinfo {volume} {483}},\ \bibinfo
  {pages} {L28} (\bibinfo {year} {2018})}\BibitemShut {NoStop}%
\bibitem [{\citenamefont {Hendriks}\ and\ \citenamefont
  {Aerts}(2019)}]{Hendriks2019a}%
  \BibitemOpen
  \bibfield  {author} {\bibinfo {author} {\bibfnamefont {L.}~\bibnamefont
  {Hendriks}}\ and\ \bibinfo {author} {\bibfnamefont {C.}~\bibnamefont
  {Aerts}},\ }\href@noop {} {\bibfield  {journal} {\bibinfo  {journal}
  {Publications of the Astronomical Society of the Pacific}\ }\textbf {\bibinfo
  {volume} {131}},\ \bibinfo {pages} {108001} (\bibinfo {year}
  {2019})}\BibitemShut {NoStop}%
\bibitem [{\citenamefont {{Tsang}}\ \emph {et~al.}(2022)\citenamefont
  {{Tsang}}, \citenamefont {{Vartanyan}},\ and\ \citenamefont
  {{Burrows}}}]{Tsang2022a}%
  \BibitemOpen
  \bibfield  {author} {\bibinfo {author} {\bibfnamefont {B.~T.~H.}\
  \bibnamefont {{Tsang}}}, \bibinfo {author} {\bibfnamefont {D.}~\bibnamefont
  {{Vartanyan}}},\ and\ \bibinfo {author} {\bibfnamefont {A.}~\bibnamefont
  {{Burrows}}},\ }\href@noop {} {\bibfield  {journal} {\bibinfo  {journal}
  {{Astrophys. J. Lett.}}\ }\textbf {\bibinfo {volume} {937}},\ \bibinfo
  {pages} {L15} (\bibinfo {year} {2022})}\BibitemShut {NoStop}%
\bibitem [{\citenamefont {Maltsev}\ \emph {et~al.}(2025)\citenamefont
  {Maltsev}, \citenamefont {Schneider}, \citenamefont {Mandel}, \citenamefont
  {M\"uller}, \citenamefont {Heger}, \citenamefont {R\"opke},\ and\
  \citenamefont {Laplace}}]{Maltsev2025a}%
  \BibitemOpen
  \bibfield  {author} {\bibinfo {author} {\bibfnamefont {K.}~\bibnamefont
  {Maltsev}}, \bibinfo {author} {\bibfnamefont {F.~R.~N.}\ \bibnamefont
  {Schneider}}, \bibinfo {author} {\bibfnamefont {I.}~\bibnamefont {Mandel}},
  \bibinfo {author} {\bibfnamefont {B.}~\bibnamefont {M\"uller}}, \bibinfo
  {author} {\bibfnamefont {A.}~\bibnamefont {Heger}}, \bibinfo {author}
  {\bibfnamefont {F.~K.}\ \bibnamefont {R\"opke}},\ and\ \bibinfo {author}
  {\bibfnamefont {E.}~\bibnamefont {Laplace}},\ }\href@noop {} {\bibfield
  {journal} {\bibinfo  {journal} {eprint arXiv:2503.23856}\ } (\bibinfo {year}
  {2025})}\BibitemShut {NoStop}%
\bibitem [{\citenamefont {{Harada}}\ \emph {et~al.}(2022)\citenamefont
  {{Harada}}, \citenamefont {{Nishikawa}},\ and\ \citenamefont
  {{Yamada}}}]{Harada2022a}%
  \BibitemOpen
  \bibfield  {author} {\bibinfo {author} {\bibfnamefont {A.}~\bibnamefont
  {{Harada}}}, \bibinfo {author} {\bibfnamefont {S.}~\bibnamefont
  {{Nishikawa}}},\ and\ \bibinfo {author} {\bibfnamefont {S.}~\bibnamefont
  {{Yamada}}},\ }\href@noop {} {\bibfield  {journal} {\bibinfo  {journal}
  {\apj}\ }\textbf {\bibinfo {volume} {925}},\ \bibinfo {pages} {117} (\bibinfo
  {year} {2022})}\BibitemShut {NoStop}%
\bibitem [{\citenamefont {{Abbar}}\ \emph {et~al.}(2024)\citenamefont
  {{Abbar}}, \citenamefont {{Wu}},\ and\ \citenamefont
  {{Xiong}}}]{abbar2024physics}%
  \BibitemOpen
  \bibfield  {author} {\bibinfo {author} {\bibfnamefont {S.}~\bibnamefont
  {{Abbar}}}, \bibinfo {author} {\bibfnamefont {M.-R.}\ \bibnamefont {{Wu}}},\
  and\ \bibinfo {author} {\bibfnamefont {Z.}~\bibnamefont {{Xiong}}},\ }\href
  {https://doi.org/10.1103/PhysRevD.109.043024} {\bibfield  {journal} {\bibinfo
   {journal} {Phys. Rev. D}\ }\textbf {\bibinfo {volume} {109}},\ \bibinfo
  {pages} {043024} (\bibinfo {year} {2024})},\ \Eprint
  {https://arxiv.org/abs/2311.15656} {2311.15656} \BibitemShut {NoStop}%
\bibitem [{\citenamefont {Richers}\ \emph {et~al.}(2024)\citenamefont
  {Richers}, \citenamefont {Froustey}, \citenamefont {Ghosh}, \citenamefont
  {Foucart},\ and\ \citenamefont {Gomez}}]{richers2024asymptoticstate}%
  \BibitemOpen
  \bibfield  {author} {\bibinfo {author} {\bibfnamefont {S.}~\bibnamefont
  {Richers}}, \bibinfo {author} {\bibfnamefont {J.}~\bibnamefont {Froustey}},
  \bibinfo {author} {\bibfnamefont {S.}~\bibnamefont {Ghosh}}, \bibinfo
  {author} {\bibfnamefont {F.}~\bibnamefont {Foucart}},\ and\ \bibinfo {author}
  {\bibfnamefont {J.}~\bibnamefont {Gomez}},\ }\href
  {https://doi.org/10.1103/PhysRevD.110.103019} {\bibfield  {journal} {\bibinfo
   {journal} {Phys. Rev. D}\ }\textbf {\bibinfo {volume} {110}},\ \bibinfo
  {pages} {103019} (\bibinfo {year} {2024})},\ \Eprint
  {https://arxiv.org/abs/2409.04405} {arXiv:2409.04405 [astro-ph.HE]}
  \BibitemShut {NoStop}%
\bibitem [{\citenamefont {Fan}\ \emph {et~al.}(2022)\citenamefont {Fan},
  \citenamefont {Willcox}, \citenamefont {DeGrendele}, \citenamefont
  {Zingale},\ and\ \citenamefont {Nonaka}}]{Fan2022a}%
  \BibitemOpen
  \bibfield  {author} {\bibinfo {author} {\bibfnamefont {D.}~\bibnamefont
  {Fan}}, \bibinfo {author} {\bibfnamefont {D.~E.}\ \bibnamefont {Willcox}},
  \bibinfo {author} {\bibfnamefont {C.}~\bibnamefont {DeGrendele}}, \bibinfo
  {author} {\bibfnamefont {M.}~\bibnamefont {Zingale}},\ and\ \bibinfo {author}
  {\bibfnamefont {A.}~\bibnamefont {Nonaka}},\ }\href@noop {} {\bibfield
  {journal} {\bibinfo  {journal} {The Astrophysical Journal}\ }\textbf
  {\bibinfo {volume} {940}},\ \bibinfo {pages} {134} (\bibinfo {year}
  {2022})}\BibitemShut {NoStop}%
\bibitem [{\citenamefont {Li}\ \emph {et~al.}(2024)\citenamefont {Li},
  \citenamefont {Sprouse}, \citenamefont {Meyer},\ and\ \citenamefont
  {Mumpower}}]{Li2024a}%
  \BibitemOpen
  \bibfield  {author} {\bibinfo {author} {\bibfnamefont {M.}~\bibnamefont
  {Li}}, \bibinfo {author} {\bibfnamefont {T.~M.}\ \bibnamefont {Sprouse}},
  \bibinfo {author} {\bibfnamefont {B.~S.}\ \bibnamefont {Meyer}},\ and\
  \bibinfo {author} {\bibfnamefont {M.~R.}\ \bibnamefont {Mumpower}},\
  }\href@noop {} {\bibfield  {journal} {\bibinfo  {journal} {Physics Letters
  B}\ }\textbf {\bibinfo {volume} {848}},\ \bibinfo {pages} {138385} (\bibinfo
  {year} {2024})}\BibitemShut {NoStop}%
\bibitem [{\citenamefont {Grichener}\ \emph {et~al.}(2025)\citenamefont
  {Grichener}, \citenamefont {Renzo}, \citenamefont {Kerzendorf}, \citenamefont
  {Farmer}, \citenamefont {de~Mink}, \citenamefont {Bellinger}, \citenamefont
  {Chan}, \citenamefont {Chen}, \citenamefont {Farag},\ and\ \citenamefont
  {Justham}}]{Grichener2025a}%
  \BibitemOpen
  \bibfield  {author} {\bibinfo {author} {\bibfnamefont {A.}~\bibnamefont
  {Grichener}}, \bibinfo {author} {\bibfnamefont {M.}~\bibnamefont {Renzo}},
  \bibinfo {author} {\bibfnamefont {W.~E.}\ \bibnamefont {Kerzendorf}},
  \bibinfo {author} {\bibfnamefont {R.}~\bibnamefont {Farmer}}, \bibinfo
  {author} {\bibfnamefont {S.~E.}\ \bibnamefont {de~Mink}}, \bibinfo {author}
  {\bibfnamefont {E.~P.}\ \bibnamefont {Bellinger}}, \bibinfo {author}
  {\bibfnamefont {C.-k.}\ \bibnamefont {Chan}}, \bibinfo {author}
  {\bibfnamefont {N.}~\bibnamefont {Chen}}, \bibinfo {author} {\bibfnamefont
  {E.}~\bibnamefont {Farag}},\ and\ \bibinfo {author} {\bibfnamefont
  {S.}~\bibnamefont {Justham}},\ }\href@noop {} {\bibfield  {journal} {\bibinfo
   {journal} {eprint arXiv:2503.00115}\ } (\bibinfo {year} {2025})}\BibitemShut
  {NoStop}%
\bibitem [{\citenamefont {{Saito}}\ \emph {et~al.}(2025)\citenamefont
  {{Saito}}, \citenamefont {{Dillmann}}, \citenamefont {{Kr{\"u}cken}},
  \citenamefont {{Mumpower}},\ and\ \citenamefont {{Surman}}}]{Saito2025a}%
  \BibitemOpen
  \bibfield  {author} {\bibinfo {author} {\bibfnamefont {Y.}~\bibnamefont
  {{Saito}}}, \bibinfo {author} {\bibfnamefont {I.}~\bibnamefont {{Dillmann}}},
  \bibinfo {author} {\bibfnamefont {R.}~\bibnamefont {{Kr{\"u}cken}}}, \bibinfo
  {author} {\bibfnamefont {M.~R.}\ \bibnamefont {{Mumpower}}},\ and\ \bibinfo
  {author} {\bibfnamefont {R.}~\bibnamefont {{Surman}}},\ }\href@noop {}
  {\bibfield  {journal} {\bibinfo  {journal} {Journal of Physics G Nuclear
  Physics}\ }\textbf {\bibinfo {volume} {52}},\ \bibinfo {pages} {055201}
  (\bibinfo {year} {2025})}\BibitemShut {NoStop}%
\bibitem [{\citenamefont {{Siegel}}\ and\ \citenamefont
  {{Metzger}}(2017)}]{Siegel2017b}%
  \BibitemOpen
  \bibfield  {author} {\bibinfo {author} {\bibfnamefont {D.~M.}\ \bibnamefont
  {{Siegel}}}\ and\ \bibinfo {author} {\bibfnamefont {B.~D.}\ \bibnamefont
  {{Metzger}}},\ }\href@noop {} {\bibfield  {journal} {\bibinfo  {journal}
  {Physical Review Letters}\ }\textbf {\bibinfo {volume} {119}},\ \bibinfo
  {pages} {231102} (\bibinfo {year} {2017})}\BibitemShut {NoStop}%
\bibitem [{\citenamefont {{Fujibayashi}}\ \emph {et~al.}(2020)\citenamefont
  {{Fujibayashi}}, \citenamefont {{Shibata}}, \citenamefont {{Wanajo}},
  \citenamefont {{Kiuchi}}, \citenamefont {{Kyutoku}},\ and\ \citenamefont
  {{Sekiguchi}}}]{Fujibayashi2020a}%
  \BibitemOpen
  \bibfield  {author} {\bibinfo {author} {\bibfnamefont {S.}~\bibnamefont
  {{Fujibayashi}}}, \bibinfo {author} {\bibfnamefont {M.}~\bibnamefont
  {{Shibata}}}, \bibinfo {author} {\bibfnamefont {S.}~\bibnamefont {{Wanajo}}},
  \bibinfo {author} {\bibfnamefont {K.}~\bibnamefont {{Kiuchi}}}, \bibinfo
  {author} {\bibfnamefont {K.}~\bibnamefont {{Kyutoku}}},\ and\ \bibinfo
  {author} {\bibfnamefont {Y.}~\bibnamefont {{Sekiguchi}}},\ }\href@noop {}
  {\bibfield  {journal} {\bibinfo  {journal} {\prd}\ }\textbf {\bibinfo
  {volume} {101}},\ \bibinfo {pages} {083029} (\bibinfo {year}
  {2020})}\BibitemShut {NoStop}%
\bibitem [{\citenamefont {{Bauswein}}\ \emph {et~al.}(2013)\citenamefont
  {{Bauswein}}, \citenamefont {{Goriely}},\ and\ \citenamefont
  {{Janka}}}]{Bauswein2013}%
  \BibitemOpen
  \bibfield  {author} {\bibinfo {author} {\bibfnamefont {A.}~\bibnamefont
  {{Bauswein}}}, \bibinfo {author} {\bibfnamefont {S.}~\bibnamefont
  {{Goriely}}},\ and\ \bibinfo {author} {\bibfnamefont {H.-T.}\ \bibnamefont
  {{Janka}}},\ }\href@noop {} {\bibfield  {journal} {\bibinfo  {journal}
  {\apj}\ }\textbf {\bibinfo {volume} {773}},\ \bibinfo {pages} {78} (\bibinfo
  {year} {2013})}\BibitemShut {NoStop}%
\bibitem [{\citenamefont {{Hotokezaka}}\ \emph {et~al.}(2013)\citenamefont
  {{Hotokezaka}}, \citenamefont {{Kiuchi}}, \citenamefont {{Kyutoku}},
  \citenamefont {{Okawa}}, \citenamefont {{Sekiguchi}}, \citenamefont
  {{Shibata}},\ and\ \citenamefont {{Taniguchi}}}]{Hotokezaka2013b}%
  \BibitemOpen
  \bibfield  {author} {\bibinfo {author} {\bibfnamefont {K.}~\bibnamefont
  {{Hotokezaka}}}, \bibinfo {author} {\bibfnamefont {K.}~\bibnamefont
  {{Kiuchi}}}, \bibinfo {author} {\bibfnamefont {K.}~\bibnamefont {{Kyutoku}}},
  \bibinfo {author} {\bibfnamefont {H.}~\bibnamefont {{Okawa}}}, \bibinfo
  {author} {\bibfnamefont {Y.-i.}\ \bibnamefont {{Sekiguchi}}}, \bibinfo
  {author} {\bibfnamefont {M.}~\bibnamefont {{Shibata}}},\ and\ \bibinfo
  {author} {\bibfnamefont {K.}~\bibnamefont {{Taniguchi}}},\ }\href@noop {}
  {\bibfield  {journal} {\bibinfo  {journal} {\prd}\ }\textbf {\bibinfo
  {volume} {87}},\ \bibinfo {pages} {024001} (\bibinfo {year}
  {2013})}\BibitemShut {NoStop}%
\bibitem [{\citenamefont {{Radice}}\ \emph {et~al.}(2016)\citenamefont
  {{Radice}}, \citenamefont {{Galeazzi}}, \citenamefont {{Lippuner}},
  \citenamefont {{Roberts}}, \citenamefont {{Ott}},\ and\ \citenamefont
  {{Rezzolla}}}]{Radice2016a}%
  \BibitemOpen
  \bibfield  {author} {\bibinfo {author} {\bibfnamefont {D.}~\bibnamefont
  {{Radice}}}, \bibinfo {author} {\bibfnamefont {F.}~\bibnamefont
  {{Galeazzi}}}, \bibinfo {author} {\bibfnamefont {J.}~\bibnamefont
  {{Lippuner}}}, \bibinfo {author} {\bibfnamefont {L.~F.}\ \bibnamefont
  {{Roberts}}}, \bibinfo {author} {\bibfnamefont {C.~D.}\ \bibnamefont
  {{Ott}}},\ and\ \bibinfo {author} {\bibfnamefont {L.}~\bibnamefont
  {{Rezzolla}}},\ }\href@noop {} {\bibfield  {journal} {\bibinfo  {journal}
  {{Mon. Not. R. Astron. Soc.}}\ }\textbf {\bibinfo {volume} {460}},\ \bibinfo
  {pages} {3255} (\bibinfo {year} {2016})}\BibitemShut {NoStop}%
\bibitem [{\citenamefont {{Foucart}}\ \emph {et~al.}(2016)\citenamefont
  {{Foucart}}, \citenamefont {{Haas}}, \citenamefont {{Duez}}, \citenamefont
  {{O'Connor}}, \citenamefont {{Ott}}, \citenamefont {{Roberts}}, \citenamefont
  {{Kidder}}, \citenamefont {{Lippuner}}, \citenamefont {{Pfeiffer}},\ and\
  \citenamefont {{Scheel}}}]{Foucart2016}%
  \BibitemOpen
  \bibfield  {author} {\bibinfo {author} {\bibfnamefont {F.}~\bibnamefont
  {{Foucart}}}, \bibinfo {author} {\bibfnamefont {R.}~\bibnamefont {{Haas}}},
  \bibinfo {author} {\bibfnamefont {M.~D.}\ \bibnamefont {{Duez}}}, \bibinfo
  {author} {\bibfnamefont {E.}~\bibnamefont {{O'Connor}}}, \bibinfo {author}
  {\bibfnamefont {C.~D.}\ \bibnamefont {{Ott}}}, \bibinfo {author}
  {\bibfnamefont {L.}~\bibnamefont {{Roberts}}}, \bibinfo {author}
  {\bibfnamefont {L.~E.}\ \bibnamefont {{Kidder}}}, \bibinfo {author}
  {\bibfnamefont {J.}~\bibnamefont {{Lippuner}}}, \bibinfo {author}
  {\bibfnamefont {H.~P.}\ \bibnamefont {{Pfeiffer}}},\ and\ \bibinfo {author}
  {\bibfnamefont {M.~A.}\ \bibnamefont {{Scheel}}},\ }\href@noop {} {\bibfield
  {journal} {\bibinfo  {journal} {\prd}\ }\textbf {\bibinfo {volume} {93}},\
  \bibinfo {pages} {044019} (\bibinfo {year} {2016})}\BibitemShut {NoStop}%
\bibitem [{\citenamefont {{Just}}\ \emph
  {et~al.}(2015{\natexlab{b}})\citenamefont {{Just}}, \citenamefont
  {{Obergaulinger}},\ and\ \citenamefont {{Janka}}}]{Just2015b}%
  \BibitemOpen
  \bibfield  {author} {\bibinfo {author} {\bibfnamefont {O.}~\bibnamefont
  {{Just}}}, \bibinfo {author} {\bibfnamefont {M.}~\bibnamefont
  {{Obergaulinger}}},\ and\ \bibinfo {author} {\bibfnamefont {H.-T.}\
  \bibnamefont {{Janka}}},\ }\href@noop {} {\bibfield  {journal} {\bibinfo
  {journal} {{Mon. Not. R. Astron. Soc.}}\ }\textbf {\bibinfo {volume} {453}},\
  \bibinfo {pages} {3386} (\bibinfo {year} {2015}{\natexlab{b}})}\BibitemShut
  {NoStop}%
\bibitem [{\citenamefont {Hix}\ and\ \citenamefont
  {Meyer}(2006)}]{hix2006thermonuclear}%
  \BibitemOpen
  \bibfield  {author} {\bibinfo {author} {\bibfnamefont {W.~R.}\ \bibnamefont
  {Hix}}\ and\ \bibinfo {author} {\bibfnamefont {B.~S.}\ \bibnamefont
  {Meyer}},\ }\href {https://doi.org/10.1016/j.nuclphysa.2004.10.009}
  {\bibfield  {journal} {\bibinfo  {journal} {Nucl. Phys. A}\ }\textbf
  {\bibinfo {volume} {777}},\ \bibinfo {pages} {188} (\bibinfo {year}
  {2006})},\ \Eprint {https://arxiv.org/abs/0509698} {arXiv:0509698 [astro-ph]}
  \BibitemShut {NoStop}%
\bibitem [{\citenamefont {Wang}\ \emph {et~al.}(2021)\citenamefont {Wang},
  \citenamefont {Huang}, \citenamefont {Kondev}, \citenamefont {Audi},\ and\
  \citenamefont {Naimi}}]{Wang.Huang.ea:2021}%
  \BibitemOpen
  \bibfield  {author} {\bibinfo {author} {\bibfnamefont {M.}~\bibnamefont
  {Wang}}, \bibinfo {author} {\bibfnamefont {W.}~\bibnamefont {Huang}},
  \bibinfo {author} {\bibfnamefont {F.}~\bibnamefont {Kondev}}, \bibinfo
  {author} {\bibfnamefont {G.}~\bibnamefont {Audi}},\ and\ \bibinfo {author}
  {\bibfnamefont {S.}~\bibnamefont {Naimi}},\ }\href
  {https://doi.org/10.1088/1674-1137/abddaf} {\bibfield  {journal} {\bibinfo
  {journal} {Chinese Phys. C}\ }\textbf {\bibinfo {volume} {45}},\ \bibinfo
  {pages} {030003} (\bibinfo {year} {2021})}\BibitemShut {NoStop}%
\bibitem [{\citenamefont {M{\"o}ller}\ \emph {et~al.}(1995)\citenamefont
  {M{\"o}ller}, \citenamefont {Nix}, \citenamefont {Myers},\ and\ \citenamefont
  {Swiatecki}}]{moller1995nuclear}%
  \BibitemOpen
  \bibfield  {author} {\bibinfo {author} {\bibfnamefont {P.}~\bibnamefont
  {M{\"o}ller}}, \bibinfo {author} {\bibfnamefont {J.~R.}\ \bibnamefont {Nix}},
  \bibinfo {author} {\bibfnamefont {W.~D.}\ \bibnamefont {Myers}},\ and\
  \bibinfo {author} {\bibfnamefont {W.~J.}\ \bibnamefont {Swiatecki}},\ }\href
  {https://doi.org/10.1006/adnd.1995.1002} {\bibfield  {journal} {\bibinfo
  {journal} {Atom. Data Nucl. Data Tabl.}\ }\textbf {\bibinfo {volume} {59}},\
  \bibinfo {pages} {185} (\bibinfo {year} {1995})},\ \Eprint
  {https://arxiv.org/abs/nucl-th/9308022} {arXiv:nucl-th/9308022} \BibitemShut
  {NoStop}%
\bibitem [{\citenamefont {{Steiner}}\ \emph {et~al.}(2013)\citenamefont
  {{Steiner}}, \citenamefont {{Hempel}},\ and\ \citenamefont
  {{Fischer}}}]{Steiner2013}%
  \BibitemOpen
  \bibfield  {author} {\bibinfo {author} {\bibfnamefont {A.~W.}\ \bibnamefont
  {{Steiner}}}, \bibinfo {author} {\bibfnamefont {M.}~\bibnamefont
  {{Hempel}}},\ and\ \bibinfo {author} {\bibfnamefont {T.}~\bibnamefont
  {{Fischer}}},\ }\href@noop {} {\bibfield  {journal} {\bibinfo  {journal}
  {\apj}\ }\textbf {\bibinfo {volume} {774}},\ \bibinfo {pages} {17} (\bibinfo
  {year} {2013})}\BibitemShut {NoStop}%
\bibitem [{\citenamefont {Obergaulinger}(2008)}]{Obergaulinger2008a}%
  \BibitemOpen
  \bibfield  {author} {\bibinfo {author} {\bibfnamefont {M.}~\bibnamefont
  {Obergaulinger}},\ }\emph {\bibinfo {title} {Astrophysical
  magnetohydrodynamics and radiative transfer}},\ \href@noop {} {\bibinfo
  {type} {Dissertation}},\ \bibinfo  {school} {Technische Universit{\"a}t
  M{\"u}nchen}, \bibinfo {address} {M{\"u}nchen} (\bibinfo {year}
  {2008})\BibitemShut {NoStop}%
\bibitem [{\citenamefont {{Timmes}}\ and\ \citenamefont
  {{Swesty}}(2000)}]{timmes2000accuracy}%
  \BibitemOpen
  \bibfield  {author} {\bibinfo {author} {\bibfnamefont {F.~X.}\ \bibnamefont
  {{Timmes}}}\ and\ \bibinfo {author} {\bibfnamefont {F.~D.}\ \bibnamefont
  {{Swesty}}},\ }\href {https://doi.org/10.1086/313304} {\bibfield  {journal}
  {\bibinfo  {journal} {Astrophys. J., Suppl.}\ }\textbf {\bibinfo {volume}
  {126}},\ \bibinfo {pages} {501} (\bibinfo {year} {2000})}\BibitemShut
  {NoStop}%
\bibitem [{\citenamefont {{Timmes}}\ and\ \citenamefont
  {{Arnett}}(1999)}]{Timmes1999}%
  \BibitemOpen
  \bibfield  {author} {\bibinfo {author} {\bibfnamefont {F.~X.}\ \bibnamefont
  {{Timmes}}}\ and\ \bibinfo {author} {\bibfnamefont {D.}~\bibnamefont
  {{Arnett}}},\ }\href@noop {} {\bibfield  {journal} {\bibinfo  {journal}
  {\apjs}\ }\textbf {\bibinfo {volume} {125}},\ \bibinfo {pages} {277}
  (\bibinfo {year} {1999})}\BibitemShut {NoStop}%
\bibitem [{\citenamefont {Goodfellow}\ \emph {et~al.}(2016)\citenamefont
  {Goodfellow}, \citenamefont {Bengio},\ and\ \citenamefont
  {Courville}}]{Goodfellow2016a}%
  \BibitemOpen
  \bibfield  {author} {\bibinfo {author} {\bibfnamefont {I.}~\bibnamefont
  {Goodfellow}}, \bibinfo {author} {\bibfnamefont {Y.}~\bibnamefont {Bengio}},\
  and\ \bibinfo {author} {\bibfnamefont {A.}~\bibnamefont {Courville}},\
  }\href@noop {} {\emph {\bibinfo {title} {Deep Learning}}}\ (\bibinfo
  {publisher} {MIT Press},\ \bibinfo {year} {2016})\BibitemShut {NoStop}%
\bibitem [{\citenamefont {Collins}\ \emph {et~al.}(2023)\citenamefont
  {Collins}, \citenamefont {Bauswein}, \citenamefont {Sim}, \citenamefont
  {Vijayan}, \citenamefont {Mart{\'{i}}nez-Pinedo}, \citenamefont {Just},
  \citenamefont {Shingles},\ and\ \citenamefont
  {Kromer}}]{collins2023radiative}%
  \BibitemOpen
  \bibfield  {author} {\bibinfo {author} {\bibfnamefont {C.~E.}\ \bibnamefont
  {Collins}}, \bibinfo {author} {\bibfnamefont {A.}~\bibnamefont {Bauswein}},
  \bibinfo {author} {\bibfnamefont {S.~A.}\ \bibnamefont {Sim}}, \bibinfo
  {author} {\bibfnamefont {V.}~\bibnamefont {Vijayan}}, \bibinfo {author}
  {\bibfnamefont {G.}~\bibnamefont {Mart{\'{i}}nez-Pinedo}}, \bibinfo {author}
  {\bibfnamefont {O.}~\bibnamefont {Just}}, \bibinfo {author} {\bibfnamefont
  {L.~J.}\ \bibnamefont {Shingles}},\ and\ \bibinfo {author} {\bibfnamefont
  {M.}~\bibnamefont {Kromer}},\ }\href {https://doi.org/10.1093/{Mon. Not. R.
  Astron. Soc.}/stad606} {\bibfield  {journal} {\bibinfo  {journal} {Mon. Not.
  R. Astron. Soc.}\ }\textbf {\bibinfo {volume} {521}},\ \bibinfo {pages}
  {1858} (\bibinfo {year} {2023})},\ \Eprint {https://arxiv.org/abs/2209.05246}
  {arXiv:2209.05246} \BibitemShut {NoStop}%
\bibitem [{\citenamefont {Mendoza-Temis}\ \emph {et~al.}(2015)\citenamefont
  {Mendoza-Temis}, \citenamefont {Wu}, \citenamefont {Langanke}, \citenamefont
  {Mart{\'{i}}nez-Pinedo}, \citenamefont {Bauswein},\ and\ \citenamefont
  {Janka}}]{mendoza2015nuclear}%
  \BibitemOpen
  \bibfield  {author} {\bibinfo {author} {\bibfnamefont {J.~D.~J.}\
  \bibnamefont {Mendoza-Temis}}, \bibinfo {author} {\bibfnamefont {M.~R.}\
  \bibnamefont {Wu}}, \bibinfo {author} {\bibfnamefont {K.}~\bibnamefont
  {Langanke}}, \bibinfo {author} {\bibfnamefont {G.}~\bibnamefont
  {Mart{\'{i}}nez-Pinedo}}, \bibinfo {author} {\bibfnamefont {A.}~\bibnamefont
  {Bauswein}},\ and\ \bibinfo {author} {\bibfnamefont {H.~T.}\ \bibnamefont
  {Janka}},\ }\href {https://doi.org/10.1103/PhysRevC.92.055805} {\bibfield
  {journal} {\bibinfo  {journal} {Phys. Rev. C}\ }\textbf {\bibinfo {volume}
  {92}},\ \bibinfo {pages} {055805} (\bibinfo {year} {2015})},\ \Eprint
  {https://arxiv.org/abs/1409.6135} {arXiv:1409.6135} \BibitemShut {NoStop}%
\bibitem [{\citenamefont {M{\"o}ller}\ \emph {et~al.}(2003)\citenamefont
  {M{\"o}ller}, \citenamefont {Pfeiffer},\ and\ \citenamefont
  {Kratz}}]{moller2003new}%
  \BibitemOpen
  \bibfield  {author} {\bibinfo {author} {\bibfnamefont {P.}~\bibnamefont
  {M{\"o}ller}}, \bibinfo {author} {\bibfnamefont {B.}~\bibnamefont
  {Pfeiffer}},\ and\ \bibinfo {author} {\bibfnamefont {K.~L.}\ \bibnamefont
  {Kratz}},\ }\href {https://doi.org/10.1103/PhysRevC.67.055802} {\bibfield
  {journal} {\bibinfo  {journal} {Phys. Rev. C}\ }\textbf {\bibinfo {volume}
  {67}},\ \bibinfo {pages} {055802} (\bibinfo {year} {2003})}\BibitemShut
  {NoStop}%
\bibitem [{\citenamefont {Goriely}\ \emph {et~al.}(2010)\citenamefont
  {Goriely}, \citenamefont {Chamel},\ and\ \citenamefont
  {Pearson}}]{goriely2010further}%
  \BibitemOpen
  \bibfield  {author} {\bibinfo {author} {\bibfnamefont {S.}~\bibnamefont
  {Goriely}}, \bibinfo {author} {\bibfnamefont {N.}~\bibnamefont {Chamel}},\
  and\ \bibinfo {author} {\bibfnamefont {J.~M.}\ \bibnamefont {Pearson}},\
  }\href {https://doi.org/10.1103/PhysRevC.82.035804} {\bibfield  {journal}
  {\bibinfo  {journal} {Phys. Rev. C}\ }\textbf {\bibinfo {volume} {82}},\
  \bibinfo {pages} {035804} (\bibinfo {year} {2010})},\ \Eprint
  {https://arxiv.org/abs/1009.3840} {arXiv:1009.3840 [nucl-th]} \BibitemShut
  {NoStop}%
\bibitem [{\citenamefont {Marketin}\ \emph {et~al.}(2016)\citenamefont
  {Marketin}, \citenamefont {Huther},\ and\ \citenamefont
  {{Mart{\'i}nez-Pinedo}}}]{marketin2016largescale}%
  \BibitemOpen
  \bibfield  {author} {\bibinfo {author} {\bibfnamefont {T.}~\bibnamefont
  {Marketin}}, \bibinfo {author} {\bibfnamefont {L.}~\bibnamefont {Huther}},\
  and\ \bibinfo {author} {\bibfnamefont {G.}~\bibnamefont
  {{Mart{\'i}nez-Pinedo}}},\ }\href
  {https://doi.org/10.1103/PhysRevC.93.025805} {\bibfield  {journal} {\bibinfo
  {journal} {Phys. Rev. C}\ }\textbf {\bibinfo {volume} {93}},\ \bibinfo
  {pages} {025805} (\bibinfo {year} {2016})},\ \Eprint
  {https://arxiv.org/abs/1507.07442} {arXiv:1507.07442 [nucl-th]} \BibitemShut
  {NoStop}%
\bibitem [{\citenamefont {Woosley}\ \emph {et~al.}(1973)\citenamefont
  {Woosley}, \citenamefont {Arnett},\ and\ \citenamefont
  {Clayton}}]{woosley1973explosive}%
  \BibitemOpen
  \bibfield  {author} {\bibinfo {author} {\bibfnamefont {S.~E.}\ \bibnamefont
  {Woosley}}, \bibinfo {author} {\bibfnamefont {W.~D.}\ \bibnamefont
  {Arnett}},\ and\ \bibinfo {author} {\bibfnamefont {D.~D.}\ \bibnamefont
  {Clayton}},\ }\href {https://doi.org/10.1086/190282} {\bibfield  {journal}
  {\bibinfo  {journal} {The Astrophysical Journal Supplement Series}\ }\textbf
  {\bibinfo {volume} {26}},\ \bibinfo {pages} {231} (\bibinfo {year}
  {1973})}\BibitemShut {NoStop}%
\bibitem [{\citenamefont {Meyer}\ \emph {et~al.}(1996)\citenamefont {Meyer},
  \citenamefont {Krishnan},\ and\ \citenamefont {Clayton}}]{meyer199648ca}%
  \BibitemOpen
  \bibfield  {author} {\bibinfo {author} {\bibfnamefont {B.~S.}\ \bibnamefont
  {Meyer}}, \bibinfo {author} {\bibfnamefont {T.~D.}\ \bibnamefont
  {Krishnan}},\ and\ \bibinfo {author} {\bibfnamefont {D.~D.}\ \bibnamefont
  {Clayton}},\ }\href {https://doi.org/10.1086/177197} {\bibfield  {journal}
  {\bibinfo  {journal} {The Astrophysical Journal}\ }\textbf {\bibinfo {volume}
  {462}},\ \bibinfo {pages} {825} (\bibinfo {year} {1996})}\BibitemShut
  {NoStop}%
\bibitem [{\citenamefont {Meyer}\ \emph {et~al.}(1998)\citenamefont {Meyer},
  \citenamefont {Krishnan},\ and\ \citenamefont {Clayton}}]{meyer1998theory}%
  \BibitemOpen
  \bibfield  {author} {\bibinfo {author} {\bibfnamefont {B.~S.}\ \bibnamefont
  {Meyer}}, \bibinfo {author} {\bibfnamefont {T.~D.}\ \bibnamefont
  {Krishnan}},\ and\ \bibinfo {author} {\bibfnamefont {D.~D.}\ \bibnamefont
  {Clayton}},\ }\href {https://doi.org/10.1086/305562} {\bibfield  {journal}
  {\bibinfo  {journal} {Astrophys. J.}\ }\textbf {\bibinfo {volume} {498}},\
  \bibinfo {pages} {808} (\bibinfo {year} {1998})}\BibitemShut {NoStop}%
\bibitem [{\citenamefont {Perego}\ \emph {et~al.}(2022)\citenamefont {Perego},
  \citenamefont {Vescovi}, \citenamefont {Fiore}, \citenamefont {Chiesa},
  \citenamefont {Vogl}, \citenamefont {Benetti}, \citenamefont {Bernuzzi},
  \citenamefont {Branchesi}, \citenamefont {Cappellaro}, \citenamefont
  {Cristallo}, \citenamefont {Fl{\"{o}}rs}, \citenamefont {Kerzendorf},\ and\
  \citenamefont {Radice}}]{perego2022production}%
  \BibitemOpen
  \bibfield  {author} {\bibinfo {author} {\bibfnamefont {A.}~\bibnamefont
  {Perego}}, \bibinfo {author} {\bibfnamefont {D.}~\bibnamefont {Vescovi}},
  \bibinfo {author} {\bibfnamefont {A.}~\bibnamefont {Fiore}}, \bibinfo
  {author} {\bibfnamefont {L.}~\bibnamefont {Chiesa}}, \bibinfo {author}
  {\bibfnamefont {C.}~\bibnamefont {Vogl}}, \bibinfo {author} {\bibfnamefont
  {S.}~\bibnamefont {Benetti}}, \bibinfo {author} {\bibfnamefont
  {S.}~\bibnamefont {Bernuzzi}}, \bibinfo {author} {\bibfnamefont
  {M.}~\bibnamefont {Branchesi}}, \bibinfo {author} {\bibfnamefont
  {E.}~\bibnamefont {Cappellaro}}, \bibinfo {author} {\bibfnamefont
  {S.}~\bibnamefont {Cristallo}}, \bibinfo {author} {\bibfnamefont
  {A.}~\bibnamefont {Fl{\"{o}}rs}}, \bibinfo {author} {\bibfnamefont {W.~E.}\
  \bibnamefont {Kerzendorf}},\ and\ \bibinfo {author} {\bibfnamefont
  {D.}~\bibnamefont {Radice}},\ }\href
  {https://doi.org/10.3847/1538-4357/ac3751} {\bibfield  {journal} {\bibinfo
  {journal} {Astrophys. J.}\ }\textbf {\bibinfo {volume} {925}},\ \bibinfo
  {pages} {22} (\bibinfo {year} {2022})},\ \Eprint
  {https://arxiv.org/abs/2009.08988} {arXiv:2009.08988} \BibitemShut {NoStop}%
\bibitem [{\citenamefont {Sneppen}\ \emph {et~al.}(2024)\citenamefont
  {Sneppen}, \citenamefont {Just}, \citenamefont {Bauswein}, \citenamefont
  {Damgaard}, \citenamefont {Watson}, \citenamefont {Shingles}, \citenamefont
  {Collins}, \citenamefont {Sim}, \citenamefont {Xiong}, \citenamefont
  {Martinez-Pinedo}, \citenamefont {Soultanis},\ and\ \citenamefont
  {Vijayan}}]{Sneppen2024e}%
  \BibitemOpen
  \bibfield  {author} {\bibinfo {author} {\bibfnamefont {A.}~\bibnamefont
  {Sneppen}}, \bibinfo {author} {\bibfnamefont {O.}~\bibnamefont {Just}},
  \bibinfo {author} {\bibfnamefont {A.}~\bibnamefont {Bauswein}}, \bibinfo
  {author} {\bibfnamefont {R.}~\bibnamefont {Damgaard}}, \bibinfo {author}
  {\bibfnamefont {D.}~\bibnamefont {Watson}}, \bibinfo {author} {\bibfnamefont
  {L.~J.}\ \bibnamefont {Shingles}}, \bibinfo {author} {\bibfnamefont {C.~E.}\
  \bibnamefont {Collins}}, \bibinfo {author} {\bibfnamefont {S.~A.}\
  \bibnamefont {Sim}}, \bibinfo {author} {\bibfnamefont {Z.}~\bibnamefont
  {Xiong}}, \bibinfo {author} {\bibfnamefont {G.}~\bibnamefont
  {Martinez-Pinedo}}, \bibinfo {author} {\bibfnamefont {T.}~\bibnamefont
  {Soultanis}},\ and\ \bibinfo {author} {\bibfnamefont {V.}~\bibnamefont
  {Vijayan}},\ }\href@noop {} {\bibinfo {title} {Helium as an indicator of the
  neutron-star merger remnant lifetime and its potential for equation of state
  constraints}} (\bibinfo {year} {2024})\BibitemShut {NoStop}%
\bibitem [{\citenamefont {{Just}}\ \emph {et~al.}(2022)\citenamefont {{Just}},
  \citenamefont {{Goriely}}, \citenamefont {{Janka}}, \citenamefont
  {{Nagataki}},\ and\ \citenamefont {{Bauswein}}}]{Just2022b}%
  \BibitemOpen
  \bibfield  {author} {\bibinfo {author} {\bibfnamefont {O.}~\bibnamefont
  {{Just}}}, \bibinfo {author} {\bibfnamefont {S.}~\bibnamefont {{Goriely}}},
  \bibinfo {author} {\bibfnamefont {H.~T.}\ \bibnamefont {{Janka}}}, \bibinfo
  {author} {\bibfnamefont {S.}~\bibnamefont {{Nagataki}}},\ and\ \bibinfo
  {author} {\bibfnamefont {A.}~\bibnamefont {{Bauswein}}},\ }\href@noop {}
  {\bibfield  {journal} {\bibinfo  {journal} {{Mon. Not. R. Astron. Soc.}}\
  }\textbf {\bibinfo {volume} {509}},\ \bibinfo {pages} {1377} (\bibinfo {year}
  {2022})}\BibitemShut {NoStop}%
\bibitem [{\citenamefont {Frank}\ \emph {et~al.}(2002)\citenamefont {Frank},
  \citenamefont {King},\ and\ \citenamefont {Raine}}]{Frank2002b}%
  \BibitemOpen
  \bibfield  {author} {\bibinfo {author} {\bibfnamefont {J.}~\bibnamefont
  {Frank}}, \bibinfo {author} {\bibfnamefont {A.~R.}\ \bibnamefont {King}},\
  and\ \bibinfo {author} {\bibfnamefont {D.~J.}\ \bibnamefont {Raine}},\
  }\href@noop {} {\emph {\bibinfo {title} {Accretion power in astrophysics}}},\
  \bibinfo {edition} {3rd}\ ed.\ (\bibinfo  {publisher} {Cambridge University
  Press},\ \bibinfo {address} {Cambridge},\ \bibinfo {year} {2002})\BibitemShut
  {NoStop}%
\bibitem [{\citenamefont {Goriely}(2015)}]{Goriely:2015}%
  \BibitemOpen
  \bibfield  {author} {\bibinfo {author} {\bibfnamefont {S.}~\bibnamefont
  {Goriely}},\ }\href {https://doi.org/10.1140/epja/i2015-15022-3} {\bibfield
  {journal} {\bibinfo  {journal} {Eur. Phys. J. A}\ }\textbf {\bibinfo {volume}
  {51}},\ \bibinfo {pages} {22} (\bibinfo {year} {2015})}\BibitemShut {NoStop}%
\bibitem [{\citenamefont {{Ardevol-Pulpillo}}\ \emph
  {et~al.}(2019)\citenamefont {{Ardevol-Pulpillo}}, \citenamefont {{Janka}},
  \citenamefont {{Just}},\ and\ \citenamefont
  {{Bauswein}}}]{Ardevol-Pulpillo2019a}%
  \BibitemOpen
  \bibfield  {author} {\bibinfo {author} {\bibfnamefont {R.}~\bibnamefont
  {{Ardevol-Pulpillo}}}, \bibinfo {author} {\bibfnamefont {H.-T.}\ \bibnamefont
  {{Janka}}}, \bibinfo {author} {\bibfnamefont {O.}~\bibnamefont {{Just}}},\
  and\ \bibinfo {author} {\bibfnamefont {A.}~\bibnamefont {{Bauswein}}},\
  }\href@noop {} {\bibfield  {journal} {\bibinfo  {journal} {{Mon. Not. R.
  Astron. Soc.}}\ }\textbf {\bibinfo {volume} {485}},\ \bibinfo {pages} {4754}
  (\bibinfo {year} {2019})}\BibitemShut {NoStop}%
\bibitem [{\citenamefont {O'Connor}(2015)}]{oconnor2015open}%
  \BibitemOpen
  \bibfield  {author} {\bibinfo {author} {\bibfnamefont {E.}~\bibnamefont
  {O'Connor}},\ }\href {https://doi.org/10.1088/0067-0049/219/2/24} {\bibfield
  {journal} {\bibinfo  {journal} {Astrophys. Journal, Suppl. Ser.}\ }\textbf
  {\bibinfo {volume} {219}},\ \bibinfo {pages} {24} (\bibinfo {year}
  {2015})}\BibitemShut {NoStop}%
\bibitem [{\citenamefont {{Metzger}}\ \emph {et~al.}(2009)\citenamefont
  {{Metzger}}, \citenamefont {{Piro}},\ and\ \citenamefont
  {{Quataert}}}]{Metzger2009b}%
  \BibitemOpen
  \bibfield  {author} {\bibinfo {author} {\bibfnamefont {B.~D.}\ \bibnamefont
  {{Metzger}}}, \bibinfo {author} {\bibfnamefont {A.~L.}\ \bibnamefont
  {{Piro}}},\ and\ \bibinfo {author} {\bibfnamefont {E.}~\bibnamefont
  {{Quataert}}},\ }\href@noop {} {\bibfield  {journal} {\bibinfo  {journal}
  {{Mon. Not. R. Astron. Soc.}}\ }\textbf {\bibinfo {volume} {396}},\ \bibinfo
  {pages} {304} (\bibinfo {year} {2009})}\BibitemShut {NoStop}%
\bibitem [{\citenamefont {{Neuweiler}}\ \emph {et~al.}(2023)\citenamefont
  {{Neuweiler}}, \citenamefont {{Dietrich}}, \citenamefont {{Bulla}},
  \citenamefont {{Chaurasia}}, \citenamefont {{Rosswog}},\ and\ \citenamefont
  {{Ujevic}}}]{Neuweiler2023a}%
  \BibitemOpen
  \bibfield  {author} {\bibinfo {author} {\bibfnamefont {A.}~\bibnamefont
  {{Neuweiler}}}, \bibinfo {author} {\bibfnamefont {T.}~\bibnamefont
  {{Dietrich}}}, \bibinfo {author} {\bibfnamefont {M.}~\bibnamefont {{Bulla}}},
  \bibinfo {author} {\bibfnamefont {S.~V.}\ \bibnamefont {{Chaurasia}}},
  \bibinfo {author} {\bibfnamefont {S.}~\bibnamefont {{Rosswog}}},\ and\
  \bibinfo {author} {\bibfnamefont {M.}~\bibnamefont {{Ujevic}}},\ }\href@noop
  {} {\bibfield  {journal} {\bibinfo  {journal} {\prd}\ }\textbf {\bibinfo
  {volume} {107}},\ \bibinfo {pages} {023016} (\bibinfo {year}
  {2023})}\BibitemShut {NoStop}%
\bibitem [{\citenamefont {Just}\ \emph {et~al.}(2022)\citenamefont {Just},
  \citenamefont {Kullmann}, \citenamefont {Goriely}, \citenamefont {Bauswein},
  \citenamefont {Janka},\ and\ \citenamefont {Collins}}]{just2022dynamical}%
  \BibitemOpen
  \bibfield  {author} {\bibinfo {author} {\bibfnamefont {O.}~\bibnamefont
  {Just}}, \bibinfo {author} {\bibfnamefont {I.}~\bibnamefont {Kullmann}},
  \bibinfo {author} {\bibfnamefont {S.}~\bibnamefont {Goriely}}, \bibinfo
  {author} {\bibfnamefont {A.}~\bibnamefont {Bauswein}}, \bibinfo {author}
  {\bibfnamefont {H.~T.}\ \bibnamefont {Janka}},\ and\ \bibinfo {author}
  {\bibfnamefont {C.~E.}\ \bibnamefont {Collins}},\ }\href
  {https://doi.org/10.1093/{Mon. Not. R. Astron. Soc.}/stab3327} {\bibfield
  {journal} {\bibinfo  {journal} {Mon. Not. R. Astron. Soc.}\ }\textbf
  {\bibinfo {volume} {510}},\ \bibinfo {pages} {2820} (\bibinfo {year}
  {2022})},\ \Eprint {https://arxiv.org/abs/2109.14617} {arXiv:2109.14617}
  \BibitemShut {NoStop}%
\bibitem [{\citenamefont {{Barnes}}\ \emph {et~al.}(2016)\citenamefont
  {{Barnes}}, \citenamefont {{Kasen}}, \citenamefont {{Wu}},\ and\
  \citenamefont {{Mart{\'\i}nez-Pinedo}}}]{Barnes2016a}%
  \BibitemOpen
  \bibfield  {author} {\bibinfo {author} {\bibfnamefont {J.}~\bibnamefont
  {{Barnes}}}, \bibinfo {author} {\bibfnamefont {D.}~\bibnamefont {{Kasen}}},
  \bibinfo {author} {\bibfnamefont {M.-R.}\ \bibnamefont {{Wu}}},\ and\
  \bibinfo {author} {\bibfnamefont {G.}~\bibnamefont
  {{Mart{\'\i}nez-Pinedo}}},\ }\href@noop {} {\bibfield  {journal} {\bibinfo
  {journal} {\apj}\ }\textbf {\bibinfo {volume} {829}},\ \bibinfo {pages} {110}
  (\bibinfo {year} {2016})}\BibitemShut {NoStop}%
\bibitem [{\citenamefont {{Dieselhorst}}\ \emph {et~al.}(2021)\citenamefont
  {{Dieselhorst}}, \citenamefont {{Cook}}, \citenamefont {{Bernuzzi}},\ and\
  \citenamefont {{Radice}}}]{Dieselhorst2021c}%
  \BibitemOpen
  \bibfield  {author} {\bibinfo {author} {\bibfnamefont {T.}~\bibnamefont
  {{Dieselhorst}}}, \bibinfo {author} {\bibfnamefont {W.}~\bibnamefont
  {{Cook}}}, \bibinfo {author} {\bibfnamefont {S.}~\bibnamefont {{Bernuzzi}}},\
  and\ \bibinfo {author} {\bibfnamefont {D.}~\bibnamefont {{Radice}}},\
  }\href@noop {} {\bibfield  {journal} {\bibinfo  {journal} {arXiv e-prints}\
  ,\ \bibinfo {pages} {arXiv:2109.02679}} (\bibinfo {year} {2021})}\BibitemShut
  {NoStop}%
\bibitem [{\citenamefont {Just}\ \emph {et~al.}(2025)\citenamefont {Just},
  \citenamefont {Xiong},\ and\ \citenamefont
  {Mart{\'\i}nez-Pinedo}}]{Just2025a}%
  \BibitemOpen
  \bibfield  {author} {\bibinfo {author} {\bibfnamefont {O.}~\bibnamefont
  {Just}}, \bibinfo {author} {\bibfnamefont {Z.}~\bibnamefont {Xiong}},\ and\
  \bibinfo {author} {\bibfnamefont {G.}~\bibnamefont {Mart{\'\i}nez-Pinedo}},\
  }\href {https://doi.org/10.5281/zenodo.15864447} {\bibinfo {title} {Rhine:
  R-process heating implementation in hydrodynamic simulations with neural
  networks}},\ \bibinfo {howpublished}
  {https://doi.org/10.5281/zenodo.15864447} (\bibinfo {year}
  {2025})\BibitemShut {NoStop}%
\bibitem [{\citenamefont {{Just}}\ \emph {et~al.}(2025)\citenamefont {{Just}},
  \citenamefont {{Xiong}},\ and\ \citenamefont
  {{Mart{\'\i}nez-Pinedo}}}]{rhine_git}%
  \BibitemOpen
  \bibfield  {author} {\bibinfo {author} {\bibfnamefont {O.}~\bibnamefont
  {{Just}}}, \bibinfo {author} {\bibfnamefont {Z.}~\bibnamefont {{Xiong}}},\
  and\ \bibinfo {author} {\bibfnamefont {G.}~\bibnamefont
  {{Mart{\'\i}nez-Pinedo}}},\ }\href {https://git.gsi.de/nucastro_public/rhine}
  {\bibinfo {title} {{RHINE}}},\ \bibinfo {howpublished}
  {\url{https://git.gsi.de/nucastro_public/rhine}} (\bibinfo {year}
  {2025})\BibitemShut {NoStop}%
\bibitem [{\citenamefont {van~der Walt}\ \emph {et~al.}(2011)\citenamefont
  {van~der Walt}, \citenamefont {Colbert},\ and\ \citenamefont
  {Varoquaux}}]{numpy}%
  \BibitemOpen
  \bibfield  {author} {\bibinfo {author} {\bibfnamefont {S.}~\bibnamefont
  {van~der Walt}}, \bibinfo {author} {\bibfnamefont {S.~C.}\ \bibnamefont
  {Colbert}},\ and\ \bibinfo {author} {\bibfnamefont {G.}~\bibnamefont
  {Varoquaux}},\ }\href {https://doi.org/10.1109/MCSE.2011.37} {\bibfield
  {journal} {\bibinfo  {journal} {Computing in Science \& Engineering}\
  }\textbf {\bibinfo {volume} {13}},\ \bibinfo {pages} {22} (\bibinfo {year}
  {2011})}\BibitemShut {NoStop}%
\bibitem [{\citenamefont {Hunter}(2007)}]{matplotlib}%
  \BibitemOpen
  \bibfield  {author} {\bibinfo {author} {\bibfnamefont {J.~D.}\ \bibnamefont
  {Hunter}},\ }\href {https://doi.org/10.1109/MCSE.2007.55} {\bibfield
  {journal} {\bibinfo  {journal} {Computing in Science \& Engineering}\
  }\textbf {\bibinfo {volume} {9}},\ \bibinfo {pages} {90} (\bibinfo {year}
  {2007})}\BibitemShut {NoStop}%
\bibitem [{\citenamefont {Virtanen}\ \emph {et~al.}(2020)\citenamefont
  {Virtanen}, \citenamefont {Gommers}, \citenamefont {Oliphant}, \citenamefont
  {Haberland}, \citenamefont {Reddy}, \citenamefont {Cournapeau}, \citenamefont
  {Burovski}, \citenamefont {Peterson}, \citenamefont {Weckesser},
  \citenamefont {Bright}, \citenamefont {{van der Walt}}, \citenamefont
  {Brett}, \citenamefont {Wilson}, \citenamefont {Millman}, \citenamefont
  {Mayorov}, \citenamefont {Nelson} \emph {et~al.}}]{scipy}%
  \BibitemOpen
  \bibfield  {author} {\bibinfo {author} {\bibfnamefont {P.}~\bibnamefont
  {Virtanen}}, \bibinfo {author} {\bibfnamefont {R.}~\bibnamefont {Gommers}},
  \bibinfo {author} {\bibfnamefont {T.~E.}\ \bibnamefont {Oliphant}}, \bibinfo
  {author} {\bibfnamefont {M.}~\bibnamefont {Haberland}}, \bibinfo {author}
  {\bibfnamefont {T.}~\bibnamefont {Reddy}}, \bibinfo {author} {\bibfnamefont
  {D.}~\bibnamefont {Cournapeau}}, \bibinfo {author} {\bibfnamefont
  {E.}~\bibnamefont {Burovski}}, \bibinfo {author} {\bibfnamefont
  {P.}~\bibnamefont {Peterson}}, \bibinfo {author} {\bibfnamefont
  {W.}~\bibnamefont {Weckesser}}, \bibinfo {author} {\bibfnamefont
  {J.}~\bibnamefont {Bright}}, \bibinfo {author} {\bibfnamefont {S.~J.}\
  \bibnamefont {{van der Walt}}}, \bibinfo {author} {\bibfnamefont
  {M.}~\bibnamefont {Brett}}, \bibinfo {author} {\bibfnamefont
  {J.}~\bibnamefont {Wilson}}, \bibinfo {author} {\bibfnamefont {K.~J.}\
  \bibnamefont {Millman}}, \bibinfo {author} {\bibfnamefont {N.}~\bibnamefont
  {Mayorov}}, \bibinfo {author} {\bibfnamefont {A.~R.~J.}\ \bibnamefont
  {Nelson}}, \emph {et~al.},\ }\href
  {https://doi.org/10.1038/s41592-019-0686-2} {\bibfield  {journal} {\bibinfo
  {journal} {Nature Methods}\ }\textbf {\bibinfo {volume} {17}},\ \bibinfo
  {pages} {261} (\bibinfo {year} {2020})}\BibitemShut {NoStop}%
\bibitem [{\citenamefont {{W}es {M}c{K}inney}(2010)}]{pandas}%
  \BibitemOpen
  \bibfield  {author} {\bibinfo {author} {\bibnamefont {{W}es {M}c{K}inney}},\
  }in\ \href {https://doi.org/10.25080/Majora-92bf1922-00a} {\emph {\bibinfo
  {booktitle} {{P}roceedings of the 9th {P}ython in {S}cience {C}onference}}},\
  \bibinfo {editor} {edited by\ \bibinfo {editor} {\bibnamefont {{S}t\'efan
  van~der {W}alt}}\ and\ \bibinfo {editor} {\bibnamefont {{J}arrod
  {M}illman}}}\ (\bibinfo {year} {2010})\ pp.\ \bibinfo {pages}
  {56--61}\BibitemShut {NoStop}%
\bibitem [{\citenamefont {Paszke}\ \emph {et~al.}(2019)\citenamefont {Paszke},
  \citenamefont {Gross}, \citenamefont {Massa}, \citenamefont {Lerer},
  \citenamefont {Bradbury}, \citenamefont {Chanan}, \citenamefont {Killeen},
  \citenamefont {Lin}, \citenamefont {Gimelshein}, \citenamefont {Antiga},
  \citenamefont {Desmaison}, \citenamefont {Kopf}, \citenamefont {Yang},
  \citenamefont {DeVito}, \citenamefont {Raison}, \citenamefont {Tejani},
  \citenamefont {Chilamkurthy}, \citenamefont {Steiner}, \citenamefont {Fang},
  \citenamefont {Bai},\ and\ \citenamefont {Chintala}}]{pytorch}%
  \BibitemOpen
  \bibfield  {author} {\bibinfo {author} {\bibfnamefont {A.}~\bibnamefont
  {Paszke}}, \bibinfo {author} {\bibfnamefont {S.}~\bibnamefont {Gross}},
  \bibinfo {author} {\bibfnamefont {F.}~\bibnamefont {Massa}}, \bibinfo
  {author} {\bibfnamefont {A.}~\bibnamefont {Lerer}}, \bibinfo {author}
  {\bibfnamefont {J.}~\bibnamefont {Bradbury}}, \bibinfo {author}
  {\bibfnamefont {G.}~\bibnamefont {Chanan}}, \bibinfo {author} {\bibfnamefont
  {T.}~\bibnamefont {Killeen}}, \bibinfo {author} {\bibfnamefont
  {Z.}~\bibnamefont {Lin}}, \bibinfo {author} {\bibfnamefont {N.}~\bibnamefont
  {Gimelshein}}, \bibinfo {author} {\bibfnamefont {L.}~\bibnamefont {Antiga}},
  \bibinfo {author} {\bibfnamefont {A.}~\bibnamefont {Desmaison}}, \bibinfo
  {author} {\bibfnamefont {A.}~\bibnamefont {Kopf}}, \bibinfo {author}
  {\bibfnamefont {E.}~\bibnamefont {Yang}}, \bibinfo {author} {\bibfnamefont
  {Z.}~\bibnamefont {DeVito}}, \bibinfo {author} {\bibfnamefont
  {M.}~\bibnamefont {Raison}}, \bibinfo {author} {\bibfnamefont
  {A.}~\bibnamefont {Tejani}}, \bibinfo {author} {\bibfnamefont
  {S.}~\bibnamefont {Chilamkurthy}}, \bibinfo {author} {\bibfnamefont
  {B.}~\bibnamefont {Steiner}}, \bibinfo {author} {\bibfnamefont
  {L.}~\bibnamefont {Fang}}, \bibinfo {author} {\bibfnamefont {J.}~\bibnamefont
  {Bai}},\ and\ \bibinfo {author} {\bibfnamefont {S.}~\bibnamefont
  {Chintala}},\ }in\ \href
  {http://papers.neurips.cc/paper/9015-pytorch-an-imperative-style-high-performance-deep-learning-library.pdf}
  {\emph {\bibinfo {booktitle} {Advances in Neural Information Processing
  Systems 32}}}\ (\bibinfo  {publisher} {Curran Associates, Inc.},\ \bibinfo
  {year} {2019})\ pp.\ \bibinfo {pages} {8024--8035}\BibitemShut {NoStop}%
\bibitem [{\citenamefont {{Plewa}}\ and\ \citenamefont
  {{M{\"u}ller}}(1999)}]{Plewa1999}%
  \BibitemOpen
  \bibfield  {author} {\bibinfo {author} {\bibfnamefont {T.}~\bibnamefont
  {{Plewa}}}\ and\ \bibinfo {author} {\bibfnamefont {E.}~\bibnamefont
  {{M{\"u}ller}}},\ }\href@noop {} {\bibfield  {journal} {\bibinfo  {journal}
  {aap}\ }\textbf {\bibinfo {volume} {342}},\ \bibinfo {pages} {179} (\bibinfo
  {year} {1999})}\BibitemShut {NoStop}%
\bibitem [{\citenamefont {{Thompson}}\ \emph {et~al.}(2001)\citenamefont
  {{Thompson}}, \citenamefont {{Burrows}},\ and\ \citenamefont
  {{Meyer}}}]{Thompson2001a}%
  \BibitemOpen
  \bibfield  {author} {\bibinfo {author} {\bibfnamefont {T.~A.}\ \bibnamefont
  {{Thompson}}}, \bibinfo {author} {\bibfnamefont {A.}~\bibnamefont
  {{Burrows}}},\ and\ \bibinfo {author} {\bibfnamefont {B.~S.}\ \bibnamefont
  {{Meyer}}},\ }\href@noop {} {\bibfield  {journal} {\bibinfo  {journal}
  {\apj}\ }\textbf {\bibinfo {volume} {562}},\ \bibinfo {pages} {887} (\bibinfo
  {year} {2001})}\BibitemShut {NoStop}%
\end{thebibliography}

%

\appendix

\section{Supplementary information about the ML models}\label{sec:addit-inform-about}

This appendix provides more details about the selection and preparation of the data used to build the ML models, the procedure of training and testing the ML models, and the methods used to ensure physical consistency of the new states implied by the source terms.

\begingroup
\begin{table*}[t]
\begin{ruledtabular}
\centering
\caption{\label{tab:parameters} Additional properties of the same ML models listed in Table~\ref{tab:mlmodels} providing the output and physical unit, number of perceptrons per hidden layer, the criteria used to pre-select the data points used for training and testing the ML models, the number of data points used for training and testing each ML model, the root-mean squared error (RMSE; in the same physical units as the output quantities) obtained from the testing data subset, and the accuracy of the three classification models for predicting true positives and true negatives.}
\begin{tabular}{clclrcc}
ML model & output $O$                     & $N_{\rm perc}$ & pre-selection criteria                                                             & dataset size & $\sqrt{\mathcal L}_{\rm test}$ & accuracy \\\hline
 1       & $\tilm$ [\MeVc]                  & 50             & $T_9\in(0.03,8)$                                                                   & 236,331      & 0.0076                         &          \\
 2       & $\lg Y_n^\Q$                   & 50             & $T_9\in(3,8)$ mixed with NSE data                                                  & 173,623      & 0.120                          &          \\
 3       & $\lg Y_p^\Q$                   & 50             & \dittostraight                                                                     & 173,623      & 0.156                          &          \\
 4       & $\lg Y_\alpha^\Q$              & 50             & \dittostraight                                                                     & 173,623      & 0.72                           &          \\
 5       & $\lg Y_h^\Q$                   & 50             & \dittostraight                                                                     & 173,623      & 0.185                          &          \\
 6       & $A_h^\Q$                       & 50             & \dittostraight                                                                     & 173,623      & 4.9                            &          \\
 7       & $z_h^\Q$                       & 50             & \dittostraight                                                                     & 173,623      & 0.0021                         &          \\
 8       & $f_\nu$                        & 50             & $T_9\in(0.03,6)$, $f_{\dot q_\nu}\in(0,0.5)$                                       & 191,577      & 0.039                          &          \\
 9       & $\lg \dot{Y}_e$ [s$^{-1}$]     & 50             & $T_9\in(0.03,6)$, $\dot Y_e > 10^{-5}\,\0{s}^{-1}$                                 & 174,033      & 0.21                           &          \\
 10      & $\lg (-\dot{Y}_n)$  [s$^{-1}$] & 50             & $T_9\in(0.03,6)$, $Y_e<0.45$, $\dot\rho<0$, $\dot Y_n<0$, $Y_n>10^{-10}$           & 142,052      & 0.188                          &          \\
 11      & $\dot{Y}_h>0$?                 & 30             & $T_9\in(0.03,6)$, $Y_e<0.45$, $\dot\rho<0$                                         & 160,558      & 0.162                          & 97\%     \\
 12      & $\lg\dot{Y}_h$  [s$^{-1}$]     & 50             & $T_9\in(0.03,6)$, $Y_e<0.45$, $\dot\rho<0$, $\dot Y_\0h/Y_\0h>10^{-6}\,\0{s}^{-1}$ & 30,973       & 0.29                           &          \\
 13      & $\dot{A}_h\neq 0$?             & 30             & $T_9\in(0.03,6)$, $Y_e<0.45$, $\dot\rho<0$                                         & 160,558      & 0.090                          & 99\%     \\
 14      & $\dot{A}_h>0$?                 & 30             & $T_9\in(0.03,6)$, $Y_e<0.45$, $\dot\rho<0$                                         & 160,558      & 0.158                          & 97\%     \\
 15      & $\lg\dot{A}_h$  [s$^{-1}$]     & 50             & $T_9\in(0.03,6)$, $Y_e<0.45$, $\dot\rho<0$, $\dot A_\0h>10^{-4}\,\0{s}^{-1}$       & 125,537      & 0.173                          &          \\
 16      & $\lg(-\dot{A}_h)$  [s$^{-1}$]  & 50             & $T_9\in(0.03,6)$, $Y_e<0.45$, $\dot\rho<0$, $-\dot A_\0h>10^{-4}\,\0{s}^{-1}$      & 7,502        & 0.44                           &          \\\hline
\end{tabular}
\end{ruledtabular}
\end{table*}
\endgroup

\subsection{Data selection and preparation}\label{sec:data-select-prep}

Our ML models have been built using post-processed trajectories from previously published NSM simulations, namely from the dynamical ejecta explored in \citet{collins2023radiative} and from the post-merger ejecta (i.e. all except the dynamical ejecta) of models sym-n1-a6 and asy-n1-a6 of \citet{just2023end}. From the time series of the original nuclear network calculations along each trajectory we extract for each time step the combinations of input and output quantities needed for the various ML models (see Table~\ref{tab:mlmodels}) keeping roughly the original time resolution in order to ensure that steep transitions in the composition are properly resolved. The rates of change of $\rho,Y_e,Y_n$~and~$A_h$ needed for some ML models in the dynamic regime are computed from linear fits of three consecutive points around a given time step. In total, about 240,000 time steps between temperatures of 0.03\,GK and 8\,GK from $\sim$5,000 trajectories of full nuclear-network calculations are used for training and testing all ML models. Since each ML model is only applied under certain conditions, not all data points are used for all ML models. See Table~\ref{tab:mlmodels} for the conditions that each ML model is used for during the simulation and Table~\ref{tab:parameters} for the pre-selection conditions used to trim the datasets and the final number of data points that are used for training and testing each ML model.

In order to ensure a smooth transition from QSE to NSE around $T_{\rm NSE}= 7\,$GK we admix an additional set of $\sim$60,000 data points to the $\sim$120,000 original data points between $3\,$GK and $8\,$GK for the training and testing of ML models 2{--}7 associated with the QSE regime. This additional data set provides the NSE composition and consists of data uniformly sampled in the following ranges: $T\in (7\,\mathrm{GK},10\,\mathrm{GK})$, $Y_e\in(0.05,0.6)$, and $\lg[(k_B T)^3 m_u/\rho/(\hbar c)^3]\in(-1,2)$.

After acquiring and trimming the datasets, they are further modified in order to avoid training dependencies that we are not interested in, e.g. for very low abundances or rates of change. For instance, we limit $Y_i$ from below to $10^{-10}$, $\lambda$ to $10^{-3}\,$s$^{-1}$, and $|\dot{Y}_n|$ to $10^{-4}\,$s$^{-1}$. 

\subsection{Training and testing}\label{sec:training-testing}

After the data preparation, each dataset is randomized and split into two subsets, one set containing $80\,\%$ of the original data points that is used for training the ML models and the remaining $20\,\%$ for the subsequent testing that is done to ensure that the ML models properly generalize to unseen data.

For training each ML model (i.e. finding optimal weights $w_n$ and biases $b$ for each perceptron that minimize the loss function, Eq.~(\ref{eq:lossfunc})) and tuning its hyper-parameters (such as the number of hidden layers, $N_{\rm hid}$, or perceptrons per layer, $N_{\rm perc}$) we use the \textsc{PyTorch} library and the ``adaptive moment estimation'' (ADAM) optimizer. The training is carried out with a constant batch size of 1000 with an initial learning rate of $10^{-3}$ that is gradually reduced to $10^{-4}$ during the training process. The training continues until the loss function saturates, which for our ML models happens after about 2,000{--}10,000 training epochs. The training of each ML model consumed about $\sim \mathcal{O}(100)$ CPU core hours.

After training, the testing subsets are used to evaluate the performance of each ML model. The square roots of the loss functions (equivalent to the root-mean squared errors, RMSE) obtained in this step are listed in Table~\ref{tab:parameters}. Overall good convergence is achieved for the trained ML models.
The best performance is obtained by the ML models predicting $\tilm$ and $z_h$, of which the RMSEs are of the order of $1\,\%$ and less of typical values attained for these quantities ($|\tilm|\sim 1$\,\MeVc and $z_h\sim 0.5$, respectively). For other ML models predicting logarithmic rates of change and abundances, the RMSEs are in many cases around 0.2 or better, corresponding to relative errors of the original quantities of $\approx 1.6$ or better. Although errors of such size are not unexpected considering that a large range of thermodynamic conditions needs to be covered by the ML models, they are not necessarily representative of the final performance of RHINE, because a significant fraction of the sampled conditions (especially for very low rates of change) may barely affect the main part of the transition of $\tilm$ from high to low values. This interpretation is supported by the overall very good agreement with full nuclear network calculations observed in our application tests (cf. Sect.~\ref{sec:applications}). For the three classification models 11,~13,~and~14, Table~\ref{tab:parameters} also provides the accuracy evaluated as the ratio of correct predictions, i.e. true positives and true negatives, over all predictions.

A relatively large RMSE of $\sim 0.7$ is found for $\lg Y_\alpha^{\rm QSE}$. This is related to the sensitive behavior of the $\alpha$-particle freeze-out, which can only poorly be resolved with the degrees of freedom of our ML models (i.e. the four-species treatment and the simplified assumption that the abundances are instantaneously determined by $\rho,T,Y_e$ in QSE). However, the error is most serious only for small, and thus less relevant, values of $Y_\alpha$. In order to verify this tendency, we evaluated the same trained ML model based on a subset of test data with $Y_\alpha<10^{-3}$ and found a higher RMSE of 1.05, while the RMSE decreases to 0.45 for the remaining subset with $Y_\alpha>10^{-3}$, and it even reduces to 0.26 for $Y_\alpha>10^{-2}$, which is close to the performance of the other neural networks.

\subsection{Ensuring physical consistency of the source terms}\label{sec:ensur-phys-cons}

The rates of change predicted from ML models in the QSE and dynamical regimes do not necessarily result in a new composition after the integration step that satisfies physical bounds and consistency relations, such as mass balance and charge balance (cf. Eqs.~(\ref{eq:mc})). This is because each rate is predicted individually and, since each ML model is carrying some intrinsic fitting error, the final rates may violate physical consistency bounds. Another source of inconsistency is connected to the possibility that in an explicit time-integration scheme the new values can exceed their physical boundaries, e.g. $X_i$ could become negative or greater than unity. Even if the level of violation is typically small, such errors may accumulate during the simulation and lead to unphysical effects or numerical problems. The way chosen here of correcting the rates is not unique and, in fact, not particularly relevant for the final simulation results because of usually very small corrections, nevertheless, we provide the prescriptions used in RHINE for completeness. We note that the corrections below only prevent the RHINE-related source terms from introducing inconsistencies. More sources of composition-related inconsistencies may be hidden in the hydrodynamics solver that RHINE is embedded in, which need to be taken care of separately\footnote{For instance, in a finite-volume scheme mass balance can be ensured by employing the Consistent Multifluid Advection (CMA) scheme by \citet{Plewa1999}, which is used in ALCAR.}.

In order to ensure that the source terms predicted by RHINE do not violate physical consistency, we first compute trial values of the composition variables,
\begin{subequations}\label{eq:trialupd}
\begin{align}
   Y_e' &= Y_e^{\rm old} +\Delta t \dot{Y}_e   \, , \\
   Y_i' &= Y_i^{\rm old} +\Delta t \dot{Y}_i   \, , \\
   A_h' &= A_h^{\rm old} +\Delta t \dot{A}_h   \, , 
\end{align}
\end{subequations}
where the time derivatives on the right-hand sides of Eqs.~(\ref{eq:trialupd}) are given by the prescriptions detailed in Sect.~\ref{sec:infer-source-terms}. The trial values are then modified to fulfill all desired physical bounds and afterwards converted back to new, corrected time derivatives. 

The first step consists of bracketing these values into physically allowed (or reasonable) ranges. To this end, $A_h'$ is limited to lie between 4 and 300, $Y_e'$ between 0 and 1, and $Y_i'$ between 0 and $1/A_i$ for $i=n,p,\alpha,h$.

The next step is to enforce mass and charge balance and depends on the thermodynamic regime that we are in. If we are in the QSE regime, we start with enforcing mass balance by adding to the four mass fractions $X_i'=A_i'Y_i'$ (with $A_i'=1,1,4,A_h'$ for $i=n,p,\alpha,h$) contributions $\Delta X_i'$ so that the resulting mass fractions $X_i''=X_i'+\Delta X_i'$ fulfill
\begin{align}\label{eq:mb1}
   \sum\limits_i X_i'' &= \sum\limits_i (X_i' +  \Delta X_i') = 1   \, .
\end{align}
As $\Delta X_i'$ are not uniquely defined, we make the ansatz that they are proportional to $X_i'^2$ with the motivation being that species with smaller mass fractions should be affected less by this correction step. The required normalization constant is computed from Eq.~(\ref{eq:mb1}). Next, we enforce charge balance for the QSE composition in the sense that $Z_h$ computed from Eq.~(\ref{eq:cb}) should be exactly equal to $Z_h^{\rm QSE}$ predicted by ML model 7 (cf. Table~\ref{tab:mlmodels}). To this end, we add another set of contributions, $\Delta X_i''$, to $X_i''$ and require the resulting $X_i'''=X_i''+\Delta X_i''$ to satisfy both charge balance and mass balance, giving
\begin{subequations}\label{eq:cb1}
\begin{align}
   \sum\limits_i z_i'' \Delta X_i'' &= \Delta Y_e  \, , \\
   \sum\limits_i \Delta X_i'' &= 0   \, ,
\end{align}
\end{subequations}
where $\Delta Y_e= Y_e- \sum_i z_i'' X_i''$ and $z_i''=0,1,0.5,Z_h^{\rm QSE}/A_h'$ for $i=n,p,\alpha,h$. We now make the ansatz $\Delta X_i''\propto \mathrm{sgn}(\Delta Y_e)\mathrm{sgn}(z_i''-Y_e)(z_i''-Y_e)^2X_i''^2$ motivated by the notion that mass fractions whose $z_i''$ is greater than $Y_e$ should increase (decrease) for positive (negative) $\Delta Y_e$ (and vice versa). We furthermore assume that positive $\Delta X_i''$ are normalized differently than negative $\Delta X_i''$. The two normalization constants are then obtained from Eqs.~(\ref{eq:cb1}). From $X_i'''$, we recover the corresponding abundances from $Y_i'''=X_i'''/A_i'$.

If the current point lies in the dynamical regime, where the composition is dominated by neutrons and heavy nuclei, a similar ansatz for $\Delta X_i'$ is made as above in order to enforce mass balance but only for $i=n,h$, while $\Delta X_i'=0$ for $i=p,\alpha$, leading to
\begin{align}\label{eq:mb2}
   X_i'' = X_i'+\Delta X_i' = X_i'+(1-\sum\limits_j X_j')\frac{X_i'^2}{X_n'^2+X_h'^2}  
\end{align}
for $i=n,h$ (while in the sum $j$ is running over all four species). For heavy nuclei, only the mass fraction is assumed to change, i.e. $A_h'' = A_h'+\Delta X_h'/Y_h'$ and $Y_h'' = Y_h'$. Since $Z_h$ is not predicted in the dynamic regime, but computed from Eq.~(\ref{eq:cb}), charge balance does not need to be enforced in this regime. However, what can happen in principle is that $Z_h$ computed from Eq.~(\ref{eq:cb}) attains unreasonably low or large values. To this end, $Z_h$ is limited to the range $0.3A_h<Z_h<0.55A_h$ and additionally, if $A_h>80$, to $Z_h<A_h(0.4+1.5/(A_h-70))$. If any of these conditions is violated and $Z_h$ is reset to the corresponding limiting value, $Y_e$ must be recomputed as well in order to satisfy charge balance (Eq.~(\ref{eq:cb})).

Once the trial values have been corrected through the above measures resulting in new quantities $Y_e^{\rm cor}, Y_i^{\rm cor}, A_h^{\rm cor}$, the time derivatives are recomputed as
\begin{subequations}\label{eq:newder}
\begin{align}
   \dot{Y}_e &= \frac{Y_e^{\rm cor}-Y_e^{\rm old}}{\Delta t}   \, , \\
   \dot{Y}_i &= \frac{Y_i^{\rm cor}-Y_i^{\rm old}}{\Delta t}   \, , \\
   \dot{A}_h &= \frac{A_h^{\rm cor}-A_h^{\rm old}}{\Delta t}   
\end{align}
\end{subequations}
before computing the source terms $R_k$ from Eq.~(\ref{eq:rkrates}).

\section{Steady-state wind solutions}\label{sec:constr-wind-solut}

In order to find the correct inflow boundary conditions corresponding to wind models with a given final outflow velocity, $v_\infty^{\rm noRHINE}$ (cf. Sect.~\ref{sec:spher-symm-winds}), we have to solve the following differential equations for the density, $\rho$, and radial velocity, $v$:
\begin{subequations}\label{eq:windeqder}
  \begin{align}
    \frac{1}{r^2}\partial_r \left(r^2\rho W v\right) &= 0 \, , \label{eq:windeqder_m}\\
    \frac{1}{r^2}\partial_r \left(r^2 \rho h W^2 c^2 v\right) &= -\rho W v \partial_r \phi \, ,
\end{align}
\end{subequations}
which result from the continuity equation and the energy equation under the assumption of stationarity and spherical symmetry. We assume Newtonian gravity with $\phi(r)=- G M_c /r$ and $M_c=2.5\,M_\odot$, a constant entropy per baryon, $s=s_0$, and electron fraction, $Y_e=Y_{e,0}$, throughout the wind, and that NSE holds everywhere. An extensive discussion of the various classes of solutions of wind equations can be found (albeit for Newtonian, not special relativistic, hydrodynamics) in the textbook by \citet{Frank2002b}, and for general relativistic solutions, additionally including neutrino-related source terms, see \citet{Thompson2001a}.

The above Eqs.~\ref{eq:windeqder} can be integrated to give:
\begin{subequations}\label{eq:windeqint}
  \begin{align}
    4\pi r^2\rho c \sqrt{W^2-1} &= \dot{M} \, , \label{eq:windeqint_m}\\
    h W + \frac{\phi}{c^2} &= B \, , \label{eq:windeqint_b} 
  \end{align}
\end{subequations}
where the outflow mass flux, $\dot{M}$, and Bernoulli parameter, $B$, are independent of radius and Eq.~(\ref{eq:windeqder_m}) was used to derive Eq.~(\ref{eq:windeqint_b}). The value of $B$ can be obtained from the asymptotic wind properties at infinity, where $\phi\rightarrow 0$, $W\rightarrow W_\infty = 1/\sqrt{1-v_\infty^2/c^2}$, and $h\rightarrow h_\infty$, with $h_\infty$ given by the EOS in the limit of low temperature and low density. The value of $\dot{M}$ cannot be computed as easily but must be determined iteratively using the condition that the solution must be transonic, i.e. the outflowing material must pass through a sonic point where the fluid velocity, $v$, equals the sound speed, $c_s$. In order to utilize this condition, we rearrange Eq.~(\ref{eq:windeqint_b}) starting by differentiation with respect to $r$ and dividing by $h W$, which leads to
\begin{align}\label{eq:trans1}
    \frac{\partial_r h}{h}+\frac{\partial_r W}{W}+\frac{\partial_r \phi}{hWc^2} = 0 \, .
\end{align}
Using the first law of thermodynamics,
\begin{align}\label{eq:trans2}
    \dd \eint = \frac{\eint+P}{\rho}\dd \rho = h c^2\dd \rho \, ,
\end{align}
for the total internal energy, $\eint=\erest+\etherm$, and the definition of the sound speed,
\begin{align}\label{eq:trans3}
  c_s=c\sqrt{\left(\frac{\dd P}{\dd \eint}\right)_{s,Y_e}} \, ,
\end{align}
the first term in Eq.~(\ref{eq:trans1}) becomes (recalling that $s$ and $Y_e$ are constant in the wind):
\begin{align}\label{eq:transfirst}
  \frac{\partial_r h}{h} = \frac{\partial_r P}{\rho h c^2} = \left(\frac{\dd P}{\dd \eint}\right)_{s,Y_e}\frac{\partial_r \eint}{\rho h c^2} = \frac{c_s^2}{c^2}\frac{\partial_r \rho}{\rho} \, .
\end{align}
The second term in Eq.~(\ref{eq:trans1}) can be modified to give
\begin{align}\label{eq:contrewr}
    \frac{\partial_r W}{W} = -\frac{v^2}{c^2}\left(\frac{\partial_r\rho}{\rho}+\frac{2}{r}\right) \, ,
\end{align}
which follows from the continuity equation, Eq.~(\ref{eq:windeqder_m}), written as
\begin{align}
    \frac{\partial_r \rho}{\rho}+\frac{\partial_r W}{W}+\frac{\partial_r v}{v}+\frac{2}{r} = 0 \, ,
\end{align}
and the identity
\begin{align}\label{eq:vWrel}
    \frac{\partial_r v}{v}=\frac{c^2-v^2}{v^2}\frac{\partial_r W}{W}  \, .
\end{align}
Equation~(\ref{eq:trans1}) now becomes
\begin{align}\label{eq:trans1new}
  \frac{\partial_r\rho}{\rho}(c_s^2-v^2)-\frac{2v^2}{r}+\frac{\partial_r\phi}{hW} = 0 \, ,
\end{align}
which implies that at the sonic point, where $v=c_s$, the following relation holds:
\begin{align}\label{eq:sonicrelation}
    2c_s^2 h W = r\partial_r \phi \, .
\end{align}
Since $W=1/\sqrt{1-c_s^2/c^2}$ at the sonic point, all quantities on the left-hand side of Eq.~(\ref{eq:sonicrelation}) are functions only of $\rho$. For a given $|\dot{M}|$, Eqs.~(\ref{eq:sonicrelation}) and (\ref{eq:windeqint_m}) can be solved for the radius, $r_s$, and density, $\rho_s=\rho(r_s)$, at the sonic point. The corresponding wind velocity at infinity, $|v_\infty|$, then follows from Eq.~(\ref{eq:windeqint_b}):
\begin{align}\label{eq:infty}
  h(\rho_s)W(\rho_s)+\phi(r_s) = h_\infty W_\infty \, .
\end{align}
We vary $|\dot{M}|$ until $|v_\infty|$ is equal to the desired value $|v_\infty^{\rm noRHINE}|$ for each wind model. Once $|\dot{M}|$ is found, Eqs.~(\ref{eq:windeqint}) are used to compute the wind profiles $\rho(r)$ and $v(r)$, while from the two possible solutions at each radius we select the one for which $v<c_s(r_s)$ at $r<r_s$ and $v>c_s(r_s)$ at $r>r_s$.

\section{Impact of training with different nuclear networks}

\begin{figure}
    \centering
    \includegraphics[width=\linewidth]{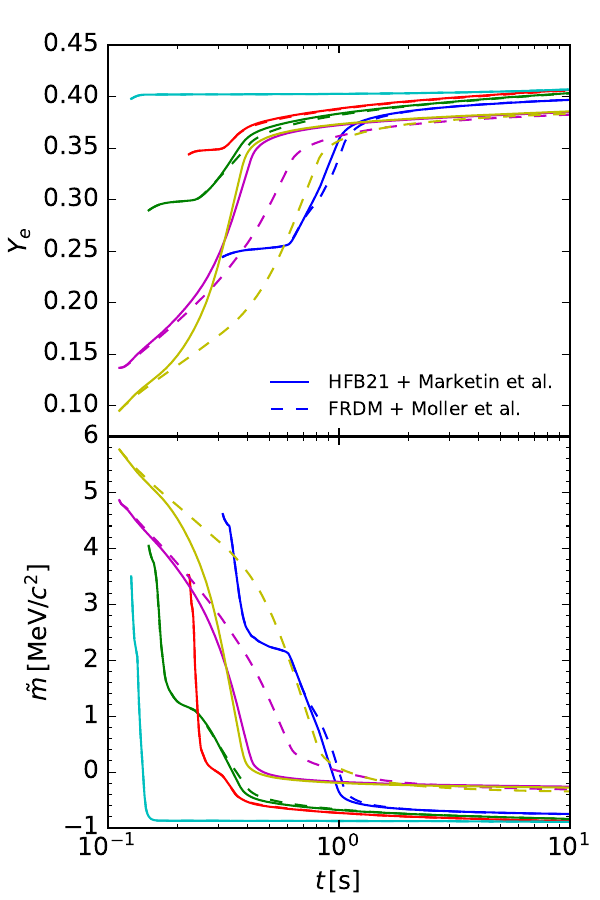}
    \caption{\label{fig:twonucinputs} Comparison of the electron fraction, $Y_e$, and mass excess per baryon, $\tilm$, as functions of simulation time between the two nuclear-physics frameworks used for training our ML models.}
\end{figure}
As mentioned in Sects.~\ref{sec:mult-perc}~and~\ref{sec:data-select-prep}, the training data for our ML models is slightly inconsistent in that two sets of trajectories post-processed with different nuclear-physics input were used. In order to get an idea of the sensitivity of r-process heating to the nuclear-physics input, we compare in Fig.~\ref{fig:twonucinputs} for trajectories with different original $Y_e$ the evolution of $Y_e$ and $\tilm$ resulting in nucleosynthesis calculations adopting these two nuclear-physics frameworks. Most importantly, as can be seen in Fig.~\ref{fig:twonucinputs}, the total amount of r-process energy released (i.e. the difference between the final and starting values of $\tilm$) is nearly insensitive to the adopted nuclear physics model. As for the detailed time evolution, the differences also remain small for trajectories with initially moderate or high $Y_e$. In contrast, for low initial values of $Y_e\lesssim 0.15$, the FRDM+Moller treatment shows a significantly slower evolution towards the corresponding asymptotic values compared to the HFB21+Marketin framework. 

Considering that low initial values of $Y_e$ are reached almost exclusively in the dynamical ejecta, for which the FRDM+Moller framework was used for the training, these results suggests that in any given hydrodynamic simulation (including the ones discussed in this study) the ML models of RHINE predicts heating rates that are in closer agreement to FRDM+Moller than to HFB21+Marketin. However, even if we had used HFB21+Marketin for our entire training data, we do not anticipate a significant change of our main results for the NS-merger models in Sect.~\ref{sec:end-end-merger}. This is because low $Y_e$ values of $0.2$ and less are reached only in the dynamical ejecta, which are so fast already that r-process heating only makes a relatively small impact.

\section{Impact of neglecting fallback in RHINE}

\begin{figure}
    \centering
    \includegraphics[width=\linewidth]{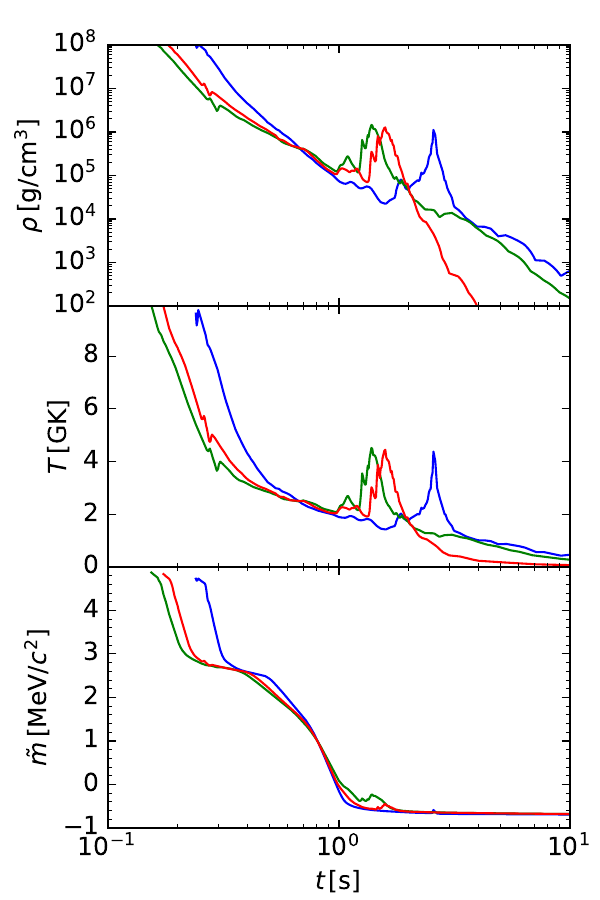}
    \caption{\label{fig:fallback} From top to bottom the density, temperature, and mass excess per baryon for three exemplary trajectories undergoing fallback.}
\end{figure}

As detailed in Sect.~\ref{sec:infer-source-terms}, the rates $\dot{Y}_n, \dot{Y}_h,$~and~$\dot{A}_h$ are assumed to vanish in situations where matter is not expanding, i.e. where $\dot{\rho}\geq 0$. This is of course only an approximation, because generally nuclear reactions continue in such conditions and may lead to compositional changes and a corresponding exchange between thermal and rest-mass energy. Figure~\ref{fig:fallback} illustrates for three selected trajectories which exhibit phases of fallback ($\dot{\rho}> 0$) the evolution of density, temperature, and released r-process heat per baryon. In terms of mass, the relative fraction of outflow trajectories undergoing fallback (at any time after the temperature drops below 5\,GK) in the merger model using RHINE (cf. Sect.~\ref{sec:end-end-merger}) is $35\,\%$, $18.3\,\%$,~and~$79\,\%$ for the dynamical, NS-torus, and BH-torus ejecta, respectively. Thus, a relatively large fraction of the ejecta experience fallback at some point. However, the total amount of energy exchanged during the fallback (computed as trajectory-mass weighted average of the time integral of $\dd \tilm/\dd t$ over periods of $\dot{\rho}>0$) is only $0.0047$,~$0.0035$,~and~$0.046$\,MeV per baryon for the three respective ejecta components. The fact that these numbers are positive means that the nuclear reactions during fallback are on average endothermic, i.e. thermal energy is consumed to create nuclei with a higher mass per baryon. However, irrespective of the sign, the absolute magnitude of these numbers is small enough to consider the impact of fallback as a relatively minor contribution to r-process heating and therefore only a small source of error in RHINE.

\end{document}